\documentclass[numberedappendix,apj]{emulateapj-rtx4}
\def\asca{{\it ASCA\/}}

\def\chandra{{\it Chandra\/}}

\def\heao{{\it HEAO 1\/}}
\def\xmm{{\it XMM-Newton\/}}

\def\ltsima{$\; \buildrel < \over \sim \;$}
\def\simlt{\lower.5ex\hbox{\ltsima}}
\def\gtsima{$\; \buildrel > \over \sim \;$}
\def\simgt{\lower.5ex\hbox{\gtsima}}
\def\kms{\ifmmode{~{\rm km~s^{-1}}}\else{~km s$^{-1}$}\fi}
\def\lsim{\lower0.3em\hbox{$\,\buildrel <\over\sim\,$}}
\def\gsim{\lower0.3em\hbox{$\,\buildrel >\over\sim\,$}}

\def\h2{H$_2$}

\def\arcsec{\mbox{$^{\prime\prime}$}}

\def\asca{{\it ASCA\/}}

\def\chandra{{\it Chandra\/}}

\def\heao1{{\it HEAO-1\/}}

\def\swift{{\it Swift\/}}

\def\xmm{{\it XMM-Newton\/}}

\begin{document}
\title{X-ray properties of the Northern Galactic Cap sources in the 58-month Swift-BAT catalog}
\author{Ranjan V. Vasudevan\altaffilmark{1,*}, William N. Brandt\altaffilmark{2,3}, Richard F. Mushotzky\altaffilmark{1}, Lisa M. Winter\altaffilmark{4}, \\ Wayne H. Baumgartner\altaffilmark{5}, Thomas T. Shimizu\altaffilmark{1}, Donald. P. Schneider\altaffilmark{2,3}, John Nousek\altaffilmark{2}}

\altaffiltext{1}{Department of Astronomy, University of Maryland, College Park, MD}
\altaffiltext{2}{Department of Astronomy \& Astrophysics, The Pennsylvania State University, 525 Davey Lab, University Park PA 16802}
\altaffiltext{3}{Institute for Gravitation and the Cosmos, The Pennsylvania State University, University Park, PA 16802, USA}
\altaffiltext{4}{Atmospheric and Environmental Research, 131 Hartwell Avenue, Lexington, MA}
\altaffiltext{5}{NASA/Goddard Space Flight Center, Astrophysics Science Division, Greenbelt, MD 20771}

\altaffiltext{*}{ranjan@astro.umd.edu}

\submitted{Accepted for publication in ApJ}
%02421-3126 Center for Astrophysics and Space Astronomy, University of Colorado, Boulder, CO

\begin{abstract}

We present a detailed X-ray spectral analysis of a complete sample of hard X-ray selected AGN in the Northern Galactic Cap of the 58-month Swift Burst Alert Telescope (Swift/BAT) catalog, consisting of 100 AGN with $b>50^{\circ}$.  This sky area has excellent potential for further dedicated study due to a wide range of multi-wavelength data that are already available, and we propose it as a low-redshift analog to the `deep field' observations of AGN at higher redshifts (e.g. CDFN/S, COSMOS, Lockman Hole).  We present distributions of luminosity, absorbing column density, and other key quantities for the catalog.  We use a consistent approach to fit new and archival X-ray data gathered from {\xmm}, Swift/XRT, ASCA and Swift/BAT.   We probe to deeper redshifts than the 9-month BAT catalog ($\langle z \rangle = 0.043$ compared to $\langle z \rangle = 0.03$ for the 9-month catalog), and uncover a broader absorbing column density distribution.  The fraction of obscured ($\rm log \thinspace N_{\rm H} \ge 22$) objects in the sample is $\sim 60$\%, and 43--56\% of the sample exhibits `complex' 0.4--10~keV spectra.  

We present the properties of iron lines, soft excesses and ionized absorbers for the subset of objects with sufficient signal-to-noise ratio.  We reinforce previous determinations of the X-ray Baldwin (Iwasawa-Taniguchi) effect for iron K-$\alpha$ lines.  We also identify two distinct populations of sources; one in which a soft excess is well-detected and another where the soft excess is undetected, suggesting that the process responsible for producing the soft excess is not at work in all AGN.   The fraction of Compton-thick sources ($\rm log \thinspace N_{\rm H} > 24.15$) in our sample is $\sim 9$\%.  We find that `hidden/buried AGN', (which may have a geometrically thick torus or emaciated scattering regions) constitute $\sim 14$\% of our sample, including seven objects previously not identified as hidden.   Compton reflection is found to be important in a large fraction of our sample using joint {\xmm}+BAT fits ($\langle R \rangle = 2.7 \pm 0.75$), indicating light bending or extremely complex absorption.  High energy cut-offs generally lie outside the BAT band ($E>200$keV) but are seen in some sources.   We present the average 1--10~keV spectrum for the sample, which reproduces the 1--10~keV X-ray background slope as found for the brighter 9-month BAT AGN sample.  The 2--10~keV log($N$)-log($S$) plot implies completeness down to fluxes $\sim 4$ times fainter than seen in the 9-month catalog.  We emphasize the utility of this Northern Galactic Cap sample for a wide variety of future studies on AGN.

\end{abstract}

\section{Introduction}
\label{sec:intro}

Active galactic nuclei (AGN) are among the most powerful energy sources in the Universe, and their luminous output is due to accretion onto supermassive black holes (`SMBHs', e.g. \citealt{1984ARA&A..22..471R}).  Strong emission from AGN has been observed across the entire spectrum, including at radio, sub-mm, infrared, optical and ultraviolet wavelengths, but an invaluable key to understanding them is provided by X-ray observations.  X-rays are not subject to the heavy host-galaxy dilution present in other bands, and can penetrate through greater amounts of absorbing material in the line of sight than is possible with observations in other wavebands.  This last feature is important in AGN studies because absorbed AGN are thought to constitute a significant proportion of the overall AGN population; it is therefore essential to have as complete a survey of AGN as possible across a range of absorbing column densities before one draws conclusions about the AGN population as whole.

%Intro to AGN

While X-ray surveys of AGN are more penetrating than optical ones (e.g., \citealt{2004ASSL..308...53M}), those AGN with heavy obscuration (with neutral Hydrogen column density $N_{H}>10^{23-24}$) still can fall out of the purview of typical X-ray imaging satellites.  Progressively larger amounts of absorption depress the fluxes at successively higher energies, and eventually the 0.4--10~keV, band explored by observatories such as {\xmm} and \emph{Chandra}, becomes insufficient to identify and constrain absorption levels in highly-absorbed AGN.  Sensitivity at $>10$~keV is required to obtain a more complete AGN census.  The Burst Alert Telescope (BAT, \citealt{2005SSRv..120..143B}) on the \emph{Swift} satellite has proven extremely useful for this purpose, and is producing an all-sky survey of AGN in the 14--195~keV band, the \emph{Swift}/BAT catalog of AGN.  The survey is augmented by detections at increasing depth as the BAT instrument continues to survey the sky.  Source lists and sample properties have been presented for the catalog after the first 9 months of surveying \citep{2008ApJ...681..113T}, 22 months \citep{2010ApJS..186..378T}, 36 months \citep{2009ApJ...699..603A}, 58 months\footnote{http://swift.gsfc.nasa.gov/docs/swift/results/bs58mon/}, 60 months \citep{2012ApJ...749...21A}, and the 70 month catalog source list is now in preparation.

Much work has been done on the 9-month BAT catalog (consisting of 153 sources), such as studies of their X-ray properties (\citealt{2009ApJ...690.1322W}, W09 hereafter), optical properties \citep{2010ApJ...710..503W}, host-galaxy properties \citep{2011ApJ...739...57K} and construction the nuclear AGN spectral energy distributions \citep{2009MNRAS.399.1553V,2010MNRAS.402.1081V}.  The X-ray absorption properties of the AGN in the 36-month catalog have been presented in \cite{2011ApJ...728...58B} (199 sources).   This work draws from the published 58-month catalog which contains 1092 sources, of which $\sim720$ are AGN candidates (i.e., have a counterpart identified as a galaxy, AGN, Seyfert, blazar, or BL Lac, but confirmed to not have a counterpart that is a Galactic black hole binary/neutron star/white dwarf/pulsar).  As the catalog becomes increasingly sensitive, the data present a great opportunity to improve and refine the conclusions drawn from the previous versions of the catalog, in particular the detailed X-ray analysis of W09.

%The catalog continues to expand in depth and number of sources, resulting in the publication of the 58-month catalog online and, recently, a source list for the 60-month catalog \citep{2012ApJ...749...21A}. 

The numbers of AGN in the BAT catalog make it prohibitive to perform pointed observations and analysis for the X-ray properties for all $\sim720$ AGN.  The fraction of AGN with good ({\xmm} quality, at $\gtrsim 4000$ counts) 0.4--10~keV data over the whole sky is much smaller for the 58-month sources than it is for the 9-month sources, requiring a more targeted approach.  The various studies on properties of BAT AGN mentioned above have concentrated on subsamples from the BAT catalog, including a recent, very thorough X-ray spectral analysis of 48 Seyfert 1--1.5 AGN in \citep{2012ApJ...745..107W}.  However, the sources in that paper were selected based on optical type and specifically to understand the prevalence of ionized absorbers, and therefore the average properties of the sample cannot be directly compared with the complete sample analysed in W09 as they miss heavily obscured sources by construction.  Our overall goal here is to update the analysis of the complete 9-month catalog from W09 and the subsequent analysis of the 36-month catalog in \cite{2011ApJ...728...58B} with a representative, unbiased subsample from the 58-month catalog.  We therefore concentrate on a more manageable sample in the Northern Galactic Cap ($b>50^{\circ}$, 4830 $\rm deg^{2}$ or 1.47 steradians, 11\% of the sky); the sample we focus on in this paper has almost exactly the same number of objects as W09's uniform sample (102 objects), so we can manageably perform our analysis to a comparable level of detail as done in W09.  

The aims of our study are threefold: firstly, to obtain the absorbing column density for this complete sample; secondly, to provide a detailed spectral analysis of a `uniform sample' akin to the one identified in W09; and thirdly, to fit the higher quality 8-channel BAT spectra alongside 0.4--10~keV data when they provide extra information on the processes at work in these AGN, not attempted previously in the W09 analysis.  We discuss the importance of each of these goals below.

Understanding the true distribution of column densities in the AGN population has been a question of particular interest in X-ray studies of AGN. Studies of the X-ray background (e.g., \citealt{2007A&A...463...79G}, \citealt{2005ARA&A..43..827B}, \citealt{2005ApJ...630..115T}, \citealt{2005MNRAS.357.1281W}, \citealt{2003MNRAS.339.1095G}) suggest that the majority of accretion in the Universe must be obscured.  If we wish to determine the true, intrinsic AGN power output, a knowledge of the amount of absorbing material is needed.  We aim to determine the true distributions of \hbox{X-ray} absorption and emission properties using a complete and representative subset of the least-biased sample of local AGNs.  We use the best-quality \hbox{X-ray} spectral data available: therefore \emph{XMM-Newton} is employed preferentially if archival data are present (supplemented by 13 new \emph{XMM-Newton} observations taken specifically for this study); other sources of \hbox{X-ray} data are detailed in \S2.   We determine the column density and nature of the \hbox{X-ray} absorption. In W09, complex \hbox{X-ray} absorption was found to be common ($\approx 55$\%) in the 9-month catalog BAT AGNs; we produce new estimates of the covering fraction of such complex absorption where it is present. Quality measurements of the level and nature of \hbox{X-ray} absorption are important for deriving reliable absorption-corrected luminosities over the broadest possible \hbox{X-ray} bandpass; the luminosities presented here will therefore be useful in constructing low-redshift luminosity functions used to assess the amount of accretion and black-hole growth in the local universe. Our absorption measurements can also serve as a useful $z\simlt 0.1$ ``anchor'' for assessments of the dependence of \hbox{X-ray} absorption upon luminosity, Eddington ratio ($L/L_{\rm Edd}$), and redshift.  The 0.4--10~keV spectra are fit jointly with \swift/BAT spectra from \hbox{14--195~keV} to constrain absorption reliably in heavily obscured AGNs \hbox{($N_{\rm H}\simgt 10^{23}$~cm$^{-2}$)}, including those with Compton-thick \hbox{X-ray} absorption \hbox{($N_{\rm H}>1.4\times 10^{24}$~cm$^{-2}$)}.

%The column density distribution can then be used to construct low-redshift luminosity functions for AGN, by allowing us to correct for the effects of absorption.  Such a low-redshift luminosity function will provide an `anchor' for work seeking to link past AGN activity with present AGN activity and SMBH mass build-up [REFs to be added].

Renewing the W09 analysis of detailed spectral features will provide further constraints on the physical processes at work in AGN.  In W09, the authors search for iron K-$\alpha$ lines at or near 6.4~keV, soft excesses (spectral excesses below $\sim1.5$~keV) and signatures of ionized absorbers or winds (edges in the spectrum).  With a knowledge of the abundance of such features and any correlations present (such as the X-ray Baldwin effect linking iron line equivalent width with 2--10~keV intrinsic luminosity (e.g, \citealt{1993ApJ...413L..15I}), we can better understand the importance of processes such as X-ray reflection (e.g., \citealt{2005MNRAS.358..211R}) or complex absorption (e.g., \citealt{2007MNRAS.377L..59D}).

%The material surrounding an AGN should be under the influence of the radiation pressure from the AGN.  Numerous works have sought to understand this influence (King 2003, Fabian et al. 2006, 2008, 2009, Beckmann et al. 2009, Raimundo et al. 2010) and the absorbing gas and dust therefore represents the first interface between the AGN and its host galaxy.  Theories connecting the radiation pressure from the AGN to the shaping of its surroundings are able to, for example, provide an explanation for the known relation between velocity dispersion of the host galaxy and the black hole mass (the $M-\sigma$ relation).  In models such as those of Fabian et al. 2008/2009,

The 14--195~keV BAT data have improved in quality significantly since the W09 study.  Only four energy channels were available in the publicly available BAT spectra for the 9-month catalog, whereas the 8-channel spectra for the 58-month catalog sources can be downloaded easily from the link provided in Footnote 1.  Knowledge of the spectrum across the entire 0.4--200~keV energy range will better constrain the absorbing column density in heavily obscured objects (as done by \citealt{2011ApJ...728...58B}) and, taking our cue from \cite{2012ApJ...745..107W}, we also use the latest BAT spectra to constrain X-ray reflection in a subset of our sources with {\xmm} data (extending the analysis to the absorbed sources in our sample).  One problem that has plagued such analyses in the past is that the BAT spectra are averaged spectra from 58 months of monitoring, whereas the 0.4--10~keV spectra are snapshots taken for periods of typically $\sim2-20$ ks.  We therefore do not have simultaneous data across the whole X-ray band, a significant concern since the X-ray emission from AGN varies rapidly throughout the 0.4--200~keV bandpass \citep{2007A&A...475..827B}, and the cross-normalization between the 0.4--10~keV data and the BAT data for a given observation is therefore not known.  We present a strategy in this paper by which the BAT spectra can be renormalized to be quasi-simultaneous with the 0.4--10~keV observations, using the publicly available BAT light curves.

We also present an analysis of `hidden' or `buried' AGN as done in W09 in order to unearth examples of obscured AGNs with a small fraction of scattered nuclear \hbox{X-rays} (e.g., \citealt{2007ApJ...664L..79U}).  These sources have small levels of scattered \hbox{X-ray} continuum.  Such objects constitute 24\% of the 9-month BAT sources in W09; they may have obscuration subtending most of the sky as seen by the \hbox{X-ray} source, or emaciated scattering regions.
   
%For performing all of these analyses, we restrict ourselves to a well-chosen complete subsample of AGN in the Northern Galactic Cap ($b > 50 ^{\circ}$, 4830 square degrees or 11\% of the sky), 

A higher fraction of good-quality 0.4--10~keV coverage is available for our chosen sky region (augmented by an {\xmm} proposal by PI Brandt to complete the 22-month catalog {\xmm} coverage) than for the whole catalog.  This region of the sky has high value for AGN researchers because it also has extensive multi-band imaging and spectroscopic coverage in the Sloan Digital Sky Survey (SDSS, \citealt{2000AJ....120.1579Y}) and other surveys including WISE (IR, \citealt{2010AJ....140.1868W}), 2MASS (IR\footnote{http://www.ipac.caltech.edu/2mass/}), GALEX (UV\footnote{http://www.galex.caltech.edu/about/overview.html}), and FIRST (radio, \citealt{1995ApJ...450..559B}), providing a ready opportunity to extend this work to construct broad-band spectral energy distributions and to understand the host-galaxy properties for a complete sample.  Using the Northern Galactic Cap also ensures low Galactic absorption and more reliable source identification, due to less potential for confusion with Galactic sources.  Our complete subsample is therefore optimally selected for its potential for future multi-wavelength studies.  
%Additionally, the sample size of $\sim 100$ objects is manageable and a comprehensive analysis can be performed for it.  

We present this subsample as a low-redshift analog to more distant samples, such as the Chandra Deep Fields (CDFS-N/CDFS-S, e.g. \citealt{2012A&A...538A..83F}, \citealt{2012ApJ...752...46L}), the COSMOS field \citep{2007ApJS..172..353B} and the Lockman Hole \citep{2011A&A...529A.135R}.  The multi-wavelength work done on the deep fields has proved very illuminating for studies of AGN, and our chosen sky region has excellent potential for similar wide-ranging multi-wavelength work, as additional observational campaigns are directed at this region of the sky.  The low redshifts offer the distinct advantage of better quality data for most objects; we outline in this work how we can use this sky region to understand AGN accretion comprehensively in the local universe with an unbiased sample, offering the potential to link our findings to those from the deep fields at higher redshifts.

One of the key features of this study is the systematic and uniform way in which we have analysed all of our X-ray spectra.  The methods implemented in our scripts and workflows perform a wide range of useful functions on data from different X-ray detectors.  All data (regardless of detector used) are fit with the same suite of models, and chi-squared comparisons are automatically performed on all of the models fit to determine the best-fit model for a particular source.  If the peculiarities of a particular instrument require a different fitting approach, this can be readily extended due to the modular, object-oriented way in which the workflows have been designed.  The fitting is semi-automated; i.e., models are initially fit to the data and can be inspected and modified if needed, then re-fit; but preliminary `first-look' fits of large numbers of spectra for many sources can also be done in the background without user interaction.  The combined properties of all sources can be readily gathered together in minutes, and the full range of properties, plots, correlations presented in this paper can be readily extended to other subsamples in the BAT catalog or indeed other catalogs.  These tools can potentially be used to produce a complete, consistent X-ray analysis of the entire BAT catalog in the longer term.

In \S\ref{sec:sampledefinition} we outline the sample and data sources used.  In \S\ref{sec:spectralanalysis} we describe the data processing and fitting of the X-ray data in detail.   In \S\ref{sec:results} we present the results from our analysis, including the absorbing column density and luminosity distributions along with analyses of features in the 0.4--10~keV data.  In \S\ref{sec:includingbat} we discuss the utility of joint 0.4--10~keV fits with the 14--195~keV BAT data. In \S\ref{sec:avgspecandlognlogs} we present the average stacked spectrum for the entire sample and the log(N)-log(S) diagram to estimate the sample completeness.  In \S\ref{sec:22monthresultscompare} we compare results for the smaller 22-month catalog source list (with $90\%$ {\xmm} coverage) with our full results for the 58-month catalog ($49\%$ {\xmm} coverage), to 1) to identify whether the higher quality data available for the 22-month subset allows more robust determination of the sample properties, and 2) quantify any differences between the properties of this deeper sample and earlier versions of the catalog.  Finally, in \S\ref{sec:discussionsandconclusions} we summarize our findings and present the conclusions.  We employ the cosmology $H_{\rm 0}=71$ $\rm km \thinspace s^{-1} \thinspace Mpc^{-1}$, $\Omega_{\rm M}=0.27$ (assuming a flat Universe, i.e., $\Omega_{\rm M}+\Omega_{\rm \Lambda} = 1$) throughout.

\section{The Sample and the Data}
\label{sec:sampledatasources}

\subsection{Sample Definition}
\label{sec:sampledefinition}

We employ the 58-month BAT catalog source list as presented online (Footnote 1) as the definitive source list for the whole, all-sky catalog.  The catalog contains 1092 entries of which we determine 720 to be AGN candidates, with a mean redshift of $\langle z \rangle = 0.16$ ($0.045$ excluding blazars and BL Lacs) providing an excellent local AGN survey for our purposes.   The signal-to-noise ratio ($S/N$) is above 4.8 for all objects in the catalog list available; \cite{2010ApJS..186..378T} identify that such a threshold yields 1.54 spurious detections in the entire 22-month catalog (i.e., there should be 1.54 BAT detections due to random fluctuations with $S/N > 4.8$), and this number should hold good for the 58-month catalog also, yielding that the number of a false detections in our region of the sky is less than 0.5, and is therefore negligible.%; scaling this number up to the 58-month catalog yields 3.51 such false matches in this latest version of the catalog. 

 We generate a potential list of targets by filtering the objects in the 58-month catalog to select only those with Galactic latitude $b>50^{\circ}$.  This criterion produces a list of 143 targets in the desired region of the sky, out of which 31 are blazars, BL Lacs, flat spectrum radio quasars (as identified from NED and the BAT catalog counterpart types) and are excluded from further analysis.   
%Scaling the number of false matches to this sky region (4,830 square degrees) yields 0.4 potential false matches in this sky region, which encourages confidence in our choice of sky region (in actuality this is likely to be even lower since that threshold was calculated including the very populated Galactic plane, which we exclude expressly). 
We also exclude 3 galaxy clusters (Coma Cluster, ABELL 1914 and ABELL 2029).  Of the remaining 109 sources, 3 lack counterpart identifications at the time of writing (SWIFT J1138.9+2529, SWIFT J1158.9+4234, SWIFT J1445.6+2702), and we leave them out of the current study in the absence of a reasonable AGN/Galaxy counterpart identification in another waveband.  The remaining 106 sources consist of Seyferts (Type 1 and 2 and intermediate types), quasars/AGN, LINERs, and the more general category of galaxies, including 2 `X-ray Bright Optically Normal Galaxies' (XBONGs).

% and we are able to obtain data for 105 of them.

It is essential to have good-quality X-ray spectral data to determine the absorbing column densities accurately for these AGN, and such data in the soft X-ray regime (0.4--2~keV) are particularly important if we wish to identify any signatures of ionized absorption such as $\rm O \thinspace VII$ and $\rm O \thinspace VIII$ edges at 0.73 and 0.87~keV, or `soft excess' components which often peak at $\sim$0.1--0.4~keV.  At present, a large fraction of the BAT AGN in the most recent 58-month catalog do not have good-quality \emph{XMM-Newton} data available in the archives.  To try to obtain the true underlying $N_{\rm H}$ distribution for local AGN from a relatively complete sample, we have therefore selected an area of the sky for which obtaining {\xmm} follow up to complete the sample is most economical (i.e., requires the fewest number of new observations to produce a complete sample).  We have already discussed the advantages and detailed properties of the chosen Northern Galactic Cap region (Galactic latitude $b > 50^{\circ}$, solid angle 4830 $\rm deg^{2}$) in the previous section.  A successful {\xmm} proposal by PI Brandt has extended the {\xmm} coverage of this region with 13 new observations.  As mentioned above, confusion with Galactic sources is minimized at this high Galactic latitude, and the Galactic neutral hydrogen absorbing column density is also low (0.61--4.08 $\times 10^{20} \rm cm^{-2}$).  

\subsection{Sources of X-ray Spectral Data}
\label{subsec:datasources}

We use the {\xmm} X-ray Science Archive (XSA) to download all the available {\xmm} data for the 106 sources, and find that 49 of these objects have {\xmm} spectra (including the new observations from the {\xmm} proposal). We therefore provide as complete an analysis as possible for the 49 AGN candidates with {\xmm} data, and turn to the \emph{Swift}-X-ray Telescope (XRT) and ASCA archives to provide basic coverage for the remaining sources.  The \emph{Swift}/XRT is of particular utility here since the XRT has been used specifically to target BAT sources, and the BAT team routinely obtains XRT exposures to gather counterpart information on BAT detections.  We prefer \emph{Swift}-XRT over \emph{Chandra} in this study since \emph{Chandra} observations are not available as extensively for BAT sources as \emph{XRT} observations are.  Additionally, for sources with the typical X-ray fluxes seen in this sample, \emph{Chandra} observations would usually exhibit pile-up. We downloaded XRT data for 45 of the remaining sources, and ASCA data for a further six sources.  We find that five sources do not have any data available at the time of writing (LBQS 1344+0233, VCC 1759, NGC 3718, Mrk 653 and 2MASXi J1313489+365358). This constitutes a final list of 100 objects for which we have obtained data from \xmm, \emph{XRT}, or \asca.   The data sources used for each object are provided in Table~\ref{table:observations}.

\section{Spectral Analysis}
\label{sec:spectralanalysis}

\begin{figure}
\centerline{
\includegraphics[width=8.0cm]{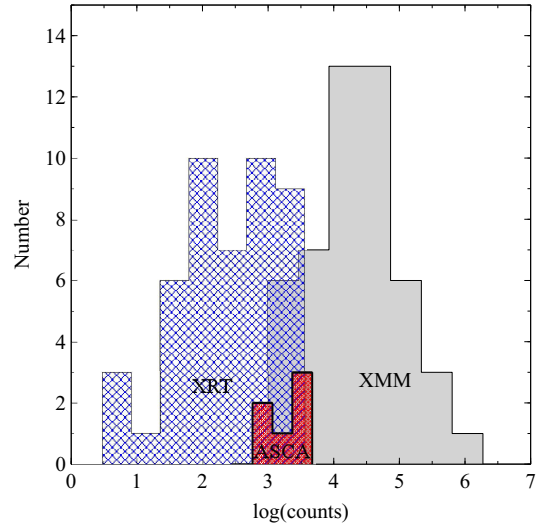}
}
\caption{Histogram of total counts (0.4--10~keV) per observation for the 100 objects in our study.  {\xmm} (49 objects) clearly shows far superior counts statistics compared to \emph{Swift-XRT} (46 objects) and \emph{ASCA} (6 objects).
\label{countshistogram}}
\end{figure}

We present the distribution of total counts per observation (in the 0.4--10~keV band) obtained using different observatories in Fig.~\ref{countshistogram}.  We can see that the {\xmm} data clearly have far better statistics than \emph{XRT} or \emph{ASCA}, but some of the {\xmm} observations at the lower-counts end of the distribution may also not be suitable for a comprehensive analysis of spectral features.  We therefore only assess the significance of iron K-$\alpha$ lines, soft excesses and warm absorber signatures for objects with $>4600$ counts in the observation used (summing over all detectors in the observatory, e.g. {\sc pn} counts + {\sc mos1} counts + {\sc mos2} counts for {\xmm} observations).  This threshold allows, for example, the detection of a 100 eV equivalent width iron line over a $\Gamma = 1.8$ continuum at the 3$\sigma$ level.   For all remaining objects, we perform more basic fits to determine fundamental properties such as luminosity and intrinsic absorbing column density.

\subsection{Data Reduction}

\subsubsection{{\xmm} data}
%XMM data

The {\xmm} data were downloaded from the {\xmm} Science Archive (XSA) and were reduced according to the standard guidelines in the {\xmm} User's Manual\footnote{http://heasarc.nasa.gov/docs/xmm/sas/USG/}, using the {\xmm} Science Analysis Software (\textsc{sas}) version 9.0.0. The tasks \textsc{epchain} and \textsc{emchain} were used to reduce the data from the pn and MOS instruments, respectively.  Initially a circular source region of radius 36{\arcsec} was used to extract a source spectrum, checking for nearby sources in the extraction region and reducing the source region size to exclude them if needed.  Background regions were either chosen to be circles near the source or annuli which exclude the central source. Additionally, the background light curves (between 10--12~keV) were inspected for flaring, and a comparison of source and background light curves in the same energy ranges was used to determine the portions of the observation in which the background was sufficiently low compared to the source; the subsequent spectra were generated from the usable portions of each observation.  The \textsc{sas} tool \textsc{epatplot} was used to determine whether pile-up was present in the observations; if strong pile-up was found, we followed the recommended approach of using annular source regions to excise the piled-up core of the source, and re-calculated spectra and light curves until the strong pile-up was removed. Response matrices and auxiliary files were generated using the tools \textsc{rmfgen} and \textsc{arfgen}, and the final spectra were grouped with a minimum of 20 counts per bin using the \textsc{grppha} tool.
%, unless the usable portion of the observation had very few counts, in which case a lower limit of 10 counts per bin were used.

\subsubsection{XRT data}

We downloaded the \emph{XRT} data from the High-Energy Astrophysics Science Archive (HEASARC)\footnote{http://heasarc.gsfc.nasa.gov/cgi-bin/W3Browse/w3browse.pl} . Pipeline-processed `level 2' FITS files were readily available from HEASARC, ready for further processing.  The pipeline-processed event files from the XRT detector were processed using the XSELECT package, as directed in the \emph{Swift}-XRT user guide\footnote{http://heasarc.nasa.gov/docs/swift/analysis/}. Source regions of 50 {\arcsec} were used, with larger accompanying background regions (average radius ∼150 {\arcsec}), and care was taken to exclude other sources in source and background regions. Background light curves were determined from the event files and inspected for flaring, but this was not found to be a problem in any of our observations. Source and background spectra were extracted, and the source spectra were grouped with a minimum of 20 counts per bin by default, or a lower limit of 10 counts per bin for observations with few counts.  The following objects had their spectra grouped to 10 counts per bin: 2E 1139.7+1040, 2MASX J13105723+0837387, 2MASX J13462846+1922432, B2 1204+34, CGCG 291-028, MCG -01-30-041, MCG +05-28-032, NGC 4939, NGC 5106 and Ark 347.  Some objects have too few counts to construct a spectrum (NGC 4180, NGC 4500, MCG -01-33-063, CGCG 102-048 and 2MASX J13542913+1328068); for these we present basic luminosity estimates and other quantities (including upper limits where appropriate) in Appendix \ref{appendix:upperlimits}.

%Chandra data

\subsubsection{ASCA data}

ASCA spectra (pre-reduced) were downloaded from the \emph{Tartarus} archive\footnote{http://tartarus.gsfc.nasa.gov/} for the sources 3C 303.0, Was 49b, Mrk 202, NGC 4941, Mrk 477 and NGC 4619, for which \emph{Swift}-XRT or {\xmm} data were not found at the time of writing.

The details of all observations used are presented in Table~\ref{table:observations}, including their sky positions, the instruments used to provide the X-ray data, observation dates, counts in the observation and optical types.  The optical types have been gathered from a variety of sources, primarily the NASA/IPAC Extragalactic Database (NED\footnote{http://ned.ipac.caltech.edu/}) and visual inspection of multiwavelength images and spectra; as a result these types are very heterogeneous.  We do not use these types further in our analysis but provide them for completeness, and urge interested readers to check the types before using them in multi-wavelength work.

%\restoregeometry
%\voffset=0.0cm
\subsection{Spectral Fitting}
\label{spectralfitting}

We consistently fit a suite of models to all the 0.4--10~keV spectral data available to determine the best-fitting model in each case, using \textsc{python} and \textsc{tcl} scripting with the \textsc{xspec} package \citep{1996ASPC..101...17A}.  For some XRT observations, or for {\xmm} observations with few counts or with large portions of the observation excluded due to flaring, we impose a lower energy limit greater than 0.4~keV.

By default, we fit only the 0.4--10~keV data in order to concentrate on the accurate determination of soft X-ray feature parameters, but where the counts in the soft band ($<10 \thinspace \rm keV$) are insufficient to obtain a good constraint on $N_{\rm H}$ or to analyse features (we adopt the threshold of 4600 counts for this purpose), we include the BAT data in the fit, extending beyond the W09 analysis.  Inclusion of the BAT data introduces its own complications, particularly due to the variability in the BAT band over the 58 months during which the spectra were constructed; we return to these issues in \S\ref{subsec:batrenormsection} and \S\ref{sec:includingbat}.  For all objects, we perform a comparison between a detailed fit in the 0.4--10~keV band and a broader, `continuum-only' fit to 0.4--200~keV before arriving on a best-fit model.  For objects with sufficient counts  where the BAT data are not included in our final best-fit model, we perform a check to ensure that inclusion of the BAT data does not modify the spectral properties recovered from 0.4--10~keV data alone; surprisingly we find that the addition of the higher energy BAT data do not significantly alter the best-fit parameters found from analysing the lower-energy {\xmm} results alone.  In summary, we find the BAT data are only required to constrain the continuum and absorption in very high column density sources, where there are insufficient counts in the 0.4--10~keV band to constrain the soft X-ray spectral features.

All models include Galactic absorption by default (determined using the \textsc{nh} tool from the \textsc{ftools} suite of utilities, \citealt{1995ASPC...77..367B}); this is by design uniformly low for this sample ($0.61<N_{\rm H}^{\rm Gal}<4.08 \times 10^{20} \thinspace \rm cm^{-2}$).  We follow W09 by classifying AGN spectra into two broad categories: those with `simple' absorbed power-law spectra (with or without features such as an iron line at 6.4~keV, ionized absorption edges, or a soft excess below $\sim$ 2~keV); and `complex' spectra for which partially covering absorbers or double power-law models must be invoked to provide a statistically acceptable fit to the spectrum.  The various model sub-types are presented in Table~\ref{table:modelcombinations}.  We differ slightly from W09's analysis by not employing a model combination that includes both partial covering and a black-body component, as used for a handful of their sources; we use the term `soft excess' in this work to refer uniformly to an excess above a clear power-law in which the spectrum shows low neutral absorption, and do not use the term to refer to the soft features often seen in heavily absorbed spectra, as these are probably due to different physical processes.  We systematically and self-consistently fit all of the model combinations using our semi-automated system which initially fits the data with a model, presents the findings to the user allowing any necessary adjustments or re-seeding/freezing of parameters, and re-fits the data with the refined model.  In this study, we also introduce the category of an `intermediate' model type, where either two models fit the data equally well (e.g., $\chi^{2}/\rm d.o.f < 1.0$ in both cases) or visual inspection of a fitted spectrum revealed that, whilst a partial covering model may fit the data better in a formal sense, the spectrum did not show obvious signatures of strong complexity.  For this class of objects, we present the results for both models in all subsequent tables and figures and indicate them clearly.  Thirteen such intermediate spectral-type objects in our sample highlight the need for better quality data (for cases where both simple or complex absorption models over-fit poorer quality data), or for further investigation of the spectra (in the cases where good-quality data reveals an indeterminate spectral shape).

Additionally, we emphasize that the complement of models used here encompasses the most common characterisations of AGN X-ray spectra, and do not represent an exhaustive list of physical scenarios.  We do not attempt to model all of the more intricate features that may be present, and as a result we do see some poor fits (null hypothesis probabilities $< 5 \times 10^{4}$), primarily due to not modelling complex iron lines and line emission at soft energies in complex-absorption sources.  These cases are briefly discussed in Appendix \ref{appendix:poorfits}. 

%\begin{landscape}
%\rotate

%\end{landscape}
%\restoregeometry

%[Describe basic subclasses: simple and complex as given in Winter et al. - describe most of them in a table]

We fit the models in Table~\ref{table:modelcombinations} to all spectra (omitting model combinations with features such as soft excesses, iron lines, or edges for XRT or ASCA data due to lower signal-to-noise ratio or lower sensitivity below 1~keV), and determine the best-fitting model by ordering the model fits based on their reduced chi-squared values.  Although we only analyze the properties of soft excesses, lines and edges for objects with $>4600$ counts, we fit models including these features to all {\xmm} datasets and later exclude objects below the counts threshold when determining the prevalence and sample-wide properties of such features. We also estimate the significance of components such as lines, edges and soft excesses by requiring a reduction in chi-squared of 4.0 per degree of freedom for a feature to be deemed `significant' (corresponding to a $\approx 95$\% confidence detection of the feature). The basic fit results are presented in Table~\ref{table:fitresults}, and the analysis of detailed features (iron lines, soft excesses, and ionized absorber edges) is presented in Table~\ref{table:fitresults_features}.  Fig~\ref{samplespectra} shows some example spectral fits, showing both the raw spectrum with the ratio of the data to the model ($\rm counts \thinspace s^{-1} \thinspace keV^{-1}$ against energy in keV in the upper panel, ratio of data-to-model against energy in keV in the lower panel) and a $\nu \thinspace F_{\nu}$ spectrum (unfolded through the response, $\rm keV^{2} \times (Photons \thinspace s^{-1} \thinspace cm^{-2} \thinspace keV^{-1})$) for 3 objects.  Table~\ref{table:upperlimits} shows basic results for objects with very few counts, including upper limiting luminosities.  In Table~\ref{table:fitresults_features}, we show upper limiting equivalent widths, soft-excess strengths and warm-absorber edge optical depths when these features produce a reduction in chi-squared, but are not significant according to the $\Delta \chi^{2} > 4.0$ criterion.

\begin{figure}
\centerline{
\includegraphics[width=8.0cm]{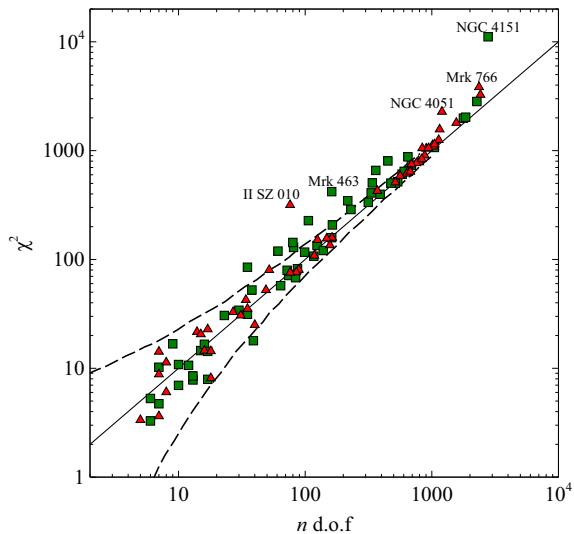}
}
\caption{Plot of $\chi^{2}$ against the number of degrees of freedom ($n \rm \thinspace d.o.f$) for the best-fit models for each source.  Red triangles represent objects with `simple' model spectral types, and green squares represent `complex' ones (as defined in Table~\ref{table:modelcombinations}).  The solid line indicates $\chi^{2}=n \rm \thinspace d.o.f$, and the dashed lines indicate the limiting $\chi^{2}$ for a given $n$ \thinspace d.o.f above or below which we expect less than a 1$\%$ probability of seeing such extremal values of $\chi^{2}$ if the model is correct (see \citealt{2010A&A...510A..35M}). \label{chisquaredvsdof}}
\end{figure}

We also present a plot of $\chi^{2}$ against the number of degrees of freedom ($n \rm \thinspace d.o.f$) for all the fits in our sample in Fig.~\ref{chisquaredvsdof} to illustrate the quality of the fits, as done by \cite{2010A&A...510A..35M} (Fig.~2 of their paper).  The solid line represents 1:1 correspondence between $\chi^{2}$ and $n \rm \thinspace d.o.f$, and the dashed lines represent the extremal values of $\chi^{2}$ above or below which we would expect less than a 1$\%$ probability of obtaining such a $\chi^{2}$ if the model is correct.  We color code the objects by spectral type, with red triangles showing `simple' spectral types and green squares showing `complex' ones.  At low $n \rm \thinspace d.o.f$ (hence low counts), almost all fits lie within these limits, but at higher $n \rm \thinspace d.o.f$, we do see some worse fits (the maximal reduced chi-sqared for the whole sample is $\chi^{2}/n \rm \thinspace d.o.f \approx 4.0$), with a majority of these objects exhibiting `complex' spectral types.  A comparison of our plot with Fig.~2 of \cite{2010A&A...510A..35M} shows that our sample extends to higher $n \rm \thinspace d.o.f$, due to a number of datasets with greater counts than those in their study.  Our plot also shows that a majority of the objects with poor fits have complex spectral types.  In the high counts regime, we encounter spectra of very high quality (with high counts statistics) exhibiting strong spectral complexity that cannot be modeled by the suite of model combinations used here, including notably NGC 4151 and Mrk 766.  Another notable outlier, II SZ 010, whilst having a `simple' spectral type, exhibits a poor fit due to the inclusion of the BAT data in the fit, and very low flux in the BAT band at the time of the \emph{XRT} observation (see \S\ref{subsec:batrenormsection}).  We find that removing the BAT data brings $\chi^{2} / n \thinspace \rm d.o.f$ much closer to 1.0 without altering the key results (photon index, column density) for this object.

\begin{figure}
%\figurenum{1}
\centerline{
\includegraphics[width=8.0cm]{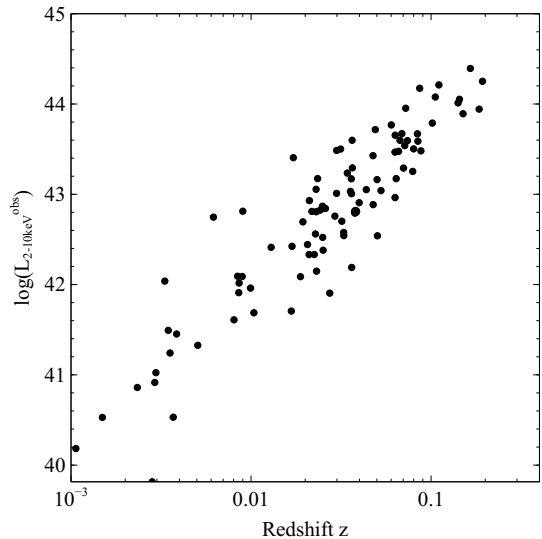}
}
\caption{
\small
Observed 2--10~keV luminosity $L_{\rm 2-10 keV}^{(obs)}$ (not corrected for absorption) against redshift for the 100 sources in our sample.\label{logLX_vs_redshift} }
\end{figure}

\begin{figure*}
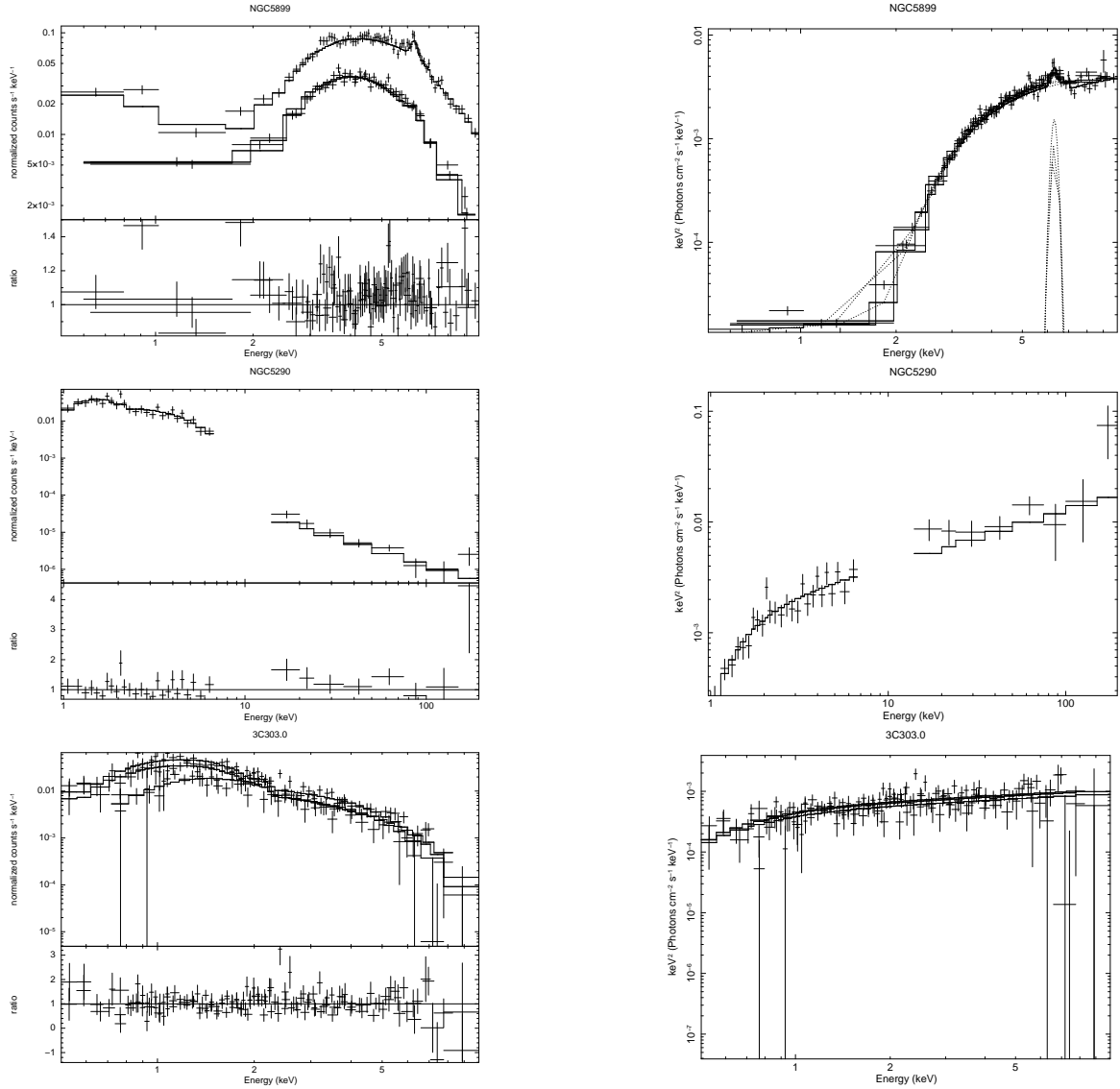

%\centerline{
\begin{minipage}{0.49\textwidth}
\includegraphics[width=5cm,angle=270]{NGC5899_WabsPcfabsPoZgauss_fitresult.txt_xrayspectrumfitRATIO0.5-10.0.ps}
\end{minipage}
\hfill
\begin{minipage}{0.49\textwidth}
\includegraphics[width=5cm,angle=270]{NGC5899_WabsPcfabsPoZgauss_fitresult.txt_xrayspectrumfitEEUF0.5-10.0.ps}
\end{minipage}
\begin{minipage}{0.49\textwidth}
\includegraphics[width=5cm,angle=270]{NGC5290_WabsWabsPowithBAT_fitresult.txt_xrayspectrumfitRATIO0.7-200.0_withBAT.ps}
\end{minipage}
\hfill
\begin{minipage}{0.49\textwidth}
\includegraphics[width=5cm,angle=270]{NGC5290_WabsWabsPowithBAT_fitresult.txt_xrayspectrumfitEEUF0.7-200.0_withBAT.ps}
\end{minipage}
\begin{minipage}{0.49\textwidth}
\includegraphics[width=5cm,angle=270]{3C303.0_WabsWabsPo_fitresult.txt_xrayspectrumfitRATIO0.4-10.0.ps}
\end{minipage}
\hfill
\begin{minipage}{0.49\textwidth}
\includegraphics[width=5cm,angle=270]{3C303.0_WabsWabsPo_fitresult.txt_xrayspectrumfitEEUF0.4-10.0.ps}
\end{minipage}
%}
\caption{
\small
Some example spectra from our sample, with the left panel plots showing the spectrum and ratio plot, and the right panel plots showing the $\nu F_{\nu}$ plots. The objects are, from top-to-bottom, NGC 5899 ({\xmm} data), NGC 5290 (XRT+BAT data) and 3C 303.0 (ASCA data).  \label{samplespectra}}
\end{figure*}

We present the rest-frame 2--10~keV observed luminosity (not corrected for absorption) against redshift in Fig.~\ref{logLX_vs_redshift}, to characterize the redshift distribution of the sample.  All sources are located at redshifts $z<0.2$, with an average of $\langle z \rangle = 0.043$, slightly higher than the average of 0.03 obtained in both W09 for the 9-month BAT catalog and in \cite{2011ApJ...728...58B} for the 36-month catalog.

%We perform two familes of fits to our data; the first set of fits is purely to the 0.4--10~keV data and the second set includes the high-energy BAT data.  In many cases the BAT data can be used to refine the estimate of the absorbing column density.  For the family of fits that include BAT data, we simply employ either a simple (absorbed power law) or complex model (power law with partially covering absorption) to the XMM/XRT/ASCA plus BAT and do not include features such as soft excesses, iron lines or warm absorbers as the key features of interest when including the BAT data are the broad-band spectral shape rather than the lower energy details, which are picked out by the more detailed fits to the 0.4--10~keV data alone.  We contrast the absorbing column densities obtained with and without inclusion of the BAT data, and discuss the merits of this approach in general, in \S\ref{sec:includingbat_nheffect}.  For a handful of objects, due to very few counts in their XRT spectra, we required the BAT data to be included to achieve a successful fit at all. These were 2E 1139.7+1040, B2 1210+33, MCG -01-30-041,MCG +06-24-008, CGCG 291-028 and 2MASX J13105723+0837387.

%We also fit versions of these models with Compton scattering included (using the plcabs model) where the preliminary $N_{\rm H}$ determination using a physically simpler model is high enough to suggest that this might be necessary.  We also present the reflection properties of a subset of these objects from 0.1--200 keV fits in a separate paper (in prep).

\subsection{On the issue of simultaneity across the entire 0.4--200~keV band}
\label{subsec:batrenormsection}

The BAT spectra have been gathered over the entire duration of the survey, and are therefore not in any sense `simultaneous' with the 0.4--10.0~keV data used from {\xmm}, \emph{Swift/XRT} or \emph{ASCA}. It is important to use simultaneous observations wherever possible when combining multi-wavelength data due to the variable nature of AGNs.  This is even more pertinent at high energies, where the short-time scale variability is reflective of rapidly changing accretion processes occurring close to the inner regions of the accretion flow.

Whilst we cannot obtain a BAT spectrum simultaneous with the 0.4--10~keV observation due to insufficient counts in the BAT instrument in such short ($1-100\rm ks$) time intervals, we can attempt to account for this effect in some measure using the BAT light curves.  These are available for each BAT catalog source, spanning the entire 58 months of the survey.  The variability displayed for the BAT AGN in these light curves indicates that the true 14--195~keV spectrum at a particular epoch within the survey may look significantly different to the final averaged 58-month spectrum.   Such variation will be composed of two components: variation in the overall normalization (i.e., flux) and spectral shape.  We aim to account for the former effect in this work. Variability in spectral shape requires particularly good signal-to-noise ratio to paramaterize properly, and a full treatment of this effect will be presented in Shimizu et al. (in prep.).  Their preliminary analysis of hardness ratios for the brightest $\sim 30$ BAT sources reveals minimal spectral variability across 14--195~keV, but this analysis is only possible on the brightest sources on variability timescales greater than 30 days.  Ideally we prefer truly simultaneous data (such as will be obtained with co-ordinated \emph{NuSTAR} and 0.4--10~keV campaigns, or \emph{ASTROSAT}) alongside the 0.4--10~keV data to be able to interpret the broad-band spectral shape fully.

Many of our soft (0.4--10~keV) X-ray observations have been taken within the timeframe of the BAT survey.  Where possible, the BAT light curve is used to estimate the variation in the overall normalization of the BAT spectrum, by considering the BAT flux ratio relative to the full 58-month average, at the date of observation of the soft X-ray data.

\begin{figure}
%\centerline{
\begin{minipage}{0.49\textwidth}
\includegraphics[width=7.0cm]{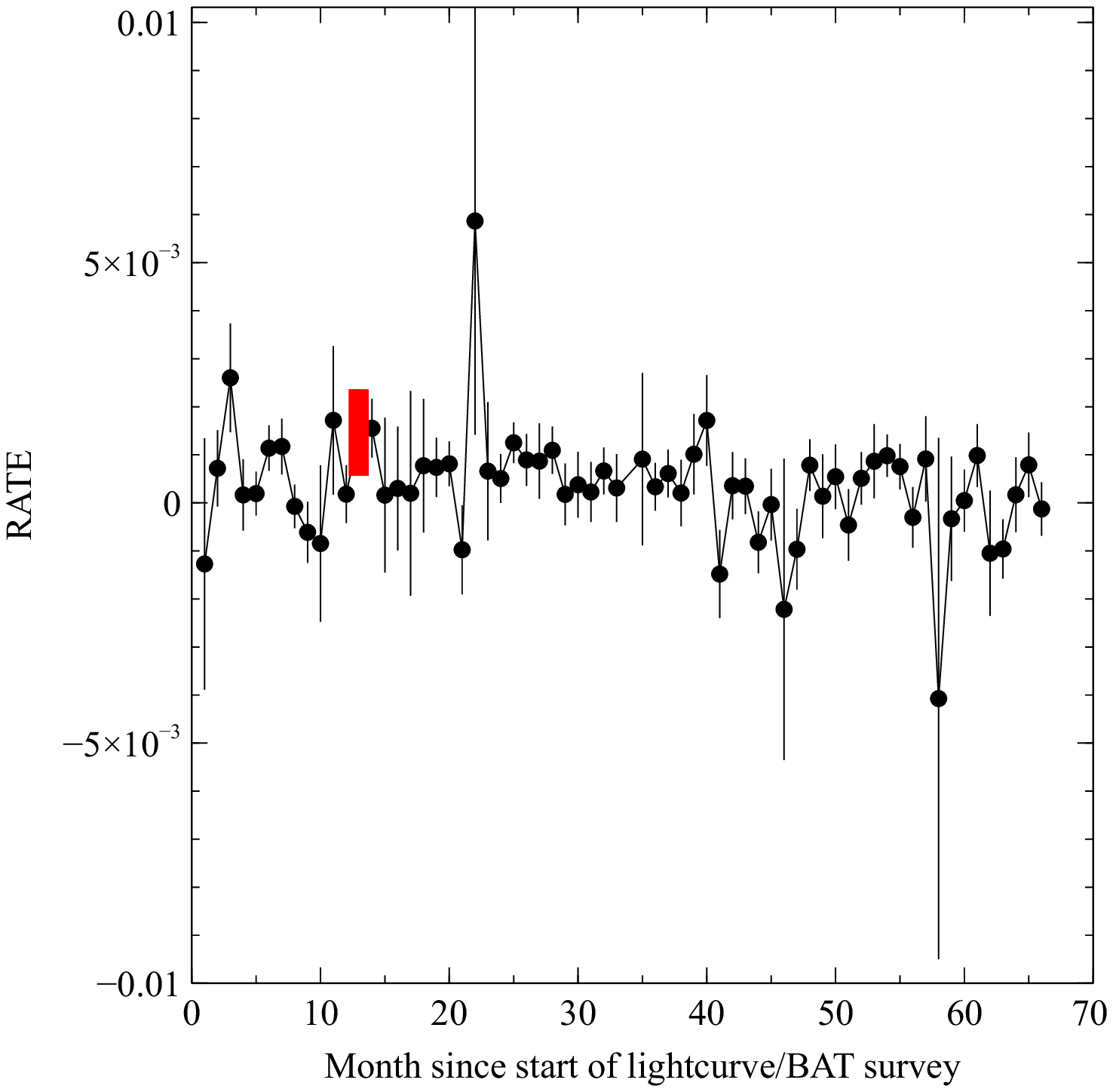}
\end{minipage}
\\
\begin{minipage}{0.49\textwidth}
\includegraphics[width=6.0cm,angle=270]{XRTplusBATMrk1310_norenorm.ps}
\end{minipage}
%\hfill
\begin{minipage}{0.49\textwidth}
\includegraphics[width=6.0cm,angle=270]{XRTplusBATMrk1310_WITHrenorm.ps}
\end{minipage}
%}
\caption{
\small
Illustration of the renormalization of the BAT spectrum using the light curve for Mrk 1310.  The top panel shows the BAT light curve, with a red bar indicating the date of observation of the 0.4--10~keV data.  The 0.4--10~keV data are from XRT in this case.  The middle panel shows the 58-month averaged BAT spectrum used as-is along with XRT data, whereas the bottom panel shows the renormalized BAT spectrum  (thereby made `quasi-simultaneous' with the XRT data).  Prior to re-normalization, the spectrum looks like a continuous unabsorbed power-law across 0.4--200~keV, but after re-normalization, the spectrum has the appearance of strong reflection or complex absorption.\label{xrtplusbat}}
\end{figure}

%Using the \textsc{Python} language, we created a tool to renormalize the BAT spectra according to the light-curve amplitude at the time of interest. The 58-month Crab-weighted monthly light-curves were downloaded from HEASARC (they are available alongside the BAT spectra). The 8 energy channels of the spectrum are summed (weighting each channel appropriately to account for the spectrum of the Crab nebula) to form a Crab-weighted total count rate, $S$. The light-curve amplitude at the time of interest, $L$, is already recorded in Crab units in the publicly available light-curves. The spectrum is then renormalized by multiplying it by $L/S$. Because the errors on the spectrum are much greater than the error in the light-curve amplitude, the \emph{relative} count-rate errors are kept constant when renormalizing the spectrum
%on count rates for the renormalized spectrum. are kept the same as the original ones.
We then re-normalize the BAT spectrum accordingly, whenever the soft X-ray data have been taken within the span of the BAT survey.  We show an example in Fig.~\ref{xrtplusbat}.  The key improvement in this approach is produced when fitting the {\xmm}/XRT/ASCA spectra jointly with the BAT data within \textsc{xspec}: without such renormalization, we would have to allow the normalizations of the {\xmm}/XRT/ASCA and BAT data to `float' with respect to each other because of this uncertainty in the absolute normalization of the BAT spectrum due to non-simultaneity with the 0.4--10~keV data.  However, by re-normalizing, we can lock the normalizations of the BAT and 0.4--10~keV ({\xmm}/XRT/ASCA) components together, removing a degree of freedom from the fit and providing more stringent constraints on the parameters obtained from model fits.  This is particularly useful when performing fits to determine reflection parameters, which we will return to in \S\ref{subsec:reflection}.

%\rotate

\begin{figure*}
%\figurenum{1}
\centerline{
\includegraphics[width=8.0cm]{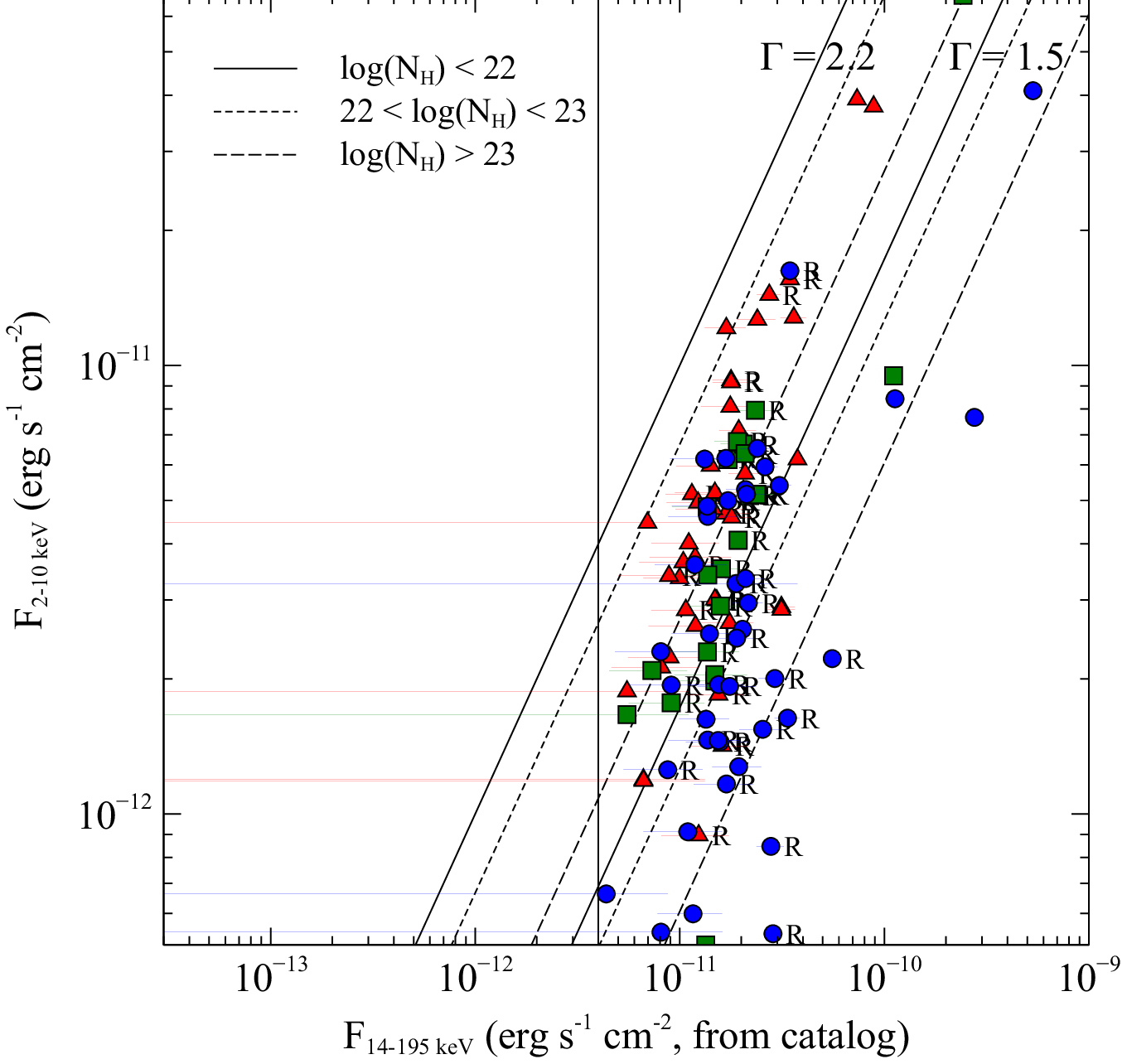}
\includegraphics[width=8.0cm]{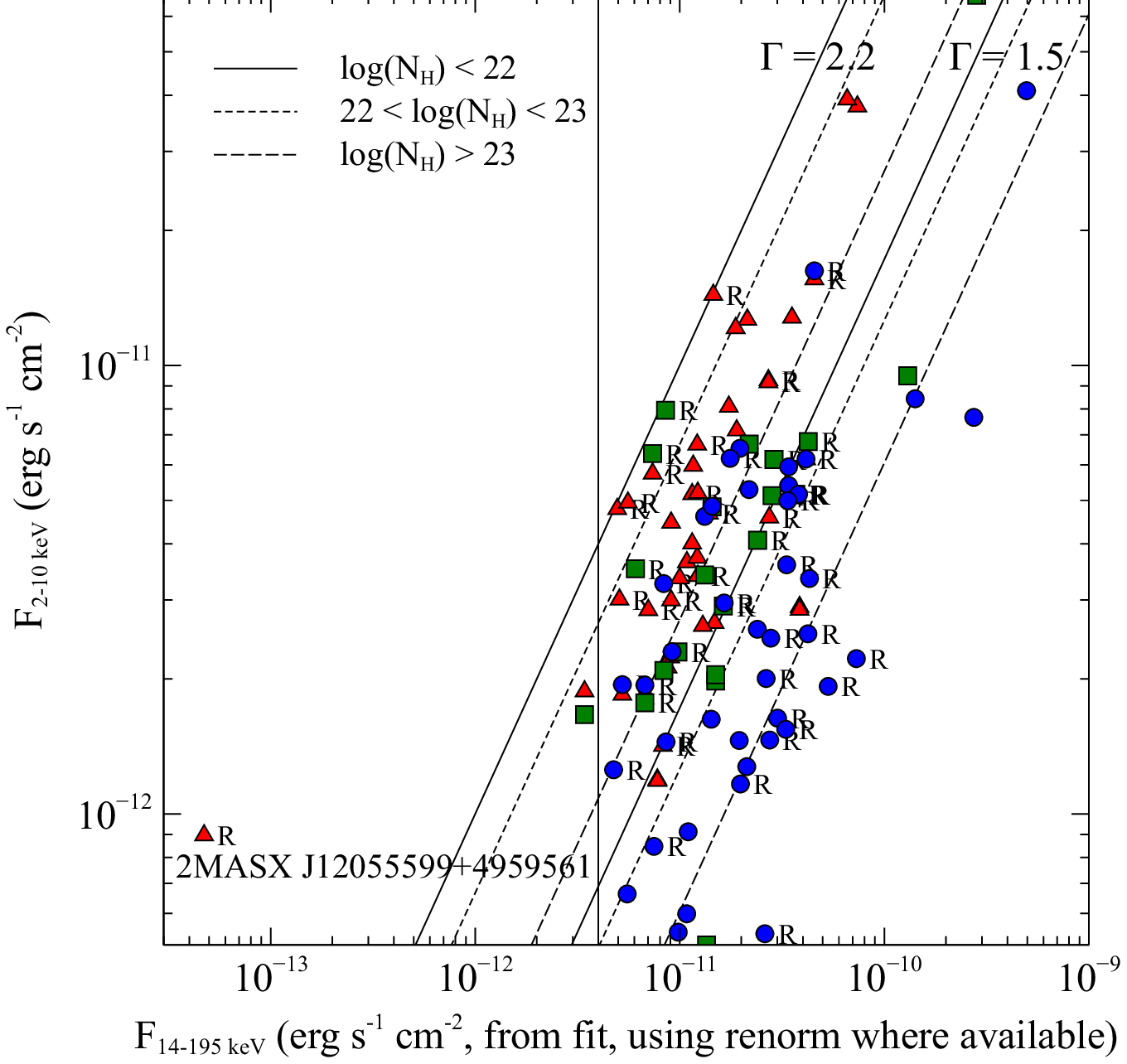}
}
\caption{\small Observed 2--10~keV flux ($F_{\rm 2-10 keV}$) against BAT flux ($F_{\rm 14-195 keV}$).  The red triangles represent sources with $N_{\rm H} < 10^{22} \rm \thinspace cm^{-2}$, while the green circles represent sources with $N_{\rm H} > 10^{23} \rm \thinspace cm^{-2}$. Blue squares indicate the intermediate column sources. Sources for which renormalization of the BAT data was done are indicated using an `R'. The left panel shows the comparison if the BAT flux is calculated from the average spectra presented in the 58-month catalog, and the right panel shows the results obtained if we re-normalize the BAT spectrum as detailed in \S\ref{spectralfitting} \label{F2to10vsFBAT}.  The vertical line indicates the flux limit of our sample based on the average fluxes reported in the 58-month catalog.}
\end{figure*}

The utility of this renormalisation is illustrated in Fig.~\ref{F2to10vsFBAT}, where we plot the 2--10~keV flux against the BAT (14--195~keV) flux, color-coding the observations based on the measured column density from spectral fitting.  We overplot lines showing the expected ratio of $F_{\rm 2-10 keV}/F_{\rm 14-195 keV}$ for different fiducial absorption levels and intrinsic photon indices, to indicate the predicted locus of objects in this plot depending on absorption and spectral slope.   In the left panel we see that before re-normalization, the fluxes cluster tightly close to the BAT flux limit for our sample, irrespective of absorption and do not lie in the regions expected for their measured column density.  After re-normalization (right panel of Fig.~\ref{F2to10vsFBAT}), the objects overwhelmingly shift into the expected regions for the three fiducial column density ranges shown.

There is one outlier in the far left of the right panel of Fig.~\ref{F2to10vsFBAT}, the low-absorption source 2MASX J12055599+4959561 for which the re-normalization does not appear to work well.  When re-normalization is applied, this object lies far from the expected position in $F_{\rm 2-10 keV}-F_{\rm 14-195 keV}$.  This behaviour is contrary to expectations, since we would expect that re-normalizing the BAT data to be contemporaneous with the 2--10~keV data would improve the congruence between the two luminositites.  This object also displays an unusual ratio of $L_{\rm 14-195 keV}$/$L_{\rm 2-10 keV}$ (Fig.~\ref{SoftHR_vs_HardHR}).  Inspection of the joint XRT+BAT spectrum reveals a renormalized BAT spectrum that lies below the XRT spectrum in flux.  If we use the raw BAT spectrum without re-normalization, the BAT and XRT spectra link continuously in a $\nu F_{\nu}$ plot, with a hard, simple power-law spectrum with $\Gamma=1.5$ and negligible $N_{\rm H}^{\rm intrinsic}$.  Inspection of the BAT light curve shows that there is a pronounced dip at the time the XRT data were taken, and that the source is very faint in the BAT band. This might imply that the re-normalization occasionally fails for very faint sources, but for all other objects the re-normalization produces results consistent with the lines of constant $\Gamma$ and $N_{\rm H}$ in Fig.~\ref{F2to10vsFBAT}.

\section{Results}
\label{sec:results}

\subsection{Average Sample Properties and Distributions}

\begin{figure}
%\figurenum{1}
\centerline{
\includegraphics[width=8.0cm]{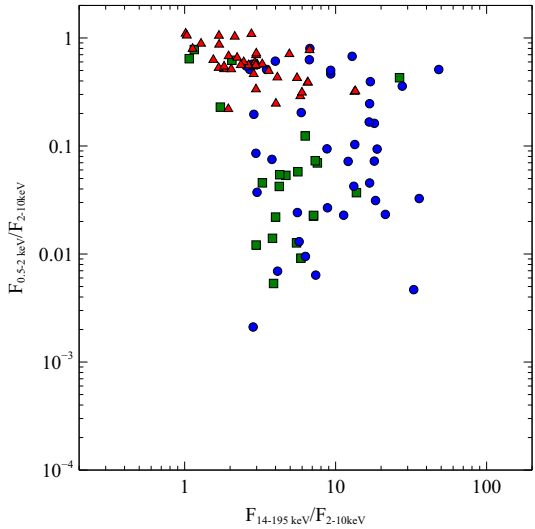}
}
\caption{
\small The 58-month BAT AGNs with $b>50^{\circ}$ with soft and hard fluxes plotted on the color-color plot initially presented in \cite{2008ApJ...674..686W}, and later in W09. The red triangles represent sources with $N_{\rm H} < 10^{22} \rm \thinspace cm^{-2}$, while the circles represent sources with $N_{\rm H} > 10^{23} \rm \thinspace cm^{-2}$. Squares indicate the intermediate column sources. \label{SoftHR_vs_HardHR} 
%Finally, diamonds are used to represent the 17 blazar/BL Lac sources, which all have low measured column densities. In the text, we describe Cyg A (the circle labeled in the plot) and five other sources (NGC 1365 and NGC 5728 (circles), Mrk 3 (square), and NGC 4945 and NGC 6814 (triangles), within the box) to have unusual positions. For variable sources, particularly the BL Lac sources, it is important to note that the BAT and softer X-ray measurements are not coeval.
}
\end{figure}

The average BAT luminosity for our sample is $\langle \rm log(L_{\rm 14-195 keV}) \rangle= 43.5, \sigma_{L_{\rm 14-195keV}}=1.1$.  This result is similar to the average BAT luminosity of $\langle \rm log(L_{\rm 14-195 keV}) \rangle= 43.7, \sigma_{L_{\rm 14-195keV}}=0.8$ from the 9-month catalog (using W09's results), and for consistency, we exclude any objects with jets analyzed in W09 in calculating this average. We also present plots of the BAT-X-ray colors/hardness ratios as done in W09 for easy comparison of our present, deeper sample to the 9-month catalog results.  In Fig.~\ref{SoftHR_vs_HardHR} we see the soft color $F_{\rm 0.5-2keV}/F_{\rm 2-10 keV}$ plotted against the hard color $F_{\rm 14-195 keV}/F_{\rm 2-10 keV}$.  The range of colors spanned in the 58-month catalog appears larger on both axes than that seen in W09.  In the same region of parameter space spanned by the 9-month catalog, we see the same division into regimes occupied by high, intermediate, and low absorption sources, but there are three sources with extreme values (outside the range of the plot): these are 2MASX J12055599+4959561 (for which $\rm log \thinspace N_{\rm H}\thinspace <22$, $F_{\rm 14-195 keV}/F_{\rm 2-10 keV} < 0.1$, and $F_{\rm 0.5-2keV}/F_{\rm 2-10 keV} \approx 0.4$), 2MASX J13105723+0837387 and CGCG 291$-$028 ($\rm log \thinspace N_{\rm H} \thinspace>23$, $F_{\rm 14-195 keV}/F_{\rm 2-10 keV} \approx 10$, and $F_{\rm 0.5-2keV}/F_{\rm 2-10 keV} < 10^{-5}$).  For CGCG 291$-$028 and 2MASX J13105723+0837387, the model fits have very little 0.5--2~keV flux, but this is likely due to their poor XRT data quality, since these objects required the inclusion of BAT data to obtain a fit at all.  For 2MASX J12055599+4959561, the highly unusual ratio of BAT flux to 2--10~keV flux indicates a renormalized BAT spectrum that lies below the XRT spectrum in flux, due to a dip in the BAT light curve at that date, as discussed in \S\ref{subsec:batrenormsection}.  
%B2 1210+33 is the highest redshift ($z=2.51$) object in our sample, and while not classed as a blazar or BL Lac, it does exhibit radio emission and has one classification as a steep-spectrum radio source on NED.  The effect of renormalization of the BAT spectrum does not significantly alter the colors, as in the case of 2MASX J12055599+4959561: we find that the extremely high column density attributed to this source is robust, regardless of whether we lock the 0.4--10~keV normalization to the BAT normalization.  However, if there is a significant jet component as suggested by the radio emission, it may not be appropriate to interpret the spectrum with an absorption model.

\begin{figure}
%\figurenum{1}
\centerline{
\includegraphics[width=8.0cm]{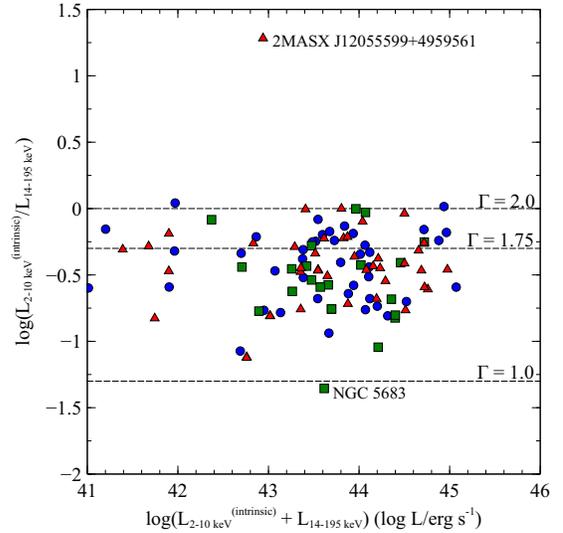}
}

\caption{
\small 
Right: ratio of $L^{\rm intrinsic}_{\rm 2–10 keV}/L_{\rm 14-195 keV}$ vs. the total luminosity in the 2--10~keV and 14--195~keV bands. The dotted lines show values of constant $\Gamma$ between both bands at constant ratios of $L^(\rm intrinsic)_{\rm 2–10 keV}/L_{\rm 14-195 keV}$, following the presentation in W09.  The different absorption levels are distinguished using the same key as in Fig.~\ref{SoftHR_vs_HardHR} \label{L2to10overBAT_vs_L2to10plusBAT}
%Finally, diamonds are used to represent the 17 blazar/BL Lac sources, which all have low measured column densities. In the text, we describe Cyg A (the circle labeled in the plot) and five other sources (NGC 1365 and NGC 5728 (circles), Mrk 3 (square), and NGC 4945 and NGC 6814 (triangles), within the box) to have unusual positions. For variable sources, particularly the BL Lac sources, it is important to note that the BAT and softer X-ray measurements are not coeval.
}
\end{figure}

W09 presented the diagnostic plot $L^{\rm intrinsic}_{\rm 2-10 keV}/L_{\rm 14-195 keV}$ versus $L^{\rm intrinsic}_{\rm 2-10 keV} + L_{\rm 14-195 keV}$, showing the regions of the plot populated by objects of different photon indices $\Gamma$; we reproduce this plot in Fig.~\ref{L2to10overBAT_vs_L2to10plusBAT}.  We note the unusual location of 2MASX J12055599+4959561 again as in Fig.~\ref{SoftHR_vs_HardHR}, and additionally highlight NGC 5683, identified as a Seyfert 1.  The measured photon index for NGC 5683 is 2.15, but this source does not have a re-normalized BAT spectrum.  As a result, the ratio of the measured BAT luminosity to the 2--10~keV luminosity places it in the region expected for extremely hard sources with $\Gamma < 1.0$.
%It has an unusually high column density for a Seyfert 1 ($\rm log \thinspace N_{\rm H} \thinspace=22.56$) and is therefore part of the class of AGN for which a mismatch is seen between optical and X-ray measures of absorption.  

 We now present the distributions of key quantities including the absorbing column density (Fig.~\ref{nh_histogram_modeltype}), photon index $\Gamma$ (Figs.~\ref{gamma_histogram_splitonNH},\ref{gamma_histogram_modeltype},\ref{gamma_histogram_smallerrors}) and 2--10~keV intrinsic luminosity (Figs.~\ref{logLX_histogram_modeltype},\ref{logLX_histogram_splitonNH}).

\subsection{Radio loudness}
\label{radioloudness}
%MOVE THIS EARLIER SINCE WE SHOULD EXCLUDE HB 0945....

\begin{figure}
%\figurenum{1}
\centerline{
\includegraphics[width=8.0cm]{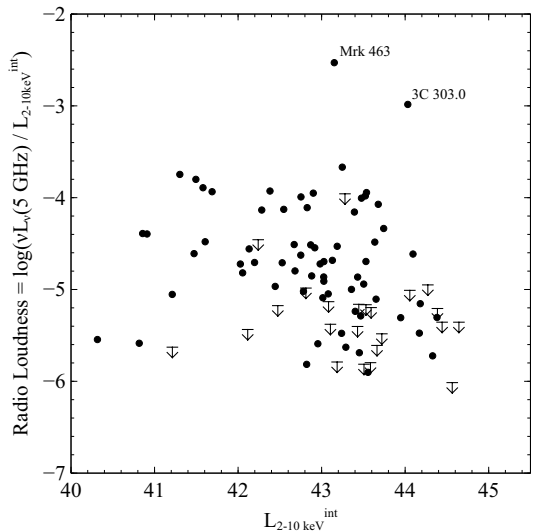}
}
\caption{
\small
\label{radioloudness_vs_lx} Radio-loudness ($\rm \nu L_{\nu} (\nu = 5 GHz)/L_{2-10 keV}$) against intrinsic 2--10~keV luminosity.  Downward pointing arrows show upper limits where FIRST radio detections were not available (assuming a flux limit of 0.75 mJy to calculate the luminosities).}
\end{figure}

We present the radio loudness values for our sample (defined as $\rm \nu L_{\nu}(\nu = \rm 5 \thinspace GHz)/L^{\rm int}_{\rm 2-10 keV}$, \citealt{2003ApJ...583..145T}) plotted against the intrinsic 2--10~keV luminosity in Fig.~\ref{radioloudness_vs_lx}, to provide an indication of the radio properties of the sample.  The radio luminosities are taken from the FIRST survey at 1.4~GHz, and we convert the fluxes to 5~GHz using a standard spectral index of $\alpha=0.7$ typical for synchrotron emission (for flux density $f_{\nu} \thinspace \propto \thinspace \nu^{-\alpha}$; see e.g., \citealt{2010MNRAS.406..493M} for a discussion of radio spectral indices).  Where no match is found within 5 {\arcsec} for a given source, we assume a flux limit of 0.75 mJy for the survey and use it to calculate an upper limiting radio luminosity.  The large angular size of the FIRST beam is likely to introduce significant contamination from the host galaxy, but our main purpose here is to catch significant outliers where the nuclear radio emission is heavily boosted by a jet.  Such objects will easily stand out from the rest of the distribution.

All of our sources have radio-loudness values below $-2$, and all but two of our sources are below the values typically seen for strongly-beamed sources such as BL Lacs or flat-spectrum radio quasars \citep{2003ApJ...583..145T}.  The two objects with the highest radio-loudness parameters are Mrk 463 and 3C 303.0.  The object 3C 303.0 is expected to be radio loud based on its inclusion in the 3C catalog.  Further investigation of Mrk 463 reveals that it contains two nuclei at very close separation (3.8 kpc, \citealt{2008MNRAS.386..105B}), with the eastern source Mrk 463E dominating the X-ray emission by a factor of $\sim 4$.  The brighter Mrk 463E nucleus was previously found to have a Seyfert 2-type optical spectrum, but a fuller consideration of the radio morphology reveals that it is a `hidden' Seyfert 1 nucleus \citep{1999ApJ...518..117K}.  Both nuclei show moderate-to-weak radio emission (\citealt{2003AJ....126.2237D}).  Our {\xmm} and Swift/BAT analysis of this source therefore is likely to include the combined emission from both sources (as discussed in \citealt{2008MNRAS.386..105B}). We discuss these two cases in Appendix \ref{appendix:radioloud}, but a more detailed study of the X-ray properties of these objects is needed.

\subsection{Column density $N_{\rm H}$}
\label{subsec:abscolumn}

\begin{figure*}
%\figurenum{1}
\centerline{
\includegraphics[width=8.0cm]{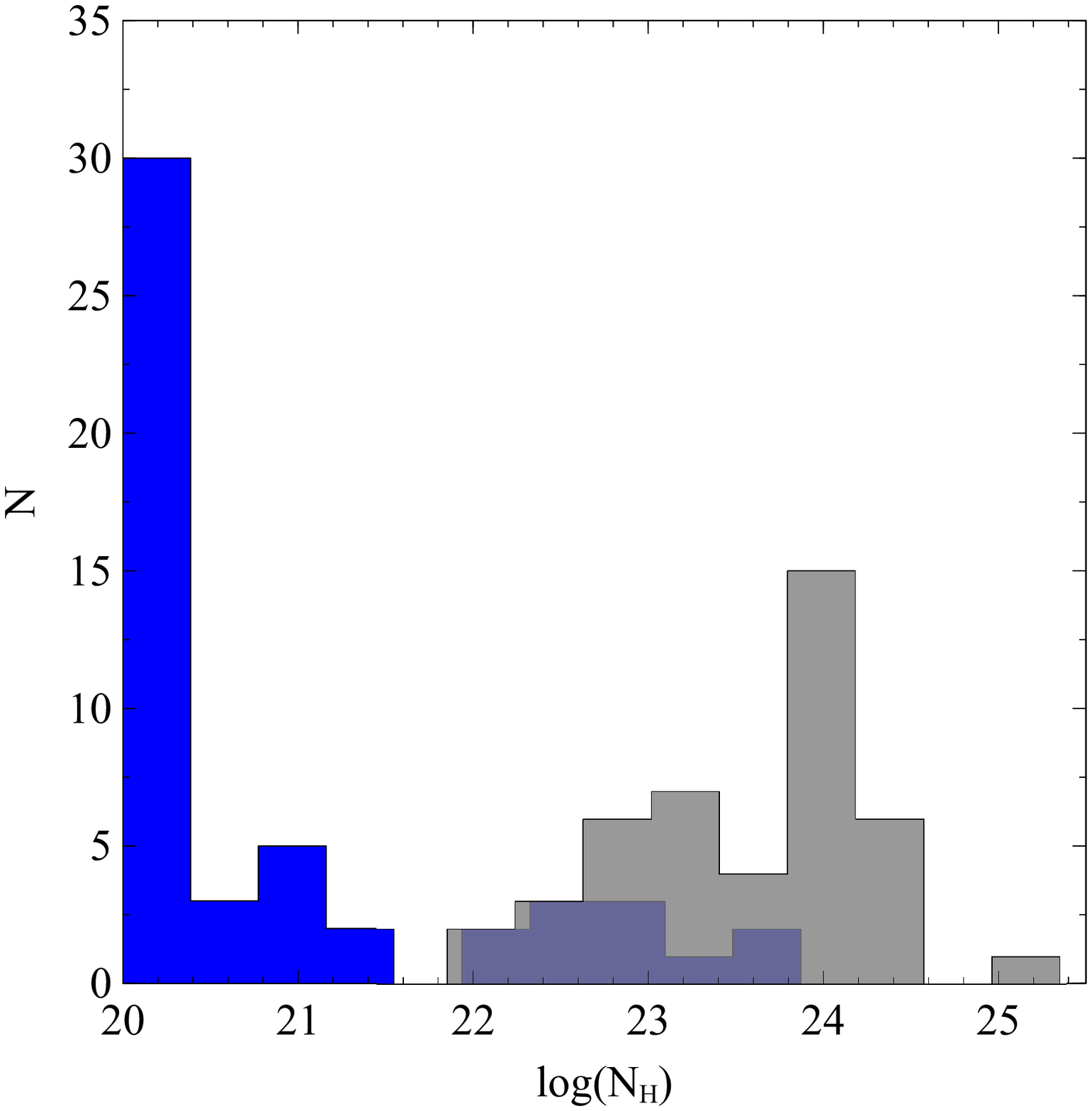}
\includegraphics[width=8.0cm]{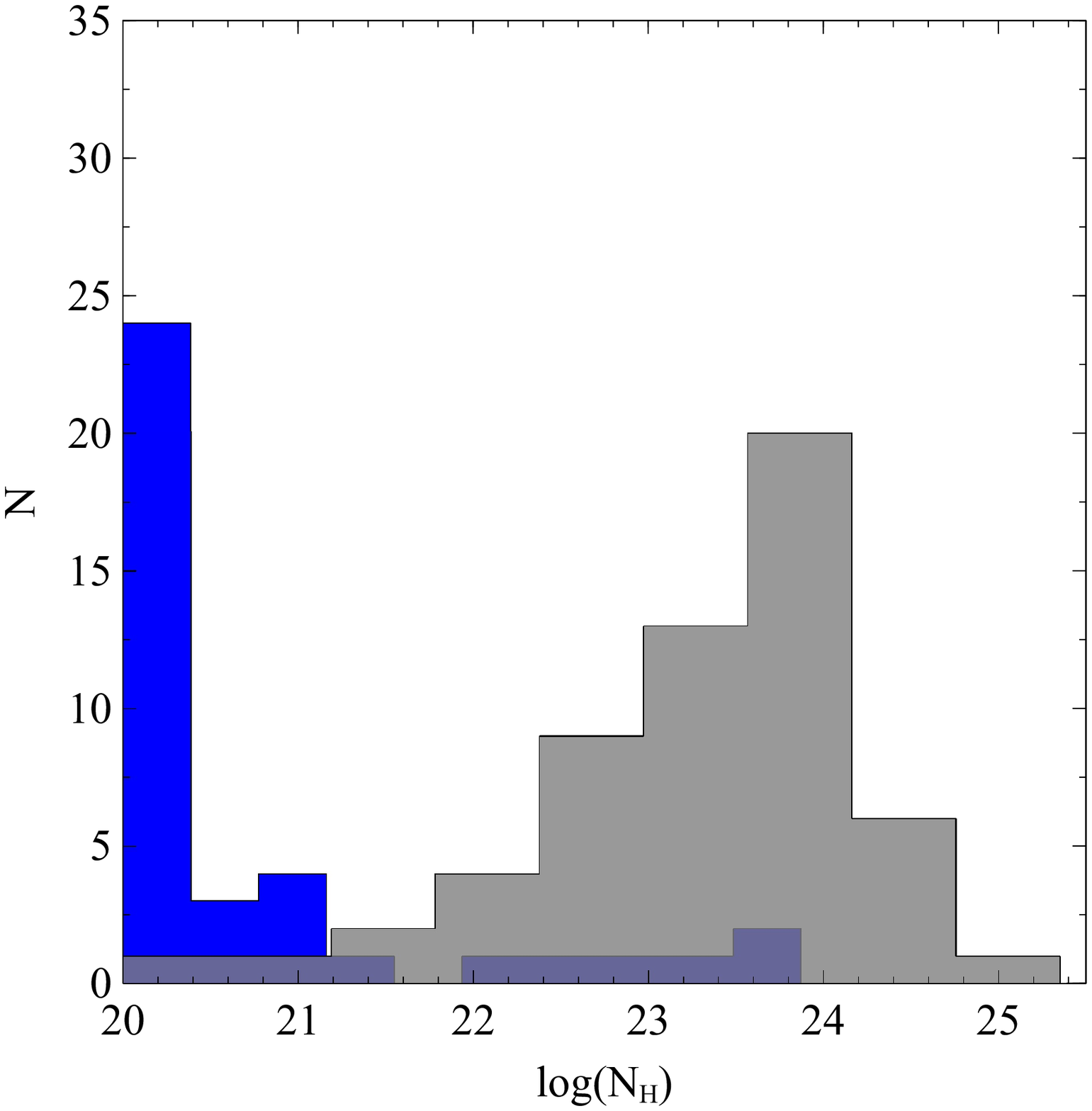}
}
\caption{
\small
Histograms of absorbing column density $\rm log(N_{\rm H})$.  The blue shaded portions represent simple spectrum objects, whereas the semi-transparent grey shaded portions represent complex spectrum objects (semi-transparent to show the underlying blue histogram when present).  In the left panel, any objects with ambiguous spectral classifications (both a simple and a complex spectral model describe the data well) have been assumed to be simple spectrum objects; in the right panel those same objects are assumed to possess complex spectra.\label{nh_histogram_modeltype}}
\end{figure*}

\begin{figure}
%\figurenum{1}
\centerline{
\includegraphics[width=8.0cm]{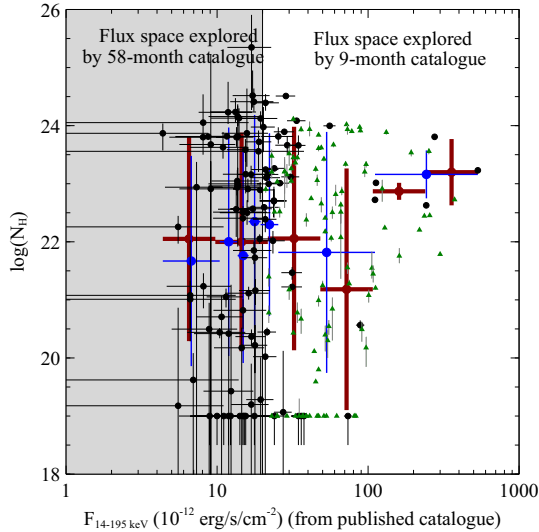}
}
\caption{
\small Absorbing column density log($N_{\rm H}/\rm cm^{-2}$) against BAT flux ($F_{\rm 14-195~keV}$); black points show the results for the 58-month catalog $b>50^{\circ}$ sources and the small green triangles show the results from the 9-month catalog (from W09). The blue (thin) and brown (thick) error bars and points show the results binned by $F_{\rm~14-195 keV}$ from two different approaches; 1) using a constant number of objects in each bin, or 2) using a constant interval in $F_{\rm 14-195 keV}$ for each bin.  The mean $N_{\rm H}/\rm cm^{-2}$ in each bin is calculated, and  the error bars are the standard deviation to show the degree of spread (\emph{not} the error on the mean).  The distribution of absorption probed in this region of the sky appears to be independent of BAT flux, with a wide range of absorbing columns seen at all flux levels.  We note the different distribution found in W09 and discuss this in the text. \label{NHvsFBAT}
}
\end{figure}

As the BAT survey increases its exposure, we expect it to uncover a more accurate reflection of the true absorption distribution for the AGN population, and to see differences from the earlier 9-month catalog analysis.  If we compare the absorption distribution seen here (Fig.~\ref{nh_histogram_modeltype}) with that seen in the 9-month catalog (W09, see Fig.~\ref{Hist_logNH_COMPARE9MONTH} of this paper), we indeed see that our distribution shows a tail at higher column densities than that seen previously, and the average absorbing columns from our distribution are $\langle \rm log \thinspace N_{\rm H} \thinspace \rangle = 20.80$, $\sigma = 1.18$ (simple, 51 objects) and $\langle \rm log \thinspace N_{\rm H} \thinspace \rangle = 23.55$, $\sigma = 0.71$ (complex, 44 objects), assuming the `simple' model type for any objects with dual best-fits.  We assume all objects with $\rm log(N_{\rm H})<20$ to have a lower-limiting absorption of $\rm log(N_{\rm H})=20$ for a consistent comparison with W09.  If we assume dual objects are by default complex, we find slightly different distributions: $\langle \rm log \thinspace N_{\rm H} \thinspace \rangle = 20.67$, $\sigma = 1.12$ (simple, 38 objects) and $\langle \rm log \thinspace N_{\rm H} \thinspace \rangle = 23.27$, $\sigma = 0.95$ (complex, 57 objects).  We contrast these results with those from W09, who find $ \langle \rm log \thinspace N_{\rm H} \thinspace \rangle = 20.58$, $\sigma = 0.74$ (simple, 46 objects) and $\langle \rm log \thinspace N_{\rm H} \thinspace \rangle = 23.03$, $\sigma = 0.71$ (complex, 56 objects), verifying a tail of higher-absorption objects in our sample.   If we split the objects based on the spectral complexity exhibited, we find that the percentage of `complex' objects is in the range 43--56\% (of the whole sample, 100 objects), with the range again due to the presence of sources with ambiguous spectral types.  However, inspection of Fig.~\ref{NHvsFBAT} shows that the absorption distributions reported in our sample do not vary appreciably depending on BAT flux.  Interestingly, in the region of flux space already covered by the 9-month catalog, we find an increase in the average absorbing column density towards the highest BAT fluxes.

%[Note: the tail of high absorption objects in this sample somewhat depends on the inclusion of B2 1210+33, so I need to decide if this source is jetted 'enough' to warrant exclusion or not.]

\begin{figure}
%\figurenum{1}
\centerline{
\includegraphics[width=8.0cm]{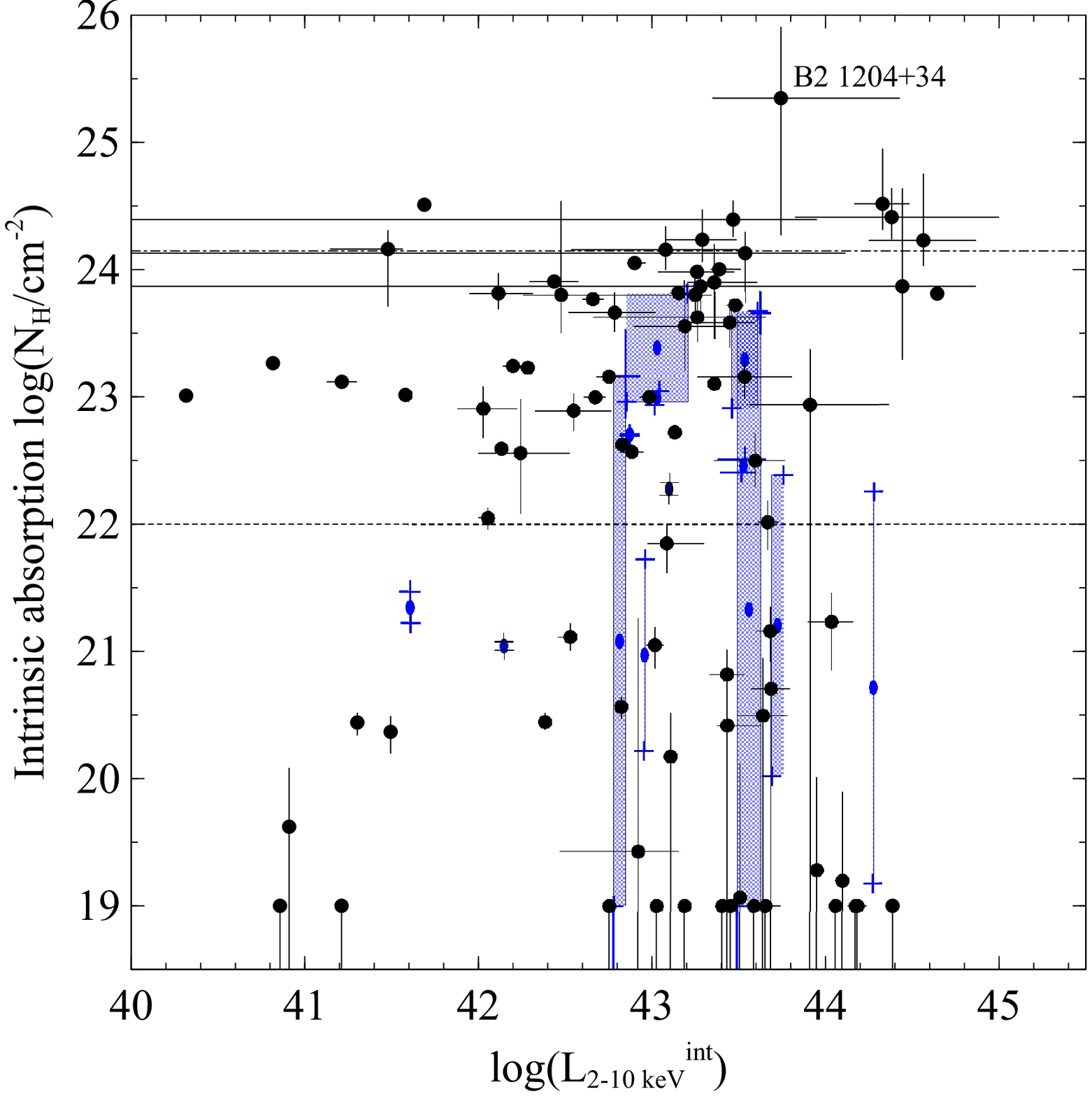}
}
\caption{
\small
Absorbing column density log($N_{\rm H}/\rm cm^{-2}$) against intrinsic 2--10~keV luminosity $L_{\rm 2-10 keV}$ (absorption-corrected).  The black circles represent objects for which a unique best-fit model was determined.  For objects where a unique best-fit model could not be determined and two `best-fit' models were identified, blue shaded areas represent the range in log($N_{\rm H}$) and $L_{\rm 2-10 keV}$ spanned by those two models (with the model fit results themselves signified by blue crosses), and the blue oval points represent the central, average values for that object.  The dashed horizontal line shows the conventional threshold between `absorbed' and `unabsorbed' objects ($\rm log \thinspace N_{\rm H} \thinspace = 22$), and the dot-dashed line shows the threshold for Compton-thick objects ($\rm log \thinspace N_{\rm H} \thinspace = 24.15$). All objects for which $N_{\rm H}$ was below $10^{19} \rm cm^{-2}$ are shown at log($N_{\rm H}$)=19; in these sources absorption by neutral gas has negligible effect on the X-ray spectrum.\label{nhvslx}}
\end{figure}

We also identify the proportion of Compton-thick ($N_{\rm H}>1.4 \times 10^{24} \rm \thinspace cm^{-2}$) sources in our sample.  In contrast to W09 who find no Compton-thick objects by this criterion, we find eight Compton-thick sources in our sample: these are NGC 4102,  2MASX J10523297+1036205, 2MASX J11491868-0416512, B2 1204+34, MCG -01-30-041, MRK 1310, PG 1138+222 and UGC 05881; additionally, NGC 4941 and NGC 5106 are very close to the threshold for being Compton-thick, and the errors on their column densities could push them over the threshold. This result suggests that $\sim 9$\% of our sample is Compton-thick.  However, at such high column densities, basic photoelectric absorption is not sufficient to model the level of absorption present and more sophisticated absorption models must be used to calculate the column density for such objects.  This exercise is beyond the scope of this paper, but we discuss alternate measures of Compton-thickness in \S\ref{subsec:comptonthick}.

%NGC 4941 pushed out when I included BAT data - only just though! NEARLY Compton thick.

%[We also find that three radio-loud sources, 4C 49.22,  B2 1210+33, [HB89] 0945+408, also have very high column densities.  Contrary to the finding from W09 beamed objects typically have low column densities, these heavily absorbed objects appear to be an exception, and we note the recent work of Behar et al. (2011) who find that a substantial fraction of high redshift blazars are absorbed.  These objects could be interesting case studies in that regard].

The histograms in Fig.~\ref{nh_histogram_modeltype} show a clear bifurcation between spectral types in terms of their absorption.  Simple spectra overwhelmingly fit objects with low absorption, and complex spectra are generally required for objects with high absorption.  However, the intermediate class of objects identified in this study introduces some uncertainty in the distributions, since for these objects the different model types yield different estimates of $N_{\rm H}$.  We therefore show the `worst-case' scenarios in Fig.~\ref{nh_histogram_modeltype}, assuming that the intermediate objects are all simple or complex in the left and right panels, respectively.  Assuming that these objects take the $N_{\rm H}$ values of their complex model fits, we see a more pronounced peak in high-absorption sources.   We also note that a substantial fraction of sources have negligible intrinsic absorption ($\rm log \thinspace N_{\rm H} \thinspace<20$).

\begin{figure}
%\figurenum{1}
\centerline{
\includegraphics[width=8.0cm]{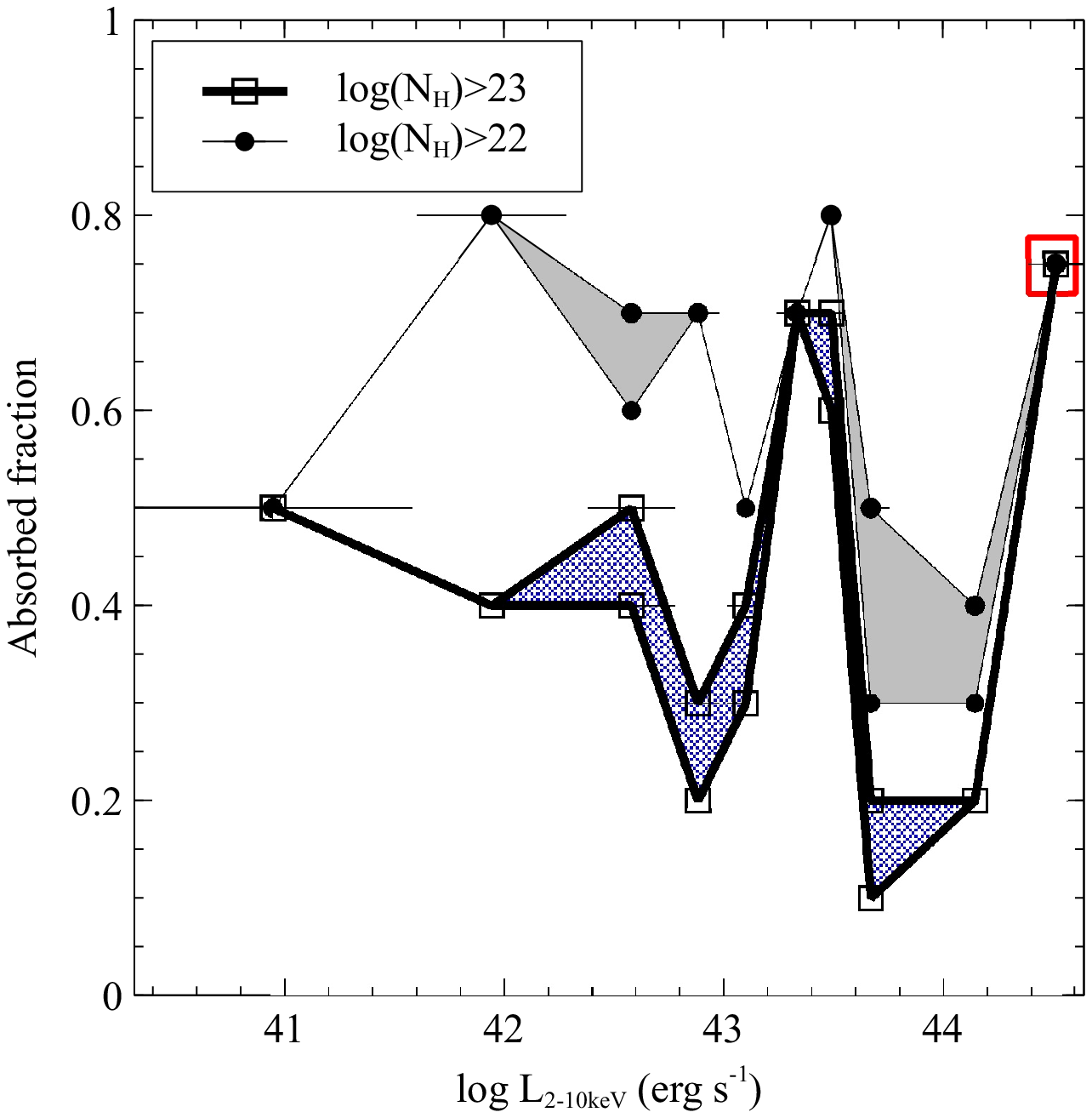}
}
\caption{
\small
Absorbed fractions against intrinsic 2--10~keV luminosity (10 objects per bin).  Filled circles connected by thin solid lines show the fraction of sources with $\rm log(N_{\rm H})>22$, whereas empty squares connected by thick lines show the fraction of sources with $\rm log(N_{\rm H})>23$.  The solid grey and blue hatched shading reveal the uncertainty in these fractions due to the thirteen sources with ambiguous spectral types (and hence two estimates for their $\rm log \thinspace N_{\rm H}$). The absorbed fraction in the highest luminosity bin (indicated by the red square) is more uncertain since it contains only four objects.}\label{absfracvslx}
\end{figure}

The relationship between absorption and intrinsic 2--10~keV luminosity is given in Fig.~\ref{nhvslx}.  This plot shows a broad absorption distribution at all luminosity levels.  The thirteen sources with ambiguous spectral types can have highly uncertain column densities (indicated by the blue shaded boxes).  We plot the absorbed fractions (using thresholds of $\rm log \thinspace N_{\rm H} \thinspace =22$ and $23$ to defined `absorbed') as a function of 2--10~keV luminosity in Fig.~\ref{absfracvslx}, using 10 objects per bin to determine the absorbed fraction.  We do not see as strong a decrease in the absorbed fraction with luminosity as previously reported in Fig.~13 of \cite{2011ApJ...728...58B} and the earlier \emph{INTEGRAL} AGN survey \citep{2009A&A...505..417B}, although a similar trend is present.  Our work reveals a more homogenous distribution of absorption throughout luminosity space, and notably three heavily absorbed sources ($\rm log \thinspace N_{\rm H} \thinspace > 23.5$) are found at the highest luminosities (3C 234, 2E 1139.7+1040 and 2MASX J11491868-0416512). The \cite{2011ApJ...728...58B} and \citep{2009A&A...505..417B} samples contain more sources at high luminosities than our sample, and it is above luminosities of $10^{44} \thinspace \rm erg \thinspace s^{-1}$ where the decrement in the absorbed fraction is more pronounced. It is therefore possible that we do not have enough sources at high luminosities to see the decrement.

\begin{figure}
%\figurenum{1}
\centerline{
\includegraphics[width=8.0cm]{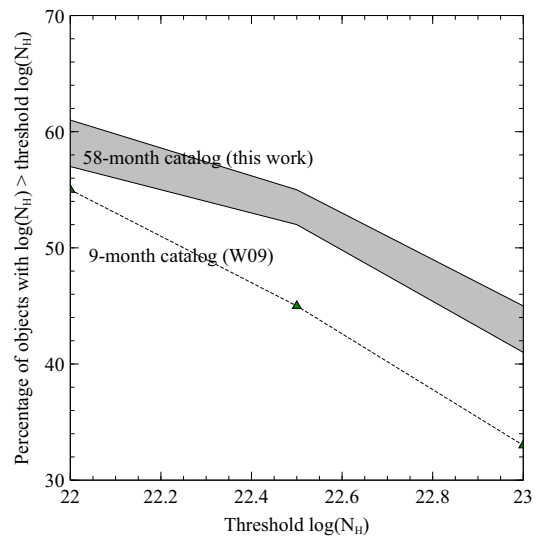}
}
\caption{
\small
Absorbed fraction vs. threshold $\rm log(N_{\rm H})$, for the 9-month catalog (W09, green triangles and dashed line) and for the 58-month catalog Northern Galactic Cap sources (solid lines and shaded area, this work).  The shaded area shows the uncertainty in the absorbed fractions in this work due to the sources with ambiguous spectral types and non-unique best-fitting column densities.\label{absfrac_vs_tresholdNH}}
\end{figure}

We see from Fig.~\ref{absfracvslx}  how the distribution of absorbed sources depends on the threshold absorption used; we quantify this effect in Fig.~\ref{absfrac_vs_tresholdNH} by employing three thresholds of $\rm log(N_{\rm H})=(22, 22.5, 23)$, and plotting the fraction of objects above these thresholds against the threshold itself. We present a comparison with the 9-month catalog results. The shaded area in the figure again shows the uncertainty due to the non-unique model fits for 13 sources.  It is clear that the absorbed fractions are uniformly higher at all thresholds in our work than in W09, and that the slower fall-off with absorption threshold indicates that a substantial proportion of our absorbed sources are heavily absorbed ($\rm log \thinspace N_{\rm H} \thinspace>23$), about 10\% more than in the 9-month catalog.  Inspection of a plot of absorbing column density against redshift (which we omit for brevity) indicates no evolution of the absorption distribution with redshift over the redshift range probed by this survey.  Therefore, the deeper flux limit must be responsible for picking up more asborbed objects. At $\rm log(N_{\rm H})=23.5$, $\sim 5$\% of the intrinsic flux in the BAT band is absorbed, so we would expect that the deepening of the survey would produce this kind of increase in the proportion of highly-absorbed objects.

%The fraction of heavily absorbed ($\rm log(N_{\rm H} \ge 23.0$) objects in the sample is in the range 41--45\%, depending on whether we assume the maximal or minimal $N_{\rm H}$ values for objects with ambiguous model types and non-unique values of $N_{\rm H}$, contrasting with a heavily absorbed fraction of {\bf 33\%} for the 9-month catalog (W09).  This emphasizes the need for longer observations of the thirteen sources with ambiguous spectral types, especially the four which straddle the $\rm log(N_{\rm H}=23.0)$ dividing line.   If we use the criterion $\rm log(N_{\rm H} \ge 22.5)$ to identify highly absorbed objects, we find that the absorbed fraction increases to 52--55\% ({\bf 45\%} of sources in W09).  Finally, adopting a criterion of $\rm log(N_{\rm H} \ge 22)$ for absorbed objects, we find that the absorbed fraction would then be 57--61\%.  We briefly contrast this with the result from W09, who find {\bf 55\%}.

\subsection{Photon Index}

\begin{figure}
%\figurenum{1}
\centerline{
\includegraphics[width=8.0cm]{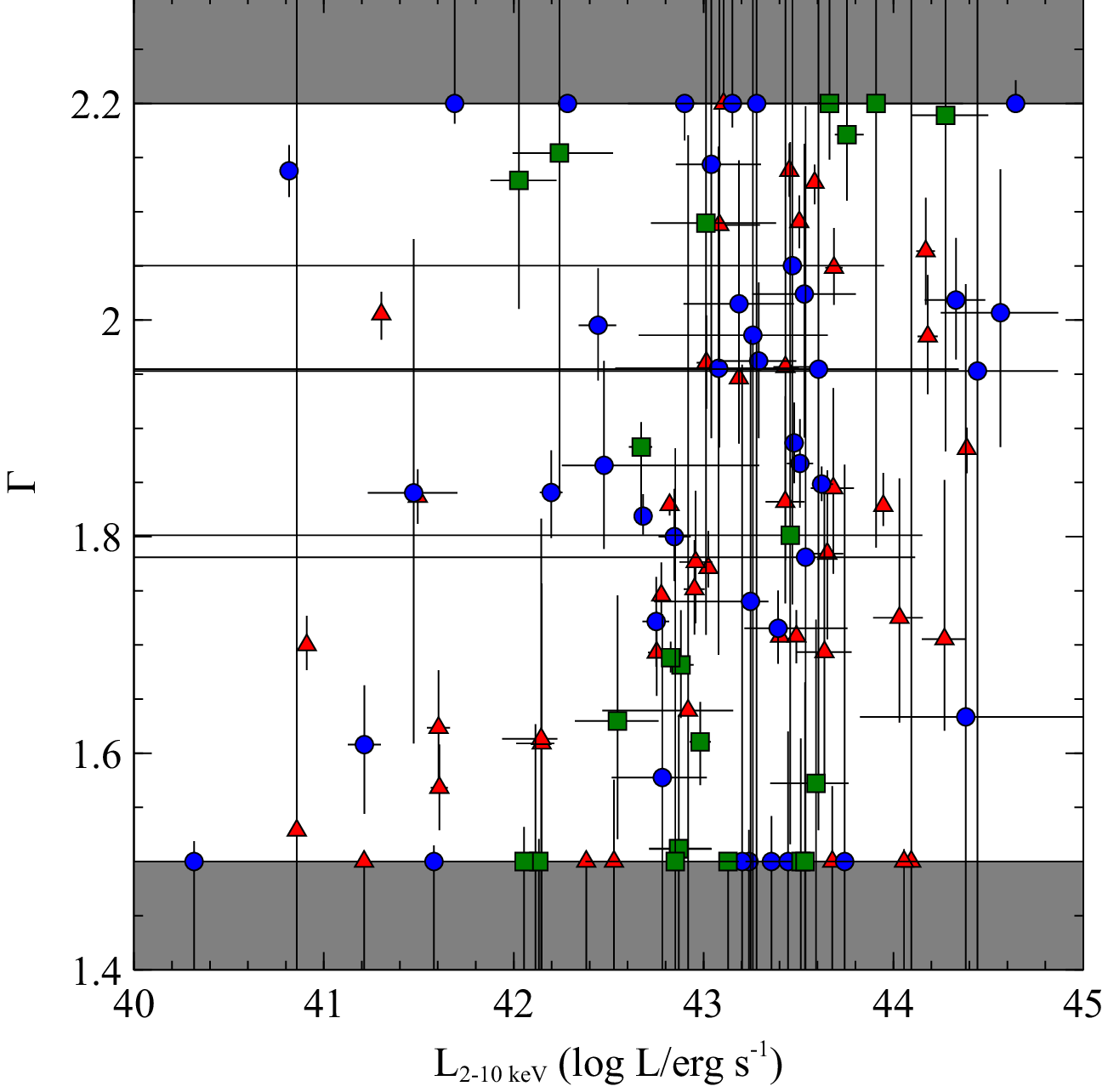}
}

\caption{
\small Photon index $\Gamma$ against intrinsic (absorption-corrected) 2--10~keV luminosity $L_{\rm 2-10 keV}$. The grey shaded areas delineate the hard limits imposed on $\Gamma$ in the fit. Red triangles depict low absorption ($\rm log \thinspace N_{\rm H} \thinspace < 22$) objects, blue circles depict high absorption objects ($\rm log \thinspace N_{\rm H} \thinspace > 22$) and green squares represent objects with intermediate absorptions between these two limits. \label{GammavsL2to10}
%Finally, diamonds are used to represent the 17 blazar/BL Lac sources, which all have low measured column densities. In the text, we describe Cyg A (the circle labeled in the plot) and five other sources (NGC 1365 and NGC 5728 (circles), Mrk 3 (square), and NGC 4945 and NGC 6814 (triangles), within the box) to have unusual positions. For variable sources, particularly the BL Lac sources, it is important to note that the BAT and softer X-ray measurements are not coeval.
}
\end{figure}

\begin{figure*}
%\figurenum{1}
\centerline{
\includegraphics[width=8.0cm]{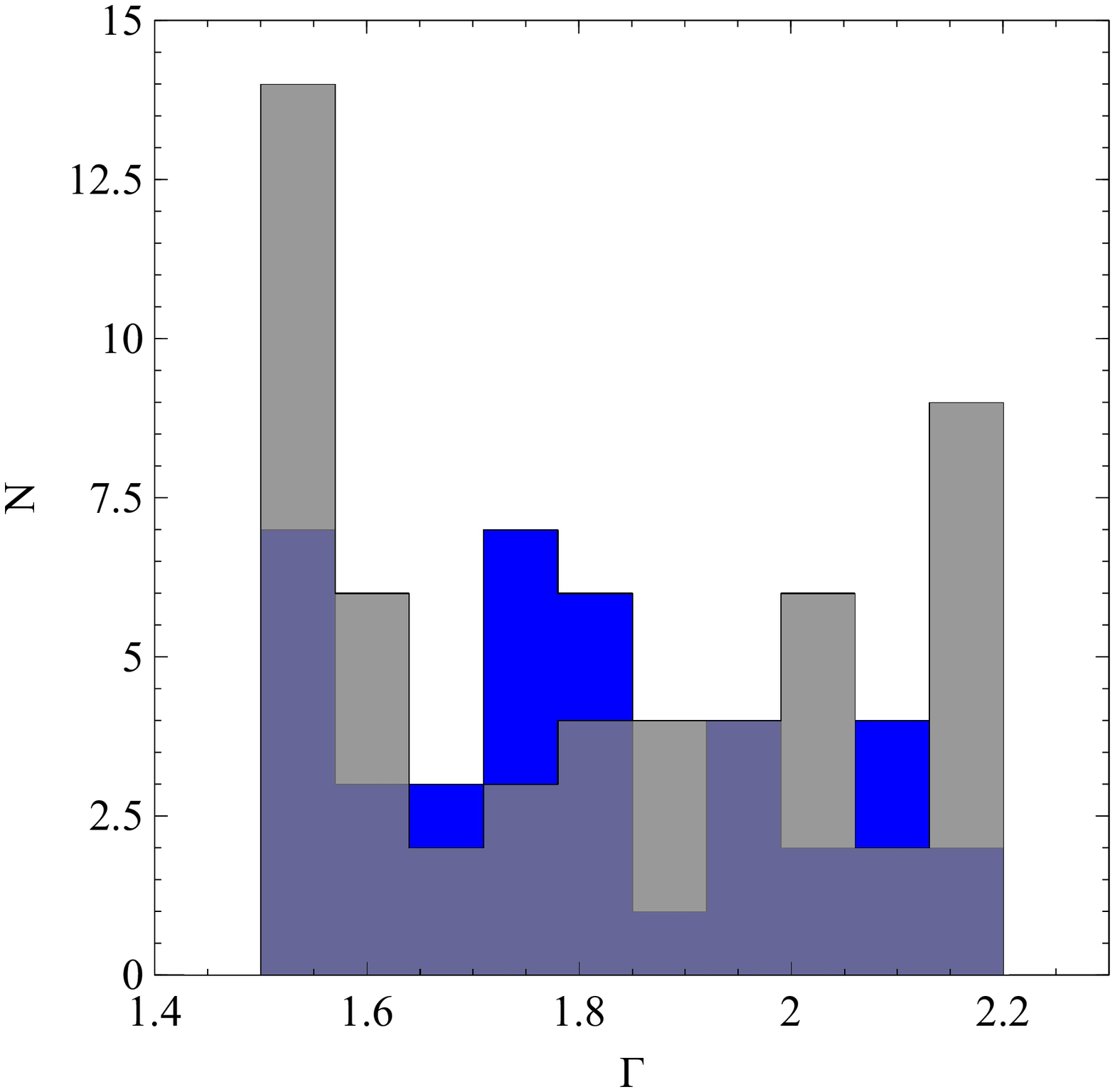}
\includegraphics[width=8.0cm]{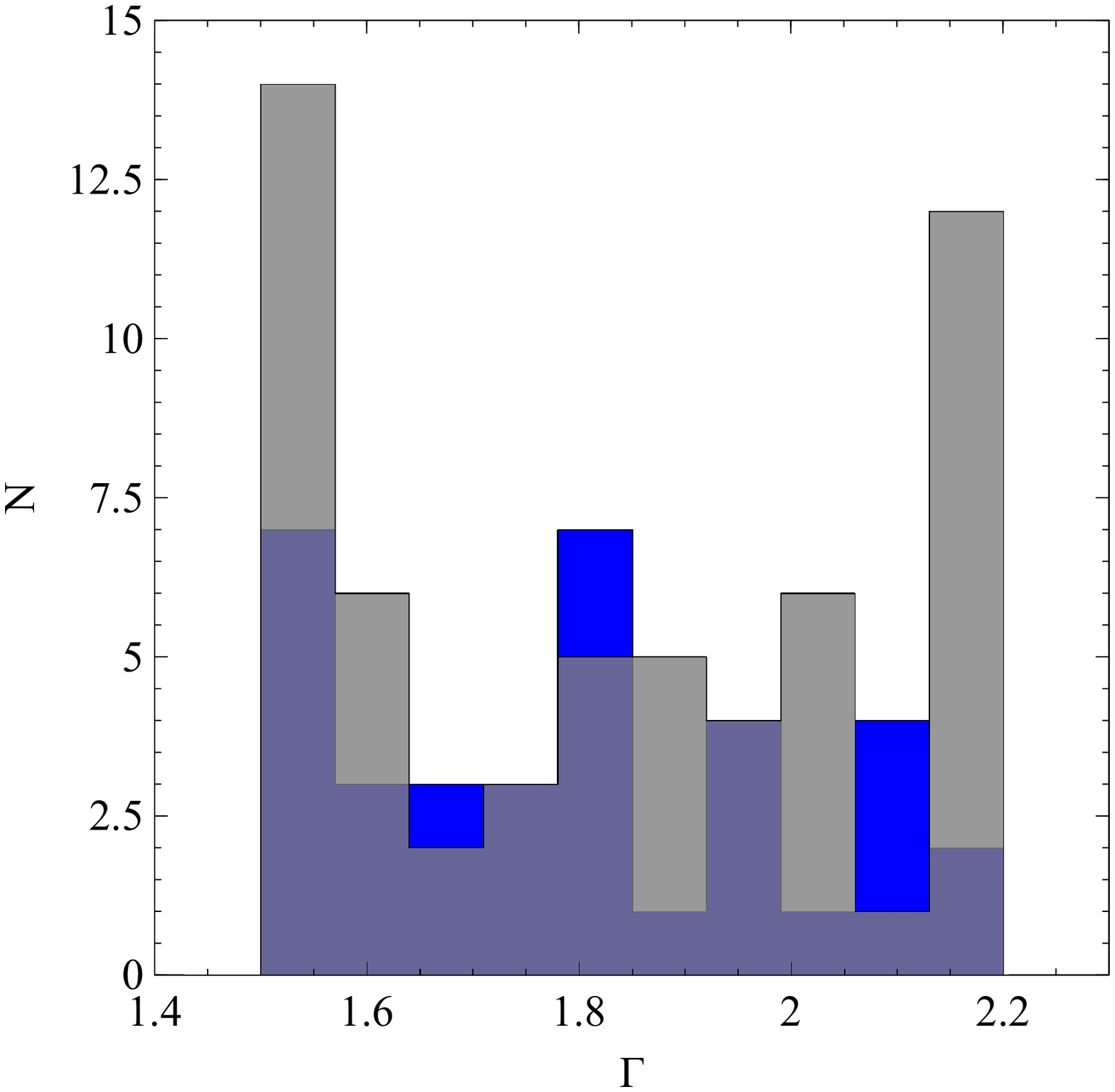}
}
\caption{
\small
Histograms of 2--10~keV photon index, $\Gamma$.  The grey shaded portions represent high-absorption objects ($\rm log \thinspace N_{\rm H} \thinspace > 22$), whereas the blue shaded portions represent low-absorption objects ($\rm log \thinspace N_{\rm H} \thinspace<22$).  In the left panel, in order to split our objects into the two absorption categories, any objects with ambiguous spectral classifications and two `best-fit' $N_{\rm H}$ estimates (both a simple and a complex spectral model describe the data well) have been assumed to have the smaller of the two $N_{\rm H}$ estimates.  In the right panel, the maximal $N_{\rm H}$ is assumed for objects with ambiguous spectral classification.\label{gamma_histogram_splitonNH}}
\end{figure*}

\begin{figure*}
%\figurenum{1}
\centerline{
\includegraphics[width=8.0cm]{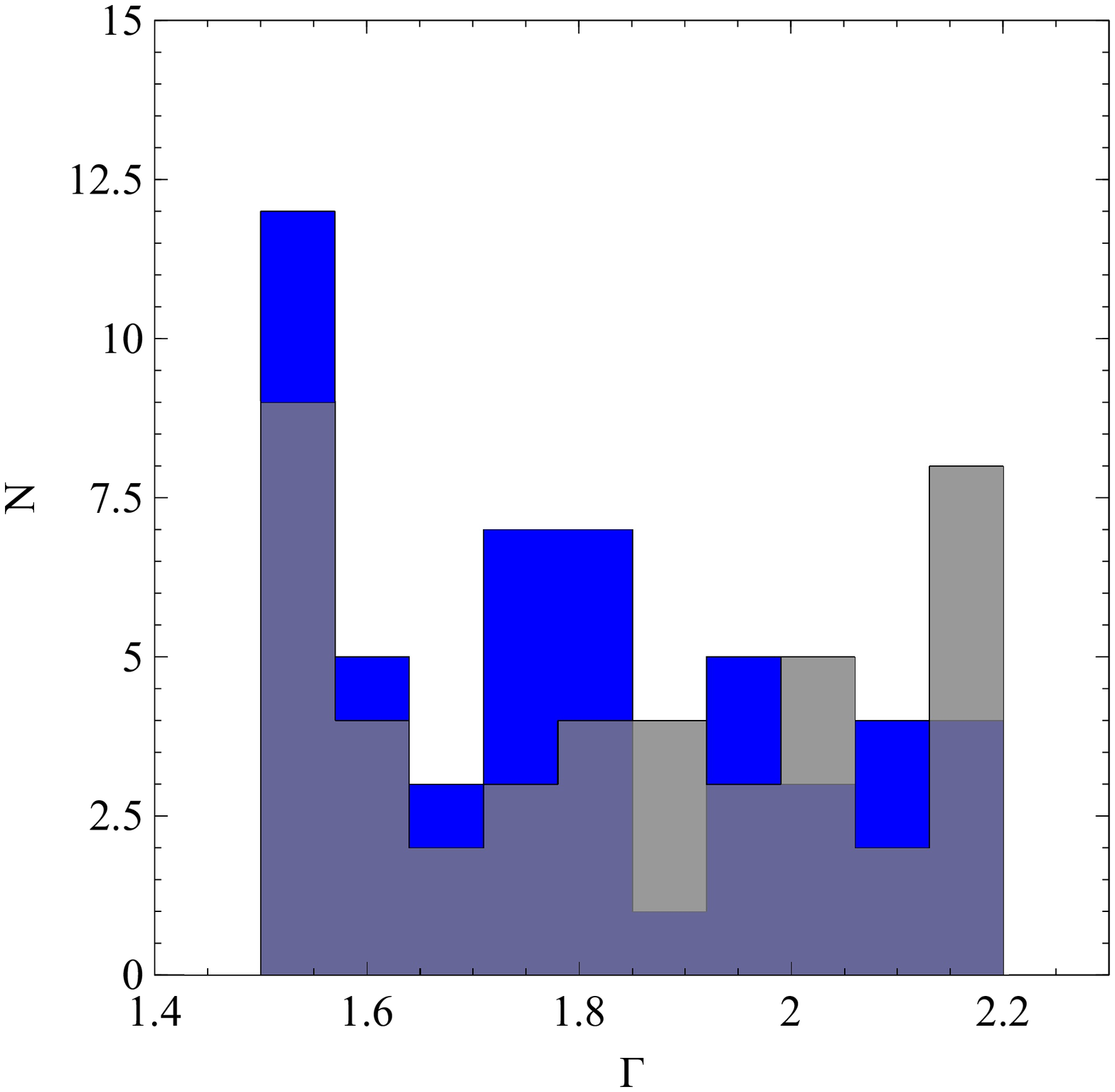}
\includegraphics[width=8.0cm]{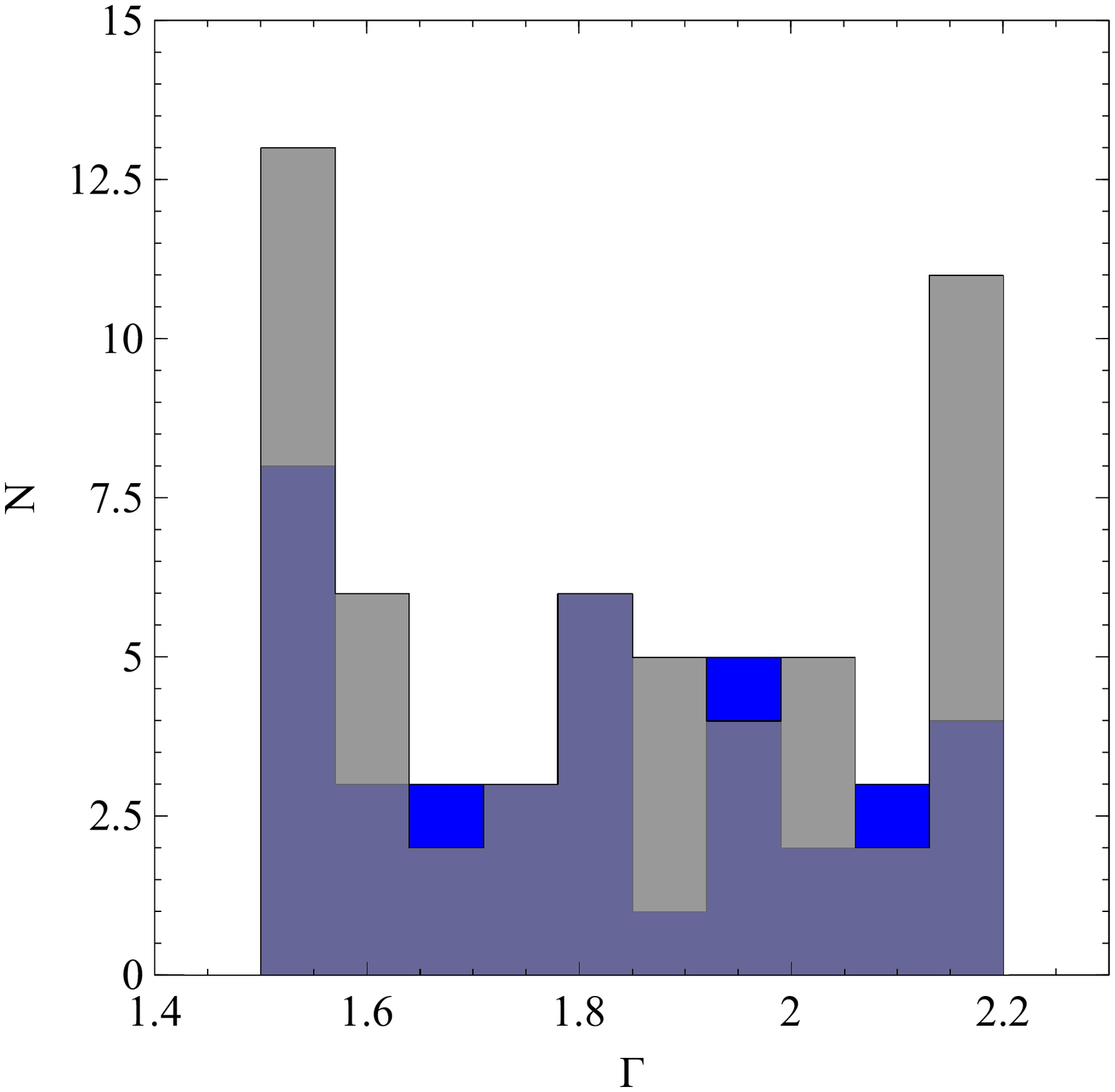}
}
\caption{
\small
Histograms of 2--10~keV photon index, $\Gamma$.  The grey shaded portions represent complex spectrum objects, whereas the blue shaded portions represent simple spectrum objects.  In the left panel, in order to split our objects into the two spectral-type categories, any objects with ambiguous spectral classifications (both a simple and a complex spectral model describe the data well) have been assumed to be simple spectrum objects; in the right panel those same objects are assumed to possess complex spectra.\label{gamma_histogram_modeltype}}
\end{figure*}

We restrict photon indices to lie between $1.5 < \Gamma < 2.2$ in our fits for observational and physical reasons: observationally, the AGN in the {\xmm} Bright Serendipitous Survey \citep{2011A&A...530A..42C} indicate 3$\sigma$ limits on their absorption-corrected photon indices consistent with this restriction; physically, photon indices much lower (harder) than 1.5 indicate unphysical Compton $y$ parameters in standard inverse-Compton scattering scenarios for modelling the X-ray power-law emission from the corona (e.g., \citealt{1990ApJ...363L...1Z}). Fig.~\ref{GammavsL2to10} is a plot of $\Gamma$ against the intrinsic X-ray luminosity $L_{\rm 2-10 keV}$, and the distribution of photon indices as histograms is presented in Figs.~\ref{gamma_histogram_splitonNH} and \ref{gamma_histogram_modeltype}.

Earlier studies have investigated whether photon index is correlated with luminosity or Eddington ratio; larger photon indices (softer X-ray spectra) are seen to accompany higher accretion rates with more prominent accretion disc components in Galactic Black Hole candidates, constraining the physics of different accretion states (for a review see e.g., \citealt{2006ARA&A..44...49R}).  This parallel has also been explored with AGN (e.g., \citealt{2006MNRAS.372.1366K}, \citealt{2008ApJ...682...81S}).  Fig.~\ref{GammavsL2to10} shows that there is no observed correlation between photon index and 2--10~keV luminosity, although the more physically relevant correlation is with Eddington ratio, which we do not explore here as we lack black hole mass estimates for the entire sample.  The limits imposed on $\Gamma$ could prevent us from seeing any correlation, if present; however, if the range $1.5 < \Gamma < 2.2$ is an appropriate physical constraint to place on $\Gamma$, then this should not be an issue.   In any case, our study reinforces the result in W09 that the photon index does not appear to be correlated with 2--10~keV luminosity, at least in our complete sample of objects, although W09 suggest that they may be correlated when comparing multiple observations of an individual object.  We also note that for the high-absorption sources, the photon index is more prone to uncertainty due to complex absorption, and that the photon indices in such sources are more likely to serve simply as an indication of the general spectral shape. If we therefore restrict our view to unabsorbed, $\rm log(N_{\rm H})<22$, sources in our search for a correlation between $\Gamma$ and $L_{\rm 2-10 keV}$, we still do not observe any obvious correlation, but note that there is a larger range of $\Gamma$ values at high luminosities, whereas low $L_{\rm 2-10~keV}$ luminosities seem to accompany lower values of $\Gamma$. We defer further investigation of this topic to a dedicated study on multi-epoch observations of BAT sources.

We find that in high-absorption objects, the photon index hits our hard limits imposed on $\Gamma$ more frequently.  In objects with $\rm log(N_{\rm H})>22$, we see a larger fraction of objects with pegged extremal (1.5 or 2.2) spectral indices (20/60, 33\%) compared to low-absorption $\rm log(N_{\rm H})<22$ objects (7/40, 18\%), indicating that the spectral complexity due to higher absorption ultimately needs a more sophisticated modelling approach.

\begin{figure*}
%\figurenum{1}
\centerline{
\includegraphics[width=8.0cm]{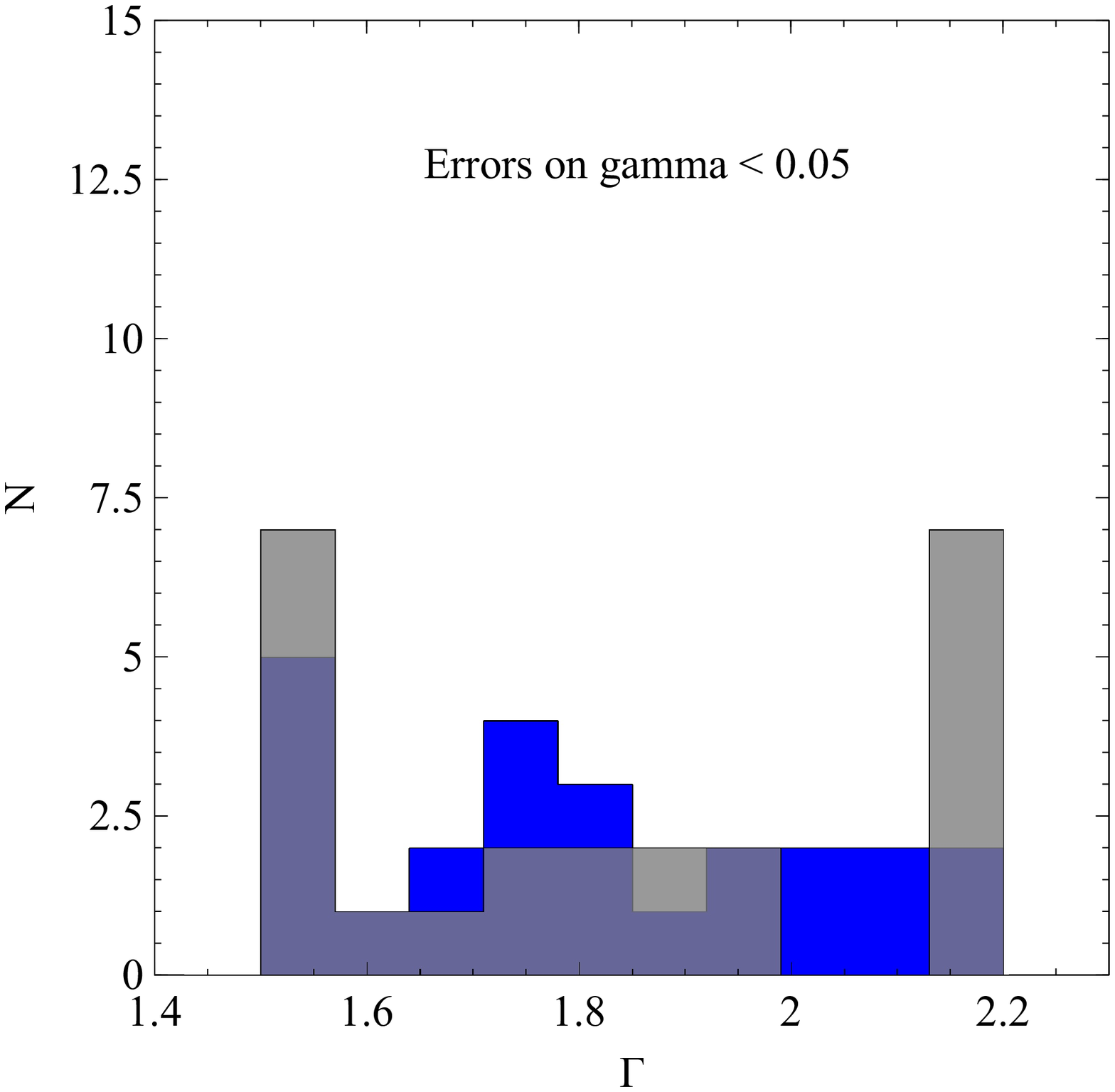}
\includegraphics[width=8.0cm]{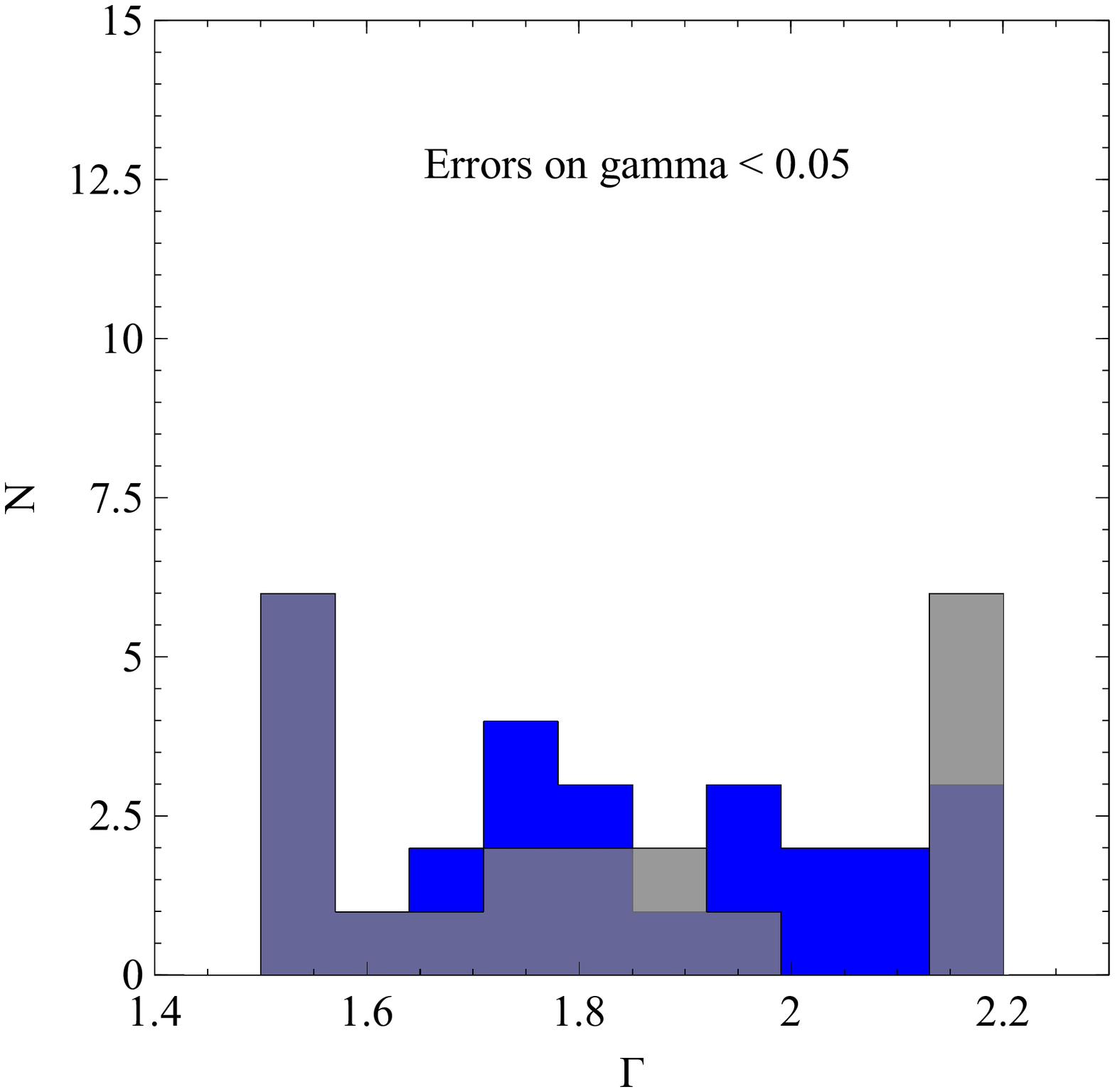}
}
\caption{
\small
Histograms of 2--10~keV photon index, $\Gamma$, with objects with large errors on $\Gamma$ removed.  The grey shaded portions represent complex spectrum objects, whereas the blue shaded portions represent simple spectrum objects.  In the left panel, the grey shading represents high absorption objects and blue shading represents low absorption (dividing line between these designations is $N_{\rm H}=10^{23} \rm cm^{-2}$ as before).  The right panel shows the same values split on model type, such that grey represents complex spectrum objects and blue represents simple spectrum objects.  For objects with dual spectral classifications, we assume simple spectral types and the smaller of the two $N_{\rm H}$ estimates. \label{gamma_histogram_smallerrors}}
\end{figure*}

We present histograms of the photon index in Figs.~\ref{gamma_histogram_splitonNH}, \ref{gamma_histogram_modeltype} and \ref{gamma_histogram_smallerrors}.  Low-absorption sources appear to peak at $\Gamma \approx 1.8$ ($\langle \Gamma_{\rm lowabs} \rangle = 1.81$, $\sigma^{\rm (lowabs)}_{\Gamma} = 0.21$), whereas high-absorption sources show a wider spread in $\Gamma$ and often hit the hard limits imposed ($\langle \Gamma_{\rm highabs} \rangle = 1.84$, $\sigma^{\rm (highabs)}_{\Gamma} = 0.26$).  Complex sources appear to have a wide range of photon indices ($\langle \Gamma_{\rm complex} \rangle = 1.87$, $\sigma^{\rm (complex)}_{\Gamma} = 0.25$), but hit the $\Gamma=2.2$ boundary more often than simple model sources ($\langle \Gamma_{\rm simple} \rangle = 1.79$, $\sigma^{\rm (simple)}_{\Gamma} = 0.22$) which, as expected, show a peak at around $\Gamma \approx 1.8$ (since they overwhelmingly overlap with low-absorption sources).  Due to the large error bars on some values of photon indices, we present histograms of the values of $\Gamma$ with absolute errors less than 0.05 in Fig.~\ref{gamma_histogram_smallerrors}, and find that these trends are borne out even when restricting ourselves to the well-determined values of photon index.  High-absorption objects with well-determined $\Gamma$ values show a distribution skewed toward slightly higher values of $\Gamma$ ($\langle \Gamma_{\rm highabs} \rangle = 1.90$, $\sigma^{\rm (highabs)}_{\Gamma} = 0.28$) than found above, as do complex objects ($\langle \Gamma_{\rm complex} \rangle = 1.90$, $\sigma^{\rm (complex)}_{\Gamma} = 0.28$), but simple spectrum or low-absorption objects with well-determined photon indices do not show significant differences in their distributions, when comparing with the whole sample of simple/low-absorption objects.  The uncertainties in the fits for the thirteen ambiguous spectral-type objects do not affect these statistics by more than 0.02 for any of the quantities presented.

%We note that many objects have large errors on $\Gamma$ (Fig.~\ref{GammavsL2to10}). We present the histograms of $\Gamma$ (split on model type and absorbing column as before) but with objects with large errors on $\Gamma$ excluded, to understand the distribution of photon indices with more stable values of $\Gamma$.  We assume the minimal $N_{\rm H}$ for objects with double-best-fit models and assume a simple model in such cases for simplicity, and present the results in Fig.~\ref{gamma_histogram_smallerrors}.

\subsection{2--10~keV intrinsic luminosity}

\begin{figure*}
%\figurenum{1}
\centerline{
\includegraphics[width=8.0cm]{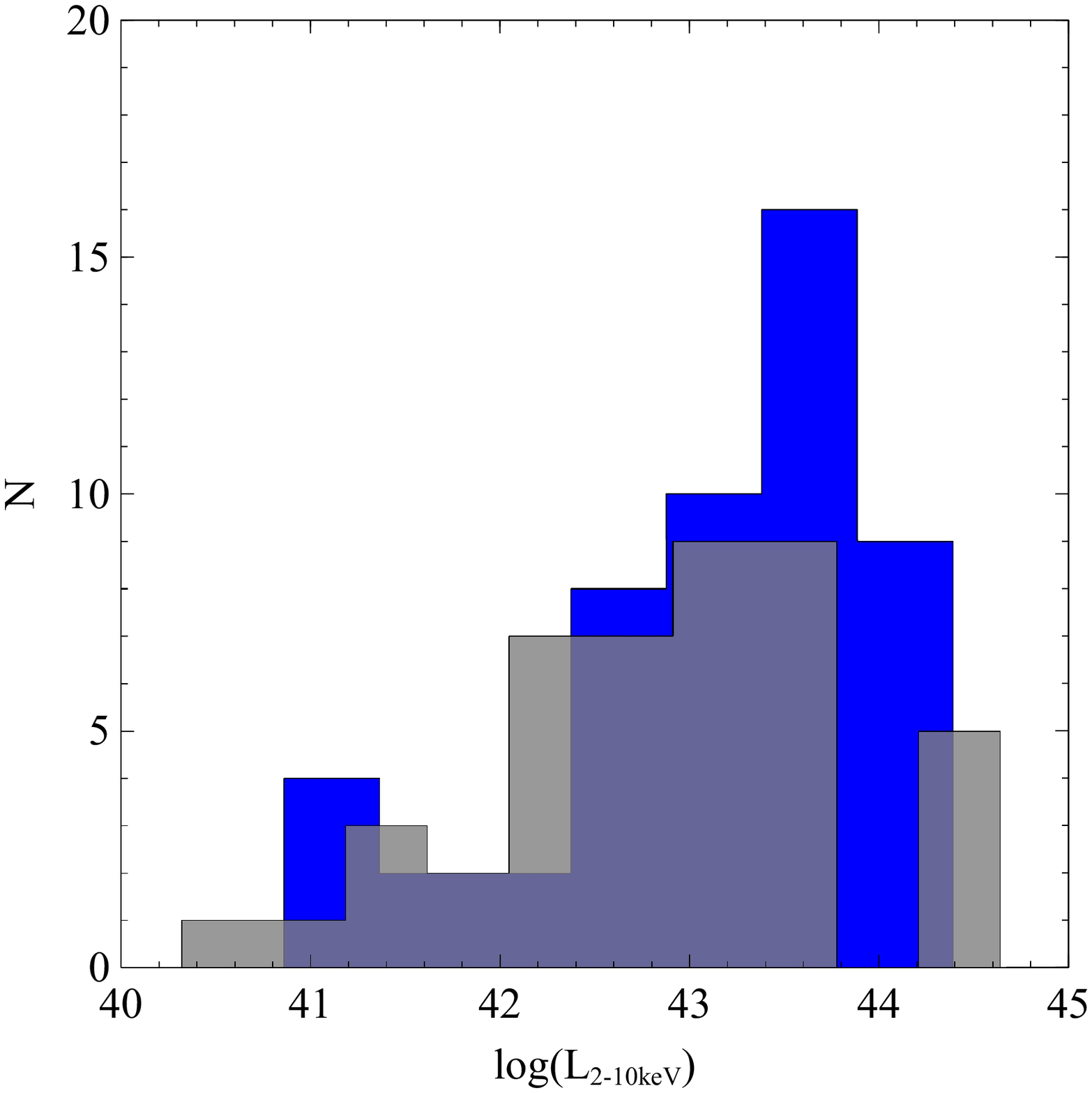}
\includegraphics[width=8.0cm]{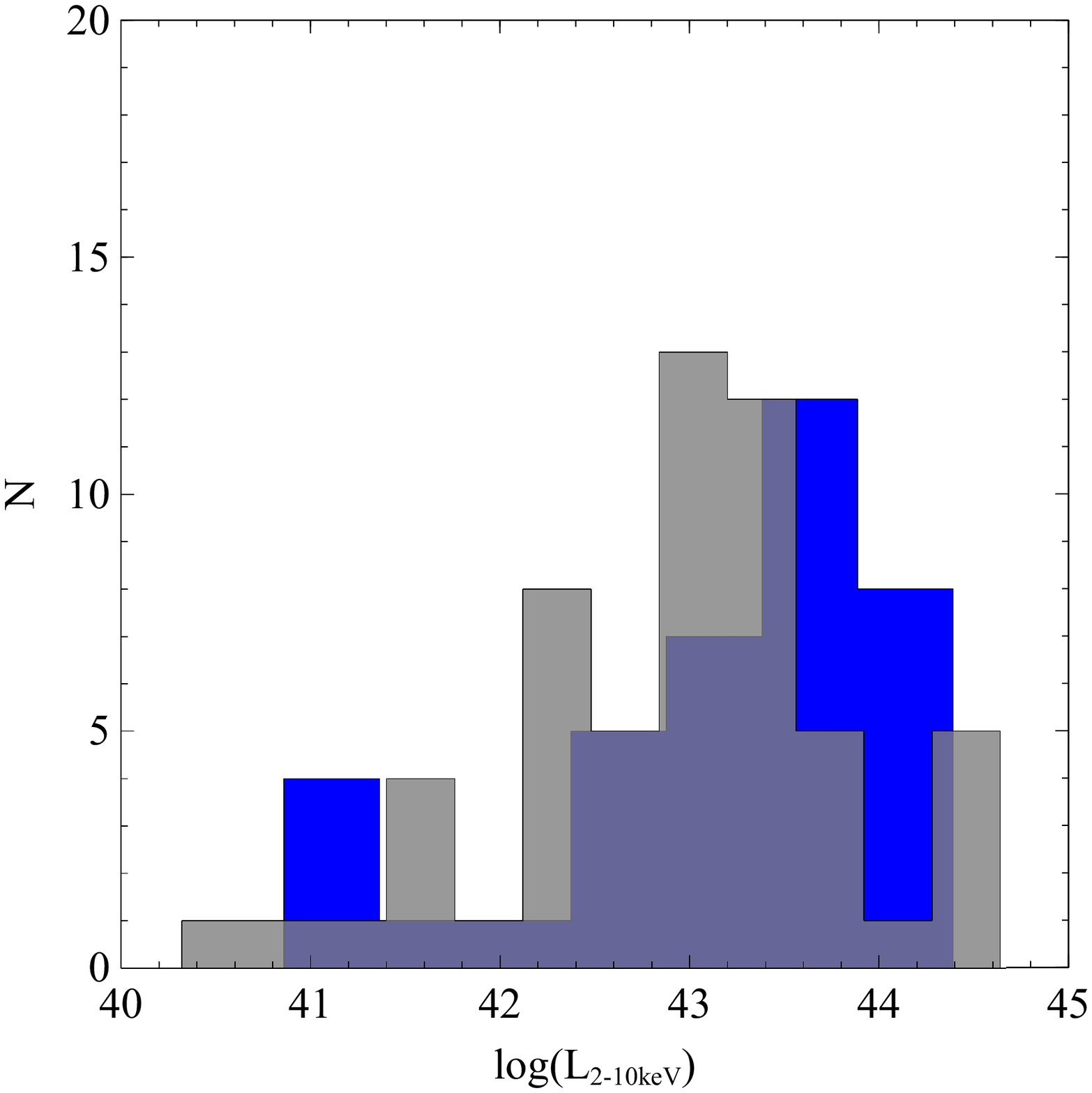}
}
\caption{
\small
Histograms of intrinsic 2--10~keV luminosity log($L_{\rm 2-10 keV}$).  The grey shaded portions represent complex spectrum objects, whereas the blue shaded portions represent simple spectrum objects.  In the left panel, any objects with ambiguous spectral classifications (both a simple and a complex spectral model describe the data well) have been assumed to be simple spectrum objects; in the right panel those same objects are assumed to possess complex spectra.\label{logLX_histogram_modeltype}}
\end{figure*}

\begin{figure*}
%\figurenum{1}
\centerline{
\includegraphics[width=8.0cm]{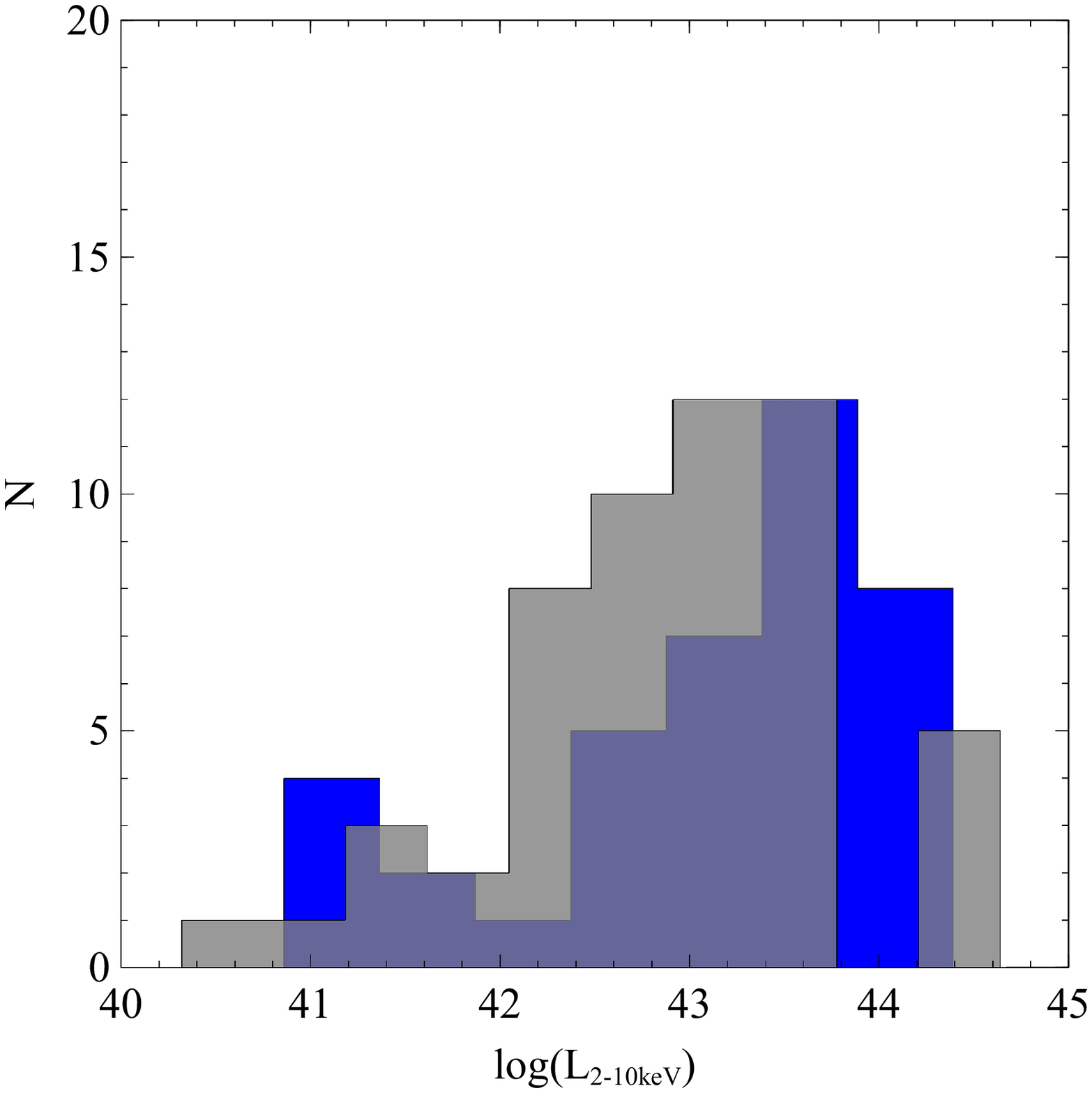}
\includegraphics[width=8.0cm]{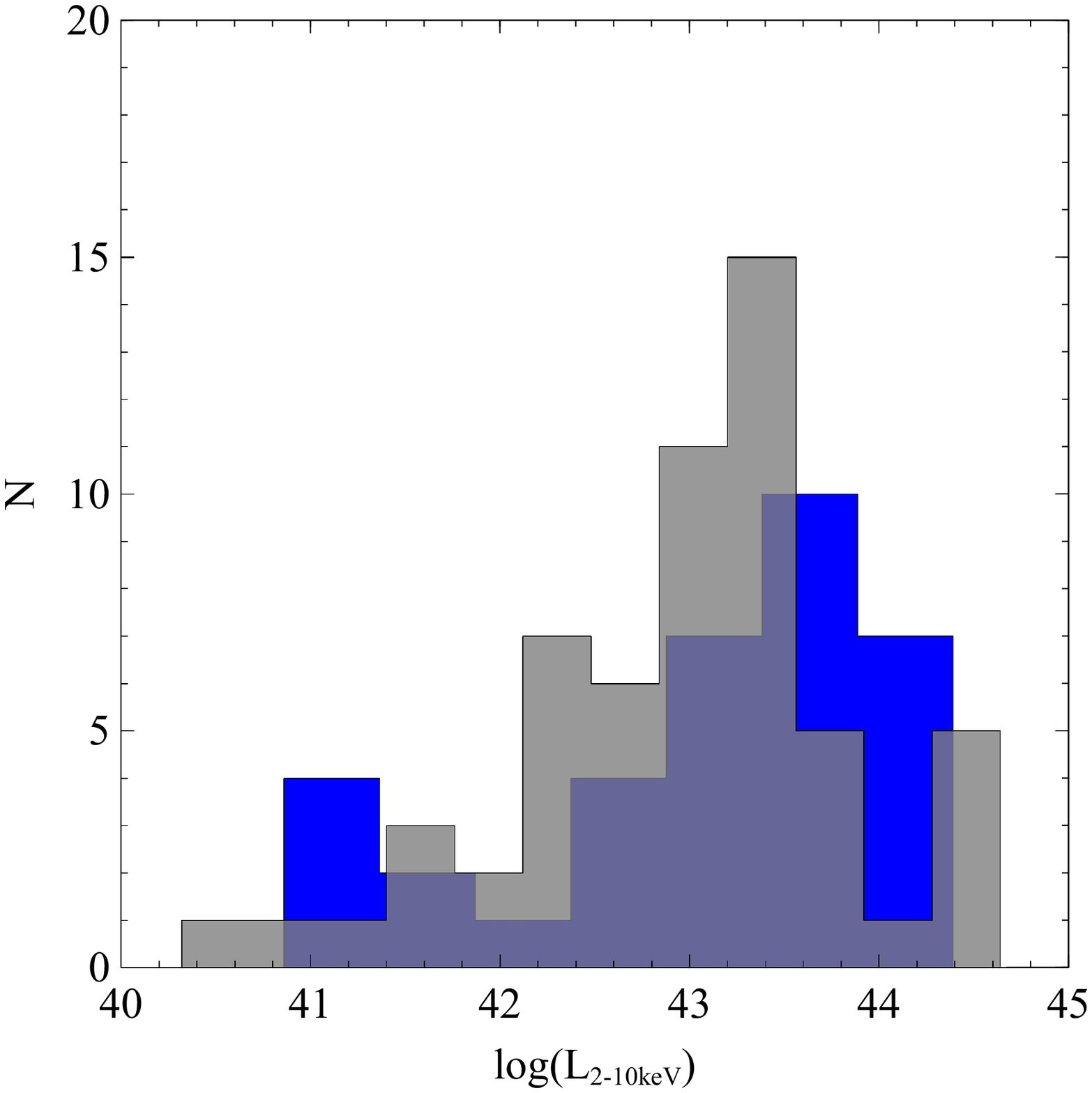}
}
\caption{
\small
Histograms of intrinsic 2--10~keV luminosity log($L_{\rm 2-10 keV}$).  The grey shaded portions represent high-absorption objects ($\rm log \thinspace N_{\rm H} \thinspace > 22$), whereas the blue shaded portions represent low-absorption objects ($\rm log \thinspace N_{\rm H} \thinspace<22$).  In the left panel, any objects with ambiguous spectral classifications and two `best-fit' $N_{\rm H}$ estimates (both a simple and a complex spectral model describe the data well) have been assumed to have the smaller of the two $N_{\rm H}$ estimates.  In the right panel, the maximal $N_{\rm H}$ is assumed for objects with ambiguous spectral classification.\label{logLX_histogram_splitonNH}}
\end{figure*}

The average 2--10~keV luminosity for the sample is $\langle \rm log(L_{\rm 2-10 keV}/\rm erg \thinspace s^{-1}) \rangle = 43.00$, with $\sigma_{\rm log L} = 0.91-0.92$, similar to the distribution seen in W09 ($\langle \rm log(L^{\rm 9month}_{\rm 2-10 keV}/\rm erg \thinspace s^{-1}) \rangle = 43.02$, with $\sigma_{\rm log L} = 0.87$), but we notice differences when we split the sample based on the absorption level or spectral complexity.  Inspecting the histograms in Figs~\ref{logLX_histogram_modeltype} and \ref{logLX_histogram_splitonNH} shows that complex/high-absorption sources appear to have a wider distribution in luminosity than simple/low-absorption sources, notably showing more of a spread to low luminosities.  Indeed, Fig.~\ref{nhvslx} demonstrates this broader distribution of absorbed sources.  The simple/complex bifurcation again closely corresponds to the low/high absorption split.  Simple/low absorption objects have an average luminosity of $\rm log(L_{\rm 2-10 keV})=43.02-43.11$, with $\sigma_{\rm log L}=0.87-0.98$ (the ranges take into account the uncertainty in spectral type for some sources).  For complex/absorbed sources, we find a slightly lower average $\rm log(L_{\rm 2-10 keV})=42.88-42.97$, $\sigma_{\rm log L}=0.89-0.96$.

%Overall distn: Average value: 43.104368932 +/- 0.107674259029
%Standard deviation: 1.08745671079
%103 objects total here. (+8 would give 111, as we have in the table)

%Simple (assuming dual types simple): 55 objects, Average value: 43.1072727273 +/- 0.116818670685
%Standard deviation: 0.858438406827

%Complex (assuming dual types simple): 43 objects, Average value: 43.1010416667 +/- 0.191284580711
%Standard deviation: 1.31138101574

%logNH<22 (assuming min NH): 43 objects, Average value: 43.0997674419 +/- 0.1447276308
%Standard deviation: 0.937942247111

%logNH>22 (assuming min NH): 58 objects, Average value: 43.0924137931 +/- 0.159834923758
%Standard deviation: 1.20672721135

%ASSUMING ALL DUAL TYPES COMPLEX:
%OVERALL: Average value: 43.1111650485 +/- 0.107742314811
%Standard deviation: 1.0881440405  (no significant change from before)

%Simple: 42 obj, Average value: 43.1307142857 +/- 0.142941975747
%Standard deviation: 0.91527522945

%Complex: 61 obj, Average value: 43.097704918 +/- 0.15491922509
%Standard deviation: 1.19999915756

%ASSUMING ALL DUAL TYPES MAX NH:

%logNH<22: 39 obj, Average value: 43.0515384615 +/- 0.155524106522
%Standard deviation: 0.958714980041

%logNH>22: 62 obj, Average value: 43.1332258065 +/- 0.151159413174
%Standard deviation: 1.18059275775

\subsection{Hidden/Buried Sources}

This class of sources was identified as a potentially important component of the X-ray background in \cite{2007ApJ...664L..79U}, and the proportion of such AGN in the BAT catalog has been discussed in \cite{2008ApJ...674..686W} and W09; the latter find that $\approx 24$\% of the 9-month BAT AGN are hidden.  The criteria employed to define such hidden AGN are that the model is complex (best fit by a partial-covering model) with a covering fraction $f \ge 0.97$ and a ratio of soft (0.5--2~keV) to hard (2--10~keV) flux $\le 0.04$.  A partial covering fraction greater than 0.97 implies a scattering fraction below 3\% and is suggestive of a geometrically thick torus or an emaciated scattering region.  We identify a total of 13-14 hidden/buried sources in our sample using these criteria.  Of these, two were previously identified as hidden sources in W09 (CGCG 041-020 and SWIFT J1309.2+1139 - NGC 4992), one narrowly missed identification as a hidden source in W09 (Mrk 417, using the same data), three were analysed in W09 but did not pass the criteria for being deemed hidden (NGC 4138, NGC 4388 and NGC 4395), and the remainder are newly identifed hidden objects (KUG 1208+386, Mrk 198, NGC 5899, NGC 4258, NGC 4686, Mrk 268, NGC 4939 and MCG +05-28-032).  For NGC 4138, we again use the same data as analyzed in W09, but the best-fit adopted by W09 is taken from the \cite{2006A&A...446..459C} study.  Cappi et al. fit NGC 4138 with a power-law plus soft-excess model, whereas our systematic model comparisons reveal that a partial covering model is clearly preferred over a soft-excess model for this particular observation.  

We point out that in this work, a uniform model-fit-comparison approach has been adopted for all objects, unlike W09, where some model fits were taken from the literature; for consistency, we perform all our analysis on the results from our own fits. For NGC 4388, we have used {\xmm} data in our study, whereas ASCA data were used previously in W09 where it narrowly escaped classification as `hidden'; we prefer our analysis of the better-quality {\xmm} data for defining the hidden status of this object.  The `hidden' classification of NGC 4395 can be called into question since it exhibits rapid variability (e.g., \citealt{2005MNRAS.356..524V}), which may argue for a reflection interpretation instead of the complex absorption seen here; we present its reflection properties in \S\ref{subsec:reflection} where we are only able to produce an upper limit on the reflection parameter ($R<0.71$).  Further detailed study of this source is needed.  MCG +05-28-032 has only XRT data and is of an intermediate model type; close inspection reveals that the covering fraction is not well determined and has a large error due to poor quality data, so it may not be hidden.
%we cannot say for sure whether this object would be classified as hidden/buried, although the soft-to-hard flux ratio would suggest so.  Better data are required on this object to make a definitive determination.

Our finding of 13--14 hidden sources indicates a percentage of hidden sources of $\approx 14$\%, lower than the 24\% fraction found in W09. However, if we assume Poissonian errors on the counts used to calculate these percentages, the proportions of hidden objects in the two samples are consistent to within $2\sigma$.  We emphasize that we have adopted a uniform strategy for fitting models to all of our data, whereas in W09 many fits were gathered from the literature.  This difference may partly explain the differing proportion of hidden sources identified. The identification of seven new hidden objects (three with newly obtained {\xmm}  data) is nevertheless interesting.  The average absorption for these hidden sources is $\rm log(N_{\rm H})=23.42$ with $\sigma_{\rm log N_{\rm H}} = 0.44$, and their average soft-to-hard flux ratio is $F_{\rm 0.5-2 keV}/F_{2-10 keV}=0.02$ with $\sigma = 0.01$, consistent with the W09 results for hidden objects.

%Hidden sources: 13 sources.

%The following are Hidden/Buried: KUG 1208+386, Mrk 198, NGC 5899, NGC 4258, NGC 4388 (in prev paper but not hidden there), NGC 4395 (not hidden there), NGC 4686, Mrk 417 (JUST failed inclusion criterion in W09), CGCG 041-020 (identified before), SWIFT J1309.2+1139 (NGC 4992 - identified before), Mrk 268, NGC 4939, MCG +05-28-032.

%How many are non-unique NH (dual model fit?)

%For hidden objects, average logNH is: 23.4145559967 +/- 0.119772035597 , stdev 0.431844215711
%Flux frac for Hidden sources, on average: 0.0200311998458 +/- 0.00311176518901 , stdev 0.0112196289462

\subsection{Detailed Features}
\label{subsec:detailedfeatures}

%We present the fraction of objects for which soft excesses, iron lines and warm absorber edges are well-detected, although we caution that since we only have sufficient signal-to-noise in our {\xmm} data to detect such features, these fractions are presented as a percentage of the total number of objects for which {\xmm} has been used (49 objects).  These results are detailed in Table~\ref{table:fitresults_features} below.  We are careful to calculate upper limiting parameters for these components in all cases where fitting with these components does not reveal a significant improvement to the fit.  This allows a more complete analysis of properties later.

We present the fraction of objects for which soft excesses, iron lines, and warm absorber edges are well detected, for the subset of objects with at least 4600 counts in the observation (39 objects).  The data sets are all from {\xmm}, and the properties of these features are detailed in Table~\ref{table:fitresults_features} below.  We are careful to calculate upper limiting parameters for these components in all cases where fitting does not reveal a significant improvement to the fit; this approach allows a more complete analysis of properties later.

\subsubsection{Iron K-$\alpha$ lines}

%We find that 64\% of these objects exhibit iron lines, compared to 81\% in W09.  For the remainder, we include a \textsc{zgauss} model fixed at an energy of 6.4~keV with a width of $\sigma = 0.1$~keV and fit the normalization.  We obtain the upper limiting equivalent width by setting the normalization to its upper limiting value as determined in \textsc{xspec}, and determining the equivalent width using the \textsc{eqwidth} command.

We find that 79\% of the objects with $>4600$ counts exhibit iron lines, which is very close to the 81\% found in W09.  For the remaining sources, we include a \textsc{zgauss} model fixed at an energy of 6.4~keV with a width of $\sigma = 0.1$~keV and fit the normalization.  We obtain the upper limiting equivalent width by setting the normalization to its upper limiting value as determined in \textsc{xspec}, and determining the equivalent width using the \textsc{eqwidth} command.

\begin{figure*}
%\figurenum{1}
\centerline{
\includegraphics[width=8.0cm]{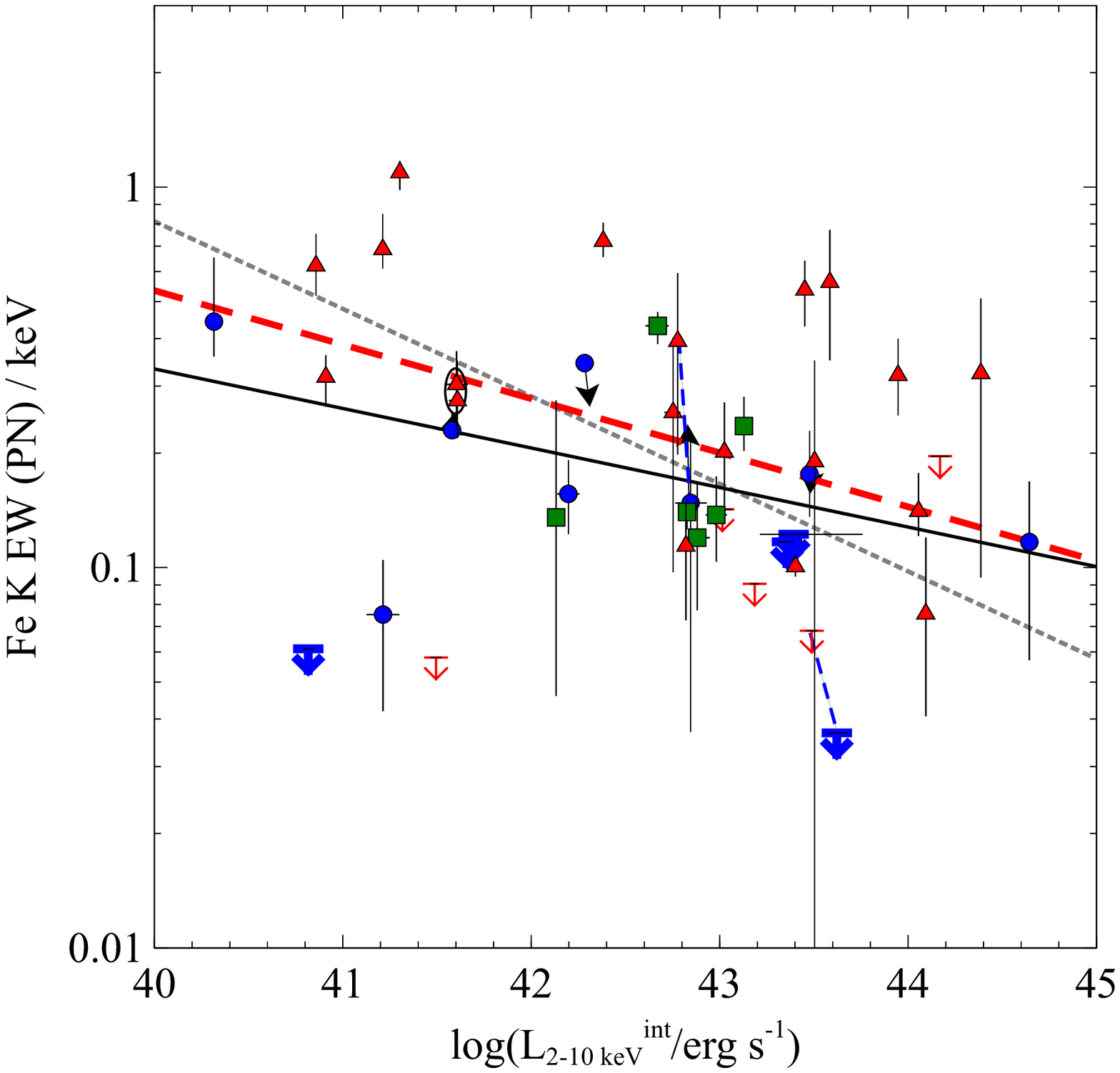}
\includegraphics[width=8.0cm]{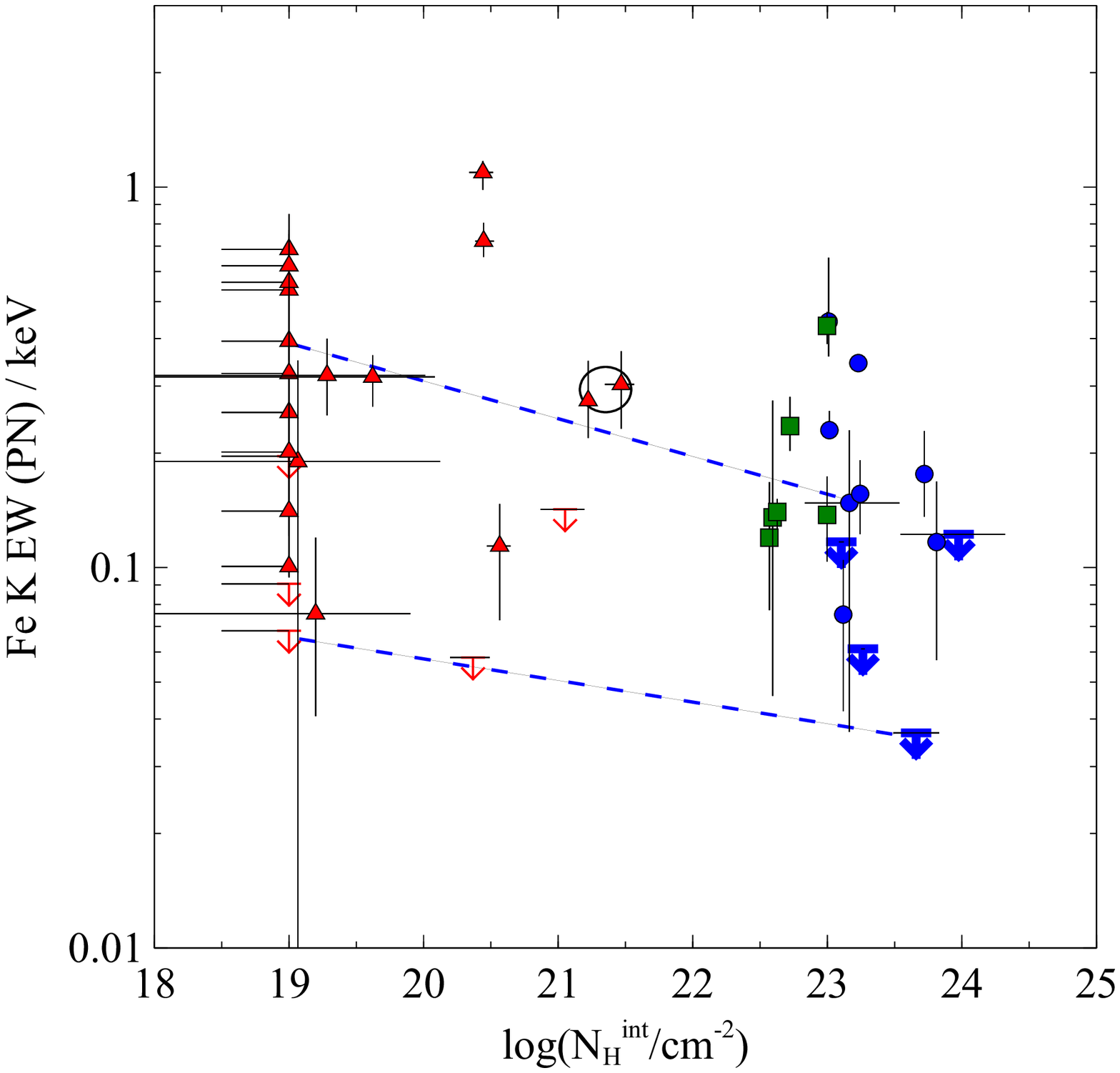}
}
\caption{
\small
Left panel: iron line equivalent width from the PN instrument (for objects with $>4600$ counts in all detectors) vs. $L_{\rm 2-10 keV}$ (absorption-corrected).  Downward pointing arrows with horizontal lines show upper limiting iron line equivalent widths wherever a source did not have a statistically significant iron line (using the \textsc{zgauss} model).  Triangles (red) show $\rm log(N_{\rm H})<22$ sources, circles (blue) show $\rm log(N_{\rm H})>23$ sources, and squares (green) show the remaining sources with intermediate absorbing columns.  Right panel: iron line equivalent width (PN) against absorbing column density; the symbols and color-coding are the same as for the left panel.  Absorbing columns below $10^{19} \thinspace \rm cm^{-2}$ are shown at $N_{\rm H}=10^{19} \thinspace \rm cm^{-2}$.\label{ironlineeqw_vs_lx}  In both figures, dashed lines or ellipses connect two different measurements for an individual source when a unique best-fit model was not found.  The black thin solid line shows a best-fit (including upper limits using {\sc asurv}) using all objects; the red dashed line shows the fit to only unabsorbed ($\rm log \thinspace N_{\rm H} \thinspace<22$) objects. The grey dotted line shows the anti-correlation found in W09, using binned data. }
\end{figure*}

From Fig.~\ref{ironlineeqw_vs_lx} (left panel), we see that there are suggestions of an anti-correlation between the iron line equivalent width and 2--10~keV absorption-corrected luminosity, known as the `X-ray Baldwin Effect' \citep{1993ApJ...413L..15I}.  However, the presence of many upper limiting equivalent widths complicates this picture.  
%If we attempt to correlate the equivalent widths with luminosity for significant iron line detections \emph{only}, we obtain a relation $\rm log (EW/\rm keV) = 3.290 - 0.091*\rm log(L_{\rm 2-10 keV})$.  This is much shallower in slope than the relation obtained in W09, and indeed the correlation is very weak, if present at all ($R^{2}=0.11$).  
Since the upper limiting equivalent widths occupy the same range as those for well-detected iron lines, we use the {\sc asurv} (\emph{Astronomy Survival Analysis}) package for censored data (from the STATCODES suite of utilities; \citealt{1985ApJ...293..192F}\footnote{http://astrostatistics.psu.edu/statcodes/sc\_censor.html}) to determine the correlation parameters. We obtain a very shallow anti-correlation of $\rm log (EW/\rm keV) = (3.678 \pm 2.406) - (0.104 \pm 0.056) \times \rm log(L_{\rm 2-10 keV})$ with a Spearman's Rank coefficient of $-0.275$ (Spearman probability 0.078).  We use the E-M algorithm whenever results from {\sc asurv} are presented; the alternative Buckley-James algorithm yields very similar results with slightly steeper slopes.  However, the definition of the equivalent width requires a good knowledge of the intrinsic continuum over which the line is detected; significant or complex absorption in many sources may make it difficult to recover the `true' iron line equivalent width in those cases. Therefore, we also check the presence of an anti-correlation for the 24 sources with $\rm log(N_{\rm H})<22$.  We find a slightly steeper relation, $\rm log (EW/\rm keV) = (5.41 \pm 3.14) - (0.14 \pm 0.07) \times \rm log(L_{\rm 2-10 keV})$ with a Spearman's Rank coefficient of $-0.356$ (Spearman probability 0.088).  These results are consistent with the \cite{2004MNRAS.347..316P} finding and later works by \cite{2006ApJ...644..725J} and \cite{2007A&A...467L..19B} for radio-quiet AGN.
%   We do not find any need to bin the data as done in W09 before a correlation is seen, at least in the low-absorption subset of objects.   Our anti-correlation is shallower than the one obtained in W09 when binning the data.}

We inspect the residuals of all of the {\xmm} fits for hints of broad iron lines that can indicate the presence of strong-gravitational processes at work in the inner part of the accretion flow near the black hole.  A systematic analysis of such lines in our sample is not presented here, but we identify four sources with {\xmm} data that display hints of complex iron lines by inspection of residuals, but have too few counts ($<4600$) to fit an iron line successfully (Mrk 417, UGC 06527, NGC 4686 and SWIFT J1309.2+1139), two sources that definitely exhibit structure in the lines beyond what is shown here using our simple \textsc{zgauss} fits (Mrk 766 and NGC 4051), and five sources that show possible signs of broad lines by inspection of residuals (NGC 5252, NGC 5273, UM 614, NGC 4151, NGC 4579).  Many of the more well-known sources have been analysed in more detail (e.g., \citealt{2007MNRAS.382..194N}, \citealt{2006A&A...446..459C}), and in other works on individual sources.

%Need to comment on Patrick, Reeves et al. (2012) paper on Suzaku Fe-K lines.

%Broad lines
%NGC 5252 - maybe?
%NGC 5273 - maybe?
%UM 614 - maybe?
%Mrk 463?
%Mrk 841?
%Mrk 417 - could not fit iron line - too few counts
%UGC 06527 - could not fit iron line - too few counts
%NGC 4051 - DEFINITELY
%NGC 4151 - POSSIBLY, PERHAPS REFL DOMINATED, BUT WE HAVE PARTIAL COVERING
%MRK 766 - DEFINITELY
%NGC 4579 - maybe
%NGC 4686 - could not model iron line - too few counts
%SWIFTJ1309.2+1139 - could not model iron line, too few counts. (NGC 4992) - Google search reveals could be Compt thick

\subsubsection{Soft excesses}

%We present soft excess properties below, using a black body component to characterise any soft excesses seen.  We find that 27 percent of the {\xmm} objects have soft excesses, compared to 41\% in W09.  This reduction of 12\% is somewhat puzzling, but could be part of the expected cosmic variance in AGN properties: we do not have any solid theory to predict what proportion of AGN should contain soft excesses based on their other properties.

Many X-ray spectra of low-absorption AGN reveal an excess at energies below $\sim$1~keV.  A number of possible origins have been suggested for this soft excess: blurred reflection (e.g., \citealt{2005MNRAS.358..211R}), complex absorption (e.g., \citealt{2007MNRAS.374..150S}) or an extension of thermal emission from the accretion disc, possibly more visible in low black hole mass AGN (e.g. Narrow Line Seyfert 1 AGN).  A definitive physical picture has not yet emerged to explain the soft excess in all AGN; we therefore employ a simple redshifted black body component (e.g., \citealt{2006MNRAS.365.1067C}) as a phenomenological description of the feature, and we do not attempt to fit the more complex models described above to the soft excess. An inventory of the observed properties of the soft excesses in our sample will allow future investigations into their physical origins. 
%We present soft-excess properties below, using a black body component (e.g., \citealt{2006MNRAS.365.1067C}) to characterize the feature.  
We find that 31--33\% of the objects with $>4600$ counts in their spectra have soft excesses (with the range due to ambiguous spectral types for a few sources), compared to 41\% in W09 (consistent within Poisson errors).  All of the objects for which soft excesses are detected have intrinsic absorbing column densities $\rm log(N_{\rm H})<21.22$, so we restrict ourselves explicitly to soft excesses seen above `unabsorbed, simple power-law' type spectra as mentioned in \S\ref{spectralfitting}.

\begin{figure}
%\figurenum{1}
\centerline{
\includegraphics[width=8.0cm]{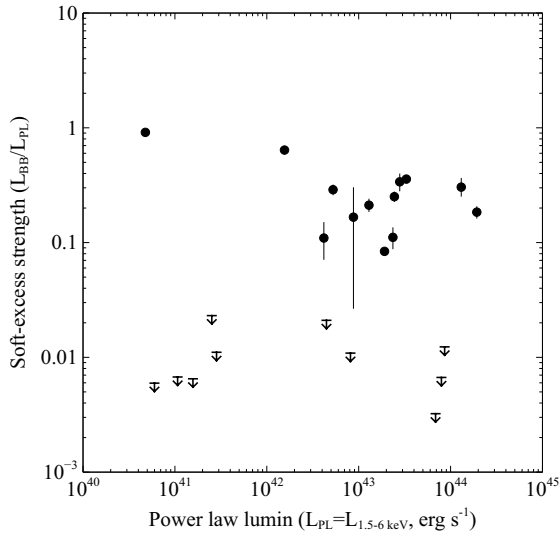}
 }
\caption{
\small
Soft-excess strength against power-law luminosity ($L_{\rm 1.5-6 keV}$).  The soft excess is modelled as a black body, and its strength is parameterized as the ratio of the luminosity in the black body component only to the luminosity in the power law between 1.5 and 6~keV (determined from the PN instrument).  Downward pointing arrows show upper limiting soft-excess strengths wherever a source did not have a statistically significant soft excess (using the \textsc{zbbody} model).\label{softx_vs_lpowerlaw}}
\end{figure}

We define the `soft-excess strength' $S_{\rm softex}$ as the ratio of the luminosity in the black body component \emph{only} (setting the power-law normalization to zero, integrating the luminosity from 0.4 to 3~keV) to the luminosity in the power-law component only (setting the black body normalization to zero), measured between 1.5 and 6~keV ($S_{\rm softex}=L_{\rm BB}/L^{\rm (PL)}_{1.5-6 keV}$); these energy ranges were selected so we could be confident of avoiding features such as edges and iron lines.  This approach is an extension of the concept presented in W09 where the power present in the soft excess was compared to that seen in the power-law component, but here we adopt the fractional measure since in some reprocessing models, we expect some fraction of the coronal power-law emission to be responsible for the soft excess.  We also extend the previous analysis by producing upper-limiting soft-excess strengths for those objects where soft excesses are not detected, and show the soft-excess strength against power-law luminosity in Fig.~\ref{softx_vs_lpowerlaw}.   The upper limits are calculated by including a black body component with a temperature fixed at the canonical soft-excess temperature 0.1~keV (e.g., as found in the sample of \citealt{2006MNRAS.365.1067C} or \citealt{2004MNRAS.349L...7G}), finding the upper limiting normalization, thereby determining the upper limiting black body luminosity.  Amongst the objects with detected soft excesses, we see a small decrease in soft-excess strength with higher power-law luminosities (in constrast to the simple proportionality seen in W09 between soft excess luminosity and power-law luminosity), but there are too few points at low luminosities to be able to constrain the slope of any such trend.  This result may suggest that a straightforward reprocessing scenario, where the power in the power-law component (due to the corona) is somehow recycled in the soft excess, is not favored.  However, this hypothesis would be better tested on a larger sample of objects.
% a simple fit yields {\bf $S_{\rm softex} = 10^{14.7} \times L_{\rm 1.5-6 keV}^{-0.36}$, $R^{2}= 0.015$} 
% {\bf We note that the power-law index of this relation is different to that expected from plotting 1/$L_{\rm PL}$ against $L_{\rm PL}$, which would yield $\sim-1.0$ instead of -0.36.

The remainder of the objects (for which upper limits have been calculated) occupy a completely different part of parameter space than those with detected soft excesses.  The objects with upper-limiting soft-excess strengths span a range of luminosities and lie well below the anti-correlation between $S_{\rm softex}$ and $L_{\rm 1.5 - 6~keV}$ for the objects with detected soft excesses. This disjoint distribution suggests that there are two classes of objects: those which show measurable, statistically significant soft excesses, in which there is a weak anti-correlation between soft-excess strength and luminosity, and those without any detectable soft excess, in which the stringent upper limits on the soft-excess strength dictate that there can be little scope for any kind of correlation or anticorrelation in those sources.  This dichotomy suggests that the physical process responsible for the soft excess occurs in some AGN only, and that a soft excess is not an intrinsic part of all AGN spectra.  We return to this issue in a companion paper on the relation between reflection/absorption and soft-excess strength in our sample (see \S\ref{subsec:reflection}).

\begin{figure}
%\figurenum{1}
\centerline{
\includegraphics[width=8.0cm]{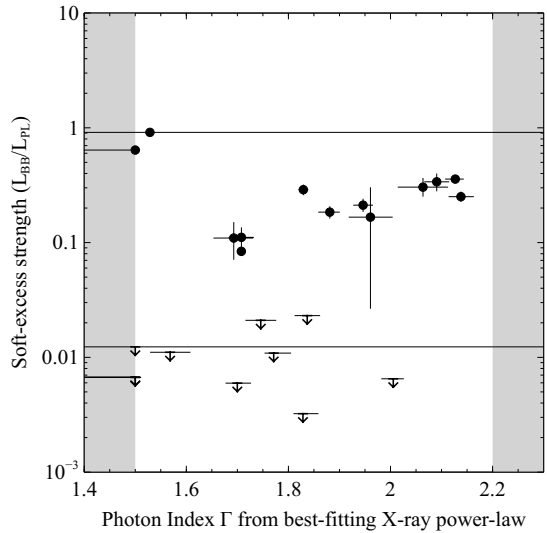}
}
\caption{
\small
Soft-excess strength against photon index $\Gamma$.  The soft-excess strength is parameterized as described in Fig.~\ref{softx_vs_lpowerlaw}.  Downward pointing arrows show upper limiting soft-excess strengths wherever a source did not have a statistically significant soft excess (using the \textsc{zbbody} model).\label{softx_vs_Gamma}}
\end{figure}

We also want to ensure that the stronger soft excesses are not biased to being found in sources with harder spectra ($\Gamma < 2.0$): a harder spectrum leaves more scope for soft features to be seen as an excess.  We see from Fig.~\ref{softx_vs_Gamma} that this is not the case; for the two objects with the largest soft-excess strengths (NGC 4051 and Mrk 766, which also happen to show pronounced spectral variability) we do indeed see that they have hard ($\Gamma = 1.5$) spectra, but for all other objects, the opposite trend seems to be evident, i.e., the strength of the soft excess increases with $\Gamma$ (as the spectrum gets softer), in line with the trend previously found in \cite{1997MNRAS.285L..25B}.

\begin{figure*}
%\figurenum{1}
\centerline{
\includegraphics[width=8.0cm]{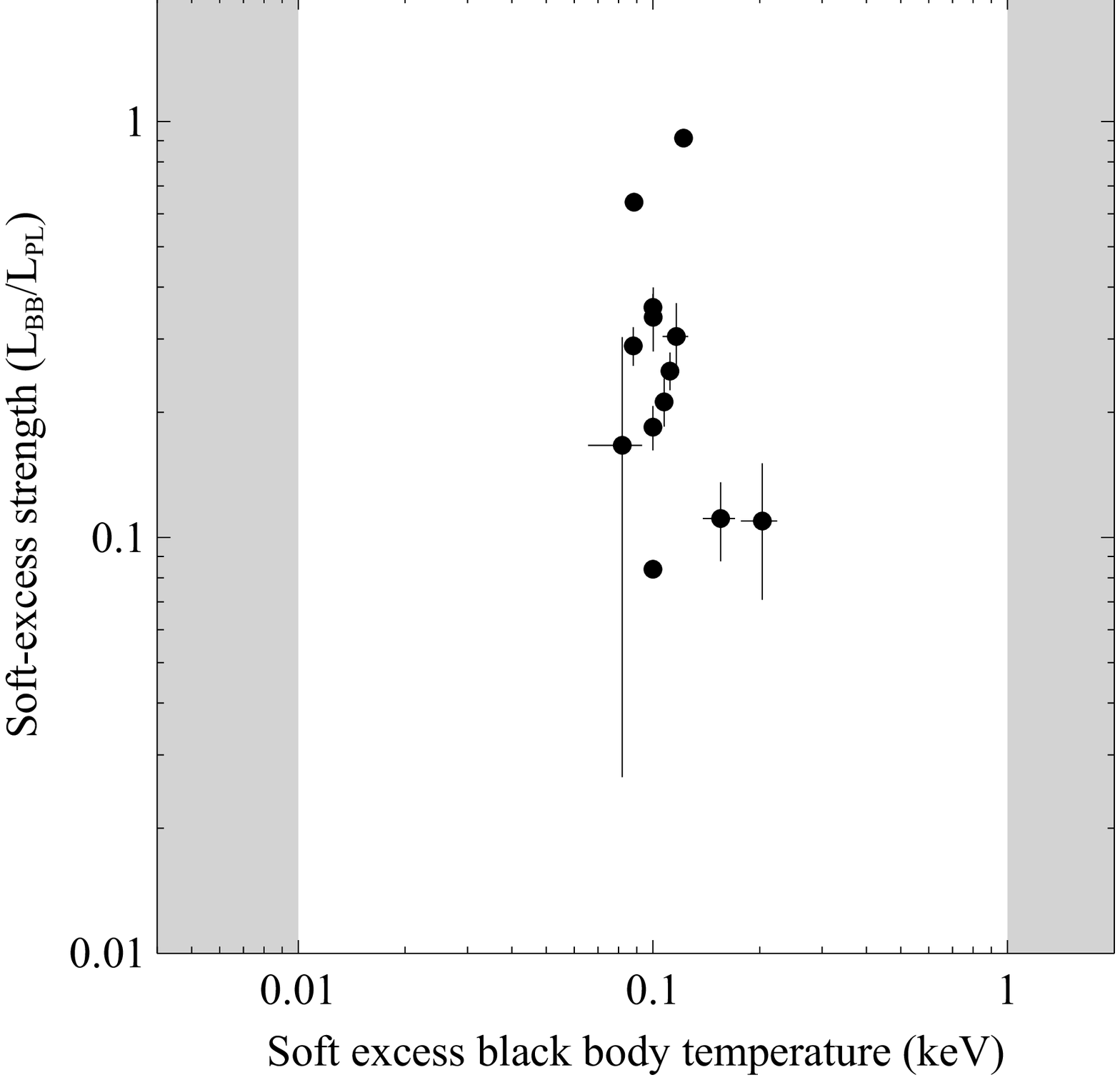}
\includegraphics[width=8.0cm]{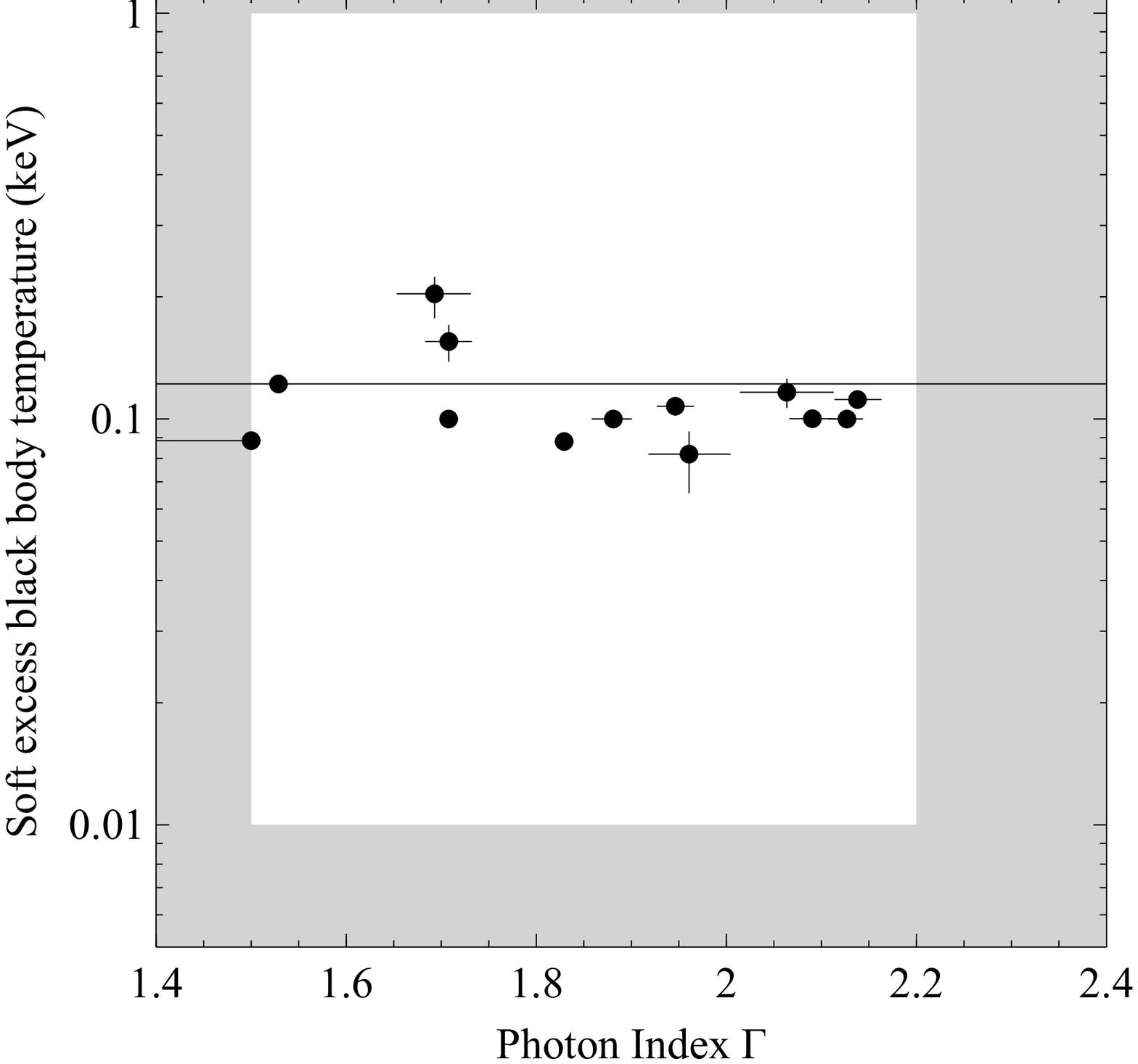}
}
\caption{
\small
Left: soft-excess strength against soft excess temperature (using the black body model as a parameterization for the soft excess).  The soft excess temperatures are clustered at $\sim 0.1$~keV, as found in previous works.  Right: soft excess black body temperature against photon index $\Gamma$.\label{softx_vs_BBtemp}}
\end{figure*}

We lastly investigate whether there is any relation between the temperature of the soft excess and the soft-excess strength or $\Gamma$.  In line with what has been found previously (e.g., W09, \citealt{2006MNRAS.365.1067C}, \citealt{2004MNRAS.349L...7G}), we find a narrow distribution of temperatures ($\langle kT_{\rm softex} \rangle = 0.113, \sigma_{\rm kT} = 0.032$) and can identify no discernible correlation with either soft-excess strength or photon index.   This result lends credence to the idea that the soft excess is not part of the direct, thermal accretion disk emission since the temperature at which it is seen is uniformly and tightly clustered around $\sim 0.1 \rm keV$.
%Black body temps:

%Average: 0.113108442857
%Stdev: 0.0315920857683
%Max: 0.203468
%Min: 0.0819341

\subsubsection{Ionized absorber edges}

We present the edge depths due to ionized absorbers in Table~\ref{table:fitresults_features}.  \cite{2011ApJ...728...28W} noted that X-ray observations do not offer the same sensitivity for picking up warm absorbers that, for example, UV spectroscopy does (such as COS spectra used in \citealt{2011ApJ...728...28W}).  However, the detection rates and properties here can at least serve as an indicator of what is detected in X-rays using an unbiased sample and can be compared with UV studies.  We find that 18\% of our high-counts subsample show warm absorber signatures in the form of an $\rm O VII$ edge at 0.73~keV.  Only 8\% exhibit a significant additional warm absorber edge at 0.87~keV.  However, this is measured as a fraction of all sources with $>4600$ counts; if we instead only consider the 21--23 unabsorbed ($\rm log \thinspace N_{\rm H} \thinspace < 22$) sources within this high-counts subset, we find that the fraction of such sources with at least one well-detected ionized absorber edge rises to $\sim 32 \%$ (two sources with ambiguous $N_{\rm H}$ values introduce an uncertainty of $\pm 1 \%$ to this figure). The study of Seyfert 1-1.5 BAT-selected AGN presented in \cite{2012ApJ...745..107W} reveals that $53 \%$ of their 48 sources exhibit a detectable ionized absorber edge using {\xmm} and \emph{Suzaku} X-ray data; the pioneering work of \cite{1997MNRAS.286..513R} finds similarly that $50\%$ of their sources exhibit such edges in \emph{ASCA} spectra, and the work of \cite{1999ApJ...516..750C} using UV spectroscopy from \emph{HST} reveals that $47 \%$ of their sources show evidence for such absorbers.  The latter two studies are not, however, from an unbiased sample, so we emphasize the utility of \cite{2012ApJ...745..107W} and this work on BAT-selected AGN.  Considering the small sample sizes involved in all the above studies, our finding that $\sim 32 \%$ of unabsorbed sources have measurable ionized absorption is broadly consistent with previous findings.

The sources that do exhibit well-detected warm absorbers are clustered around luminosities of $L_{\rm 2-10 keV}=10^{43.8} \thinspace \rm erg \thinspace s^{-1}$, with only two sources having luminosites above $10^{44} \thinspace \rm erg \thinspace s^{-1}$.  The upper limiting optical depths of the $\rm O VII$ edge for the remaining sources show increasingly stringent upper limits that indicate less scope for ionized absorption at higher luminosities.  This is consistent with a general picture where absorption of any kind (neutral or ionized) is less prevalent at high intrinsic luminosities (e.g., \citealt{2012ApJ...745..107W} and \S\ref{subsec:abscolumn} of this work).

\subsection{New {\xmm} observations}
\label{subsec:newxmmobs}

Our sample of 100 objects in this study includes 13 objects with new {\xmm} observations, gathered specifically for improving the coverage of the BAT catalog at $b>50^{\circ}$.  We highlight some of the interesting features of these 13 objects here.  Two of these {\xmm} datasets have been studied in more detail in other works already: Mrk 817 (\citealt{2011ApJ...728...28W}, a multi-wavelength study including UV spectroscopy from the Cosmic Origins Spectrograph - \emph{COS} along with \emph{HST} and \emph{IUE} archival data, looking for outflows and broad-band variability in this source) and NGC 3758 (also known as Mrk 739, \citealt{2011ApJ...735L..42K}, identifying a faint counterpart Mrk 739W using high spatial resolution {\chandra} data thereby identifying a dual AGN in this system).    The remaining new {\xmm} datasets reveal diverse properties for these 13 objects (including three hidden AGN, KUG 1208+386, Mrk 198, NGC 5899 and one Compton-thick candidate NGC 4102). One object of particular note is Mrk 50, a very bright unabsorbed Seyfert 1 galaxy with a soft excess but no measurable Iron line; this object would benefit from further detailed study.

\section{Including the BAT data}
\label{sec:includingbat}

\subsection{Compton-thick sources}
\label{subsec:comptonthick}

Using our simple \textsc{ztbabs} model to model absorption, we identify eight Compton-thick sources ($N_{\rm H}>1.4 \times 10^{24} \rm cm^{-2}$): NGC 4102, 2MASX J10523297+1036205, 2MASX J11491868-0416512, B2 1204+34, MCG -01-30-041, MRK 1310, PG 1138+222 and UGC 05881; NGC 4941 and NGC 5106 may be Compton thick within the errors on their fitted column densities.  This constitutes $\sim 9\%$ of our sample; this is a notable change from W09 (where no Compton thick sources are detected by this definition) and \cite{2011ApJ...728...58B} where 4.6\% of their sample (the 36-month BAT catalog AGN) are Compton thick.   However, we have not taken into account the effect of Compton scattering for our heavily absorbed sources; more sophisticated models such as \textsc{plcabs} or \textsc{MyTorus} \citep{2009MNRAS.397.1549M} or that presented by \cite{2011MNRAS.413.1206B} are required to model these effects.  \cite{2011ApJ...728...58B} examine the effect of using \textsc{MyTorus} for a fraction of their sources and also use the built-in \textsc{xspec} model \textsc{cabs} to take Compton scattering into account.  \textsc{MyTorus} and similar models are sufficiently complex to warrant a separate study since they have many more parameters than simple absorption models, and determining these parameters individually for each object is beyond the scope of this study.   However, a number of other measures have also been used to identify Compton-thick sources, such as a high Fe-K$\alpha$ line equivalent width, a flat photon index (here we take this to mean $\Gamma$ pegs at the minimal value of 1.5) or an unusually high reflection fraction ($R>1$) (see W09 and Table~\ref{table:reflection} for reflection values).  Using these alternative metrics, W09 note that the proportion of Compton-thick AGN in the 9-month catalog could increase to 6\%, although they do not use BAT data to constrain absorption in the highly absorbed sources. We employ these metrics with our sample and include our full 0.4--200~keV band fits and find that a few more sources may be Compton-thick: Mrk 766, NGC 4051, UM 614, Mrk 744, NGC 3227, CGCG 041-020.  We caution that we can only use the Fe-K$\alpha$ equivalent width metric on the objects with {\xmm} data since XRT data are not of sufficient quality to analyse these line properties, and in this study we restrict ourselves to reflection fits to objects with {\xmm} data.  The object Mrk 766 exhibits all three of these alternative signatures despite having a low column density ($\rm log \thinspace N_{\rm H} \thinspace <21$).  Mrk 766 is known to be highly variable and has been extensively studied in the literature (e.g., \citealt{2007A&A...475..121T}, where the pronounced variation in spectral shape between epochs is shown); this conclusion also holds for NGC 4051 which exhibits both a broad Iron line and a high reflection fraction, alongside well-studied variability favouring the reflection scenario \citep{2006MNRAS.368..903P}.  There are four sources with both high reflection and $\Gamma=1.5$ (UM 614, Mrk 744, NGC 3227 and CGCG 041-020), the last three of which have absorptions at or above log($N_{\rm H}$)=23; they may also be good Compton-thick candidates.  The object B2 1204+34, which only has XRT data, also exhibits $\Gamma=1.5$ reinforcing its status as Compton-thick.  Based on these considerations, we may add 2--6 sources to the existing list of $\sim 9$ with measurably Compton-thick column densities, to yield a Compton-thick fraction of 11-15\%.  However, what is ultimately required is the use of models that fully include Compton scattering and all the possible types of reflection, coupled with an understanding of the geometry in each source, to determine the true Compton-thick fraction.

\begin{figure*}
%\figurenum{1}
\centerline{
\includegraphics[width=8.0cm]{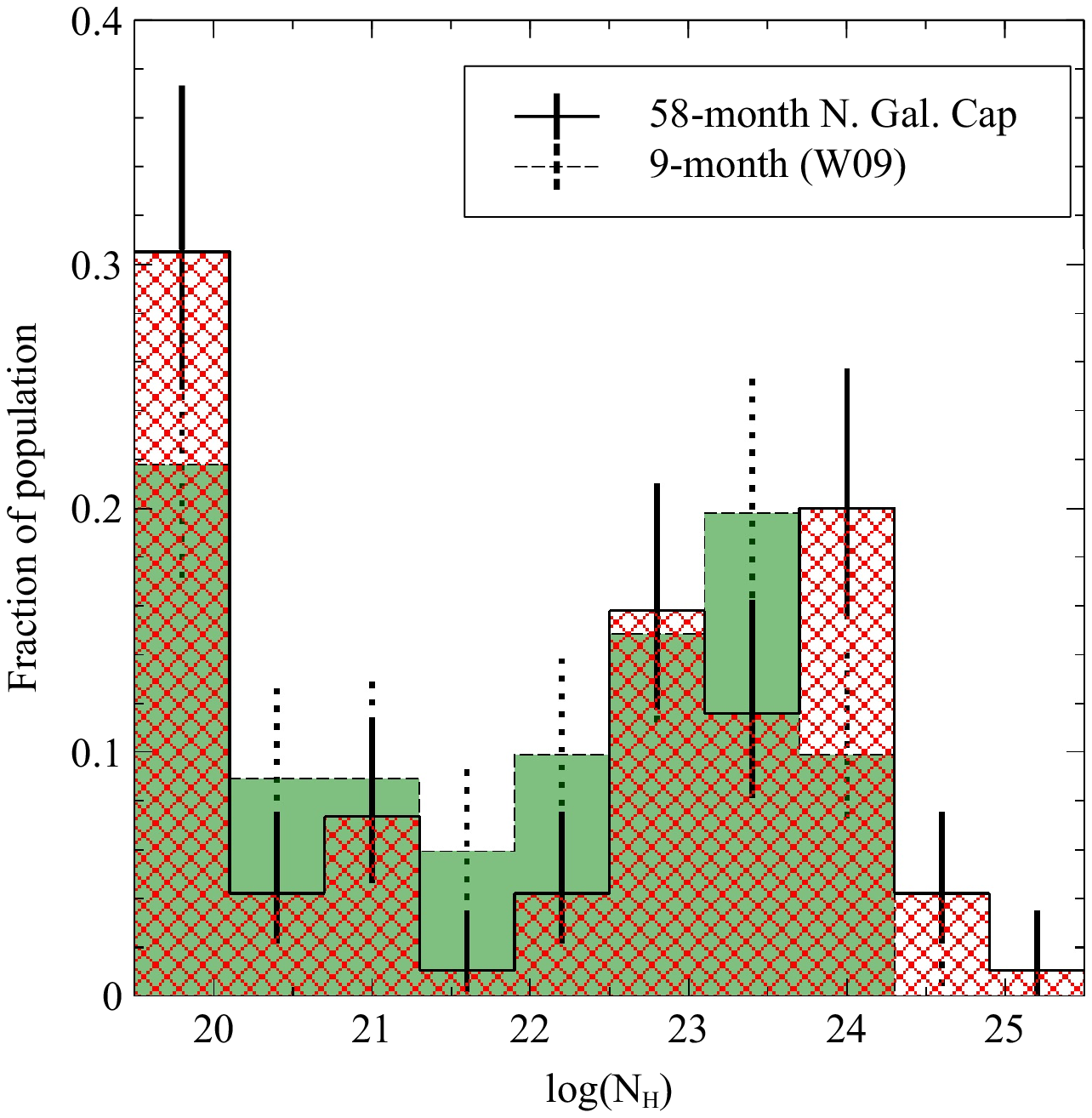}
\includegraphics[width=8.0cm]{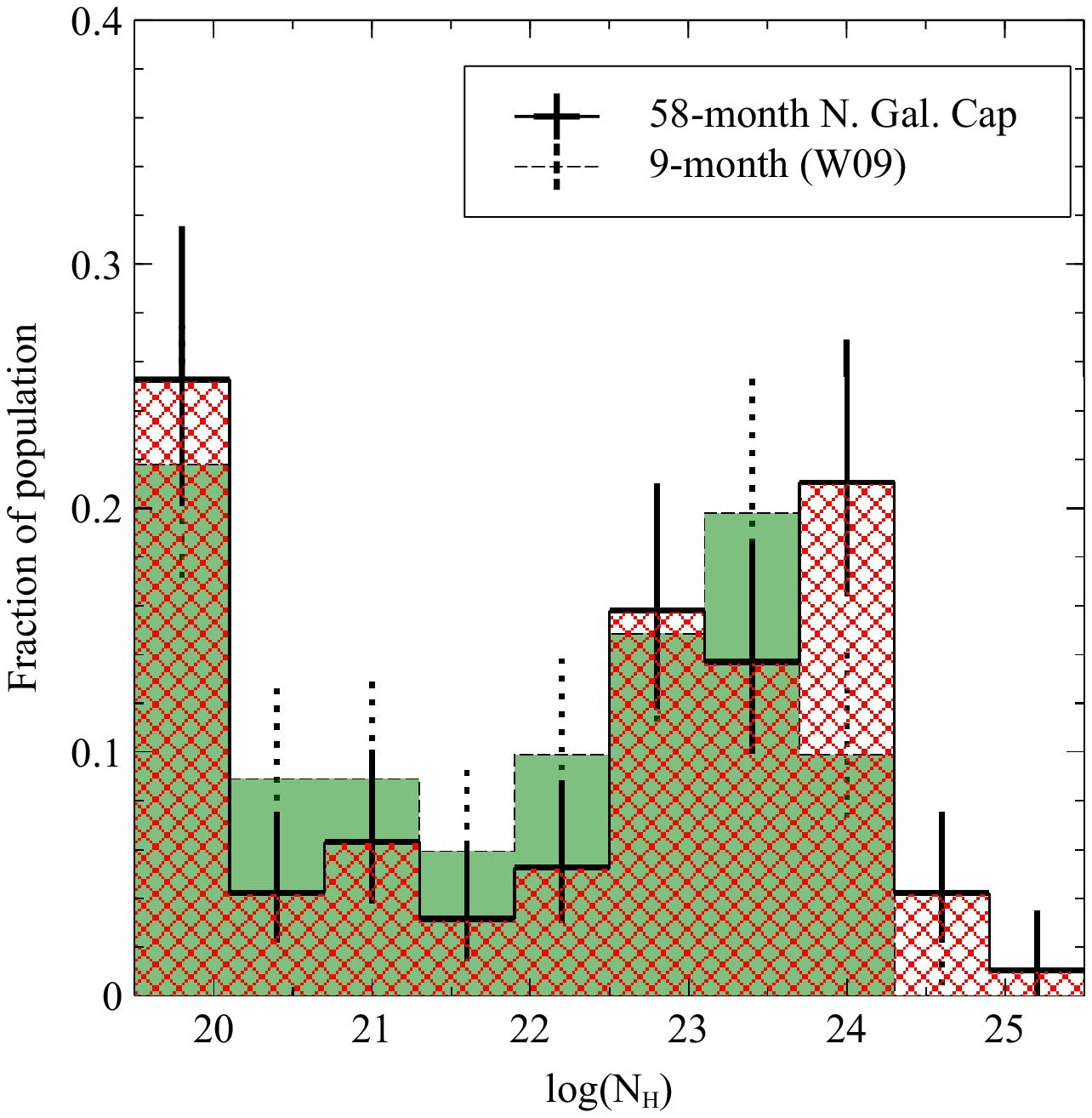}
}
\caption{
\small Comparison of log($N_{\rm H}$) distributions between the 58-month catalog (red diagonal cross-hatching shading, solid error bars) and the 9-month catalog (solid green shading, dotted error bars).  The left panel shows the 58-month column density distribution assuming that any ambiguous spectral-type objects have the `simpler' model, and the right shows the results assuming that such objects are better represented by their `complex' (partial-covering absorption) model fit.
\label{Hist_logNH_COMPARE9MONTH}}
\end{figure*}

In Fig.~\ref{Hist_logNH_COMPARE9MONTH}, we show how the log$(\rm N_{\rm H})$ distributions vary between the 9-month (W09) and 58-month ($b>50$) catalog subsamples. We present these distributions as a fraction of the total number of objects in each sample for easy comparison.  As before, we allow for the uncertainty in $N_{\rm H}$ for some sources by assuming any sources with ambiguous spectral types have the `simple' absorption fit in the left panel and the `complex' (partial covering) absorption fit in the right panel.  The distributions for the earlier catalog and the 58-month catalog are consistent within errors (calculated according to the \citealt{1986ApJ...303..336G} Poisson approximation) for all columns up to $\rm log(N_{\rm H})\approx 23.6$; the bin centered on $\rm log(N_{\rm H})\approx 24$ shows a twofold increase in objects in the 58-month catalog, and a discrepancy remains even when errors are taken into account.  Our 58-month catalog does show some objects at even higher columns, but the numbers are small.  In general, these results show how the increased sensitivity of the 58-month catalog allows detection of intrinsically bright objects with $\rm log(N_{\rm H})\gtrsim 23.6$.  As discussed in \cite{2011ApJ...728...58B}, the increased proportion of highly absorbed sources as the BAT catalog deepens in exposure suggests that the BAT hard X-ray survey is still missing a significant fraction of $\rm log(N_{\rm H})>24$ sources.  We constructed simple simulations to estimate the fraction of such missed sources at higher columns, by assuming a given underlying (intrinsic) BAT flux and $N_{\rm H}$ distribution and using the `attenuation' factors for the BAT flux for different column densities (ratios for the observed-to-intrinsic BAT fluxes calculated using the \textsc{MyTorus} model).  We attempt to recover the observed absorption distribution for different flux limits corresponding to the 9-month, 58-month, and future, deeper surveys.  However, we discover that the precise degree of improvement expected with deepening exposure is difficult to predict and relies on a realistic source input BAT flux distribution.  We defer publishing of this simulation to a future study.

\subsection{Reflection}

\label{subsec:reflection}

\begin{figure}
%\figurenum{1}
\centerline{
\includegraphics[width=8.0cm]{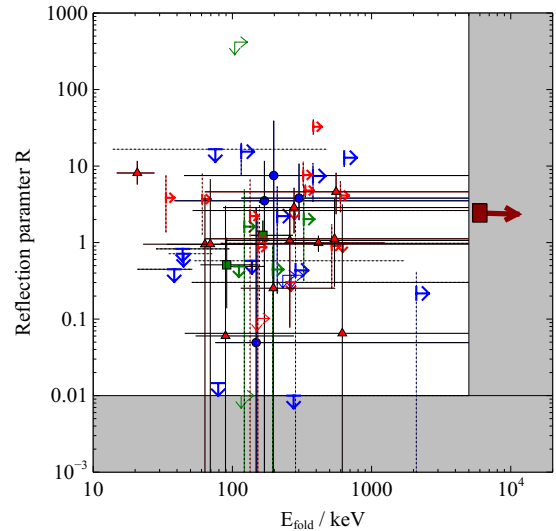}
}
\caption{\small Reflection parameter $R$ against folding energy $E_{\rm fold}$ from our \textsc{pexrav} fits.  Red triangles and small upper/lower limit arrows depict low absorption ($\rm log \thinspace N_{\rm H} \thinspace < 22$) objects, blue circles and large upper/lower limit arrows depict high absorption objects ($\rm log \thinspace N_{\rm H} \thinspace > 22$) and green squares and intermediate-sized upper/lower-limit arrows represent objects with intermediate absorptions between these two limits.  The dark red square shows the mean $R$ and $E_{\rm fold}$ for the whole sample; the `formal' mean $E_{\rm fold}$ lies well outside the BAT band.  We plot any objects with $R<0.01$ as an upper limit at $R=0.01$, and for objects $E_{\rm fold}$ above 5000~keV, we treat these fold energies as unobservable and compute a lower limit on $E_{\rm fold}$ using the negative error; these limits are indicated by the gray shaded areas. \label{Refl_vs_Efold}}
\end{figure}

\begin{figure}
%\figurenum{1}
\centerline{
\includegraphics[width=8.0cm]{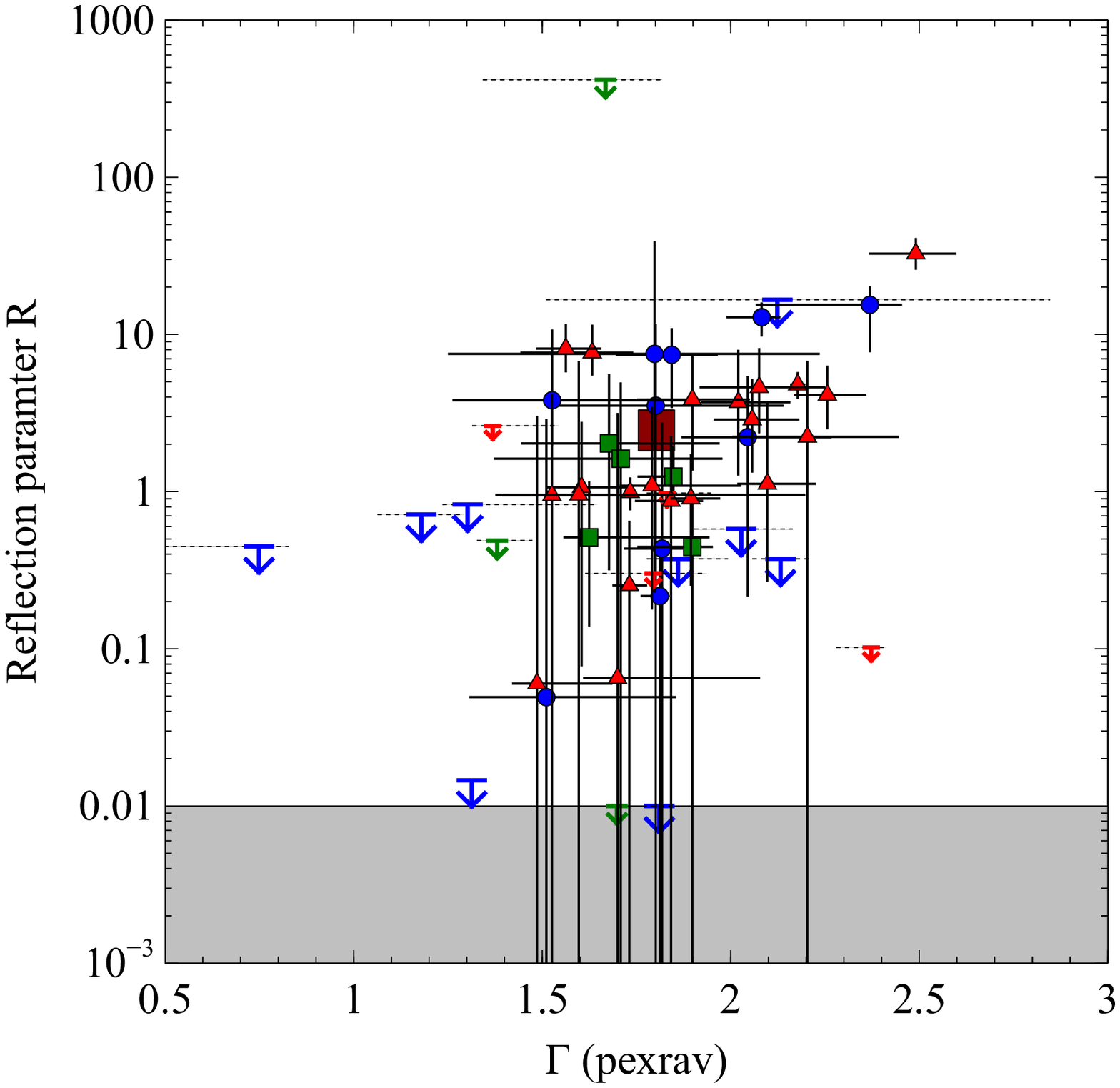}
}
\caption{
\small
Reflection parameter $R$ against photon index $\Gamma$ from our \textsc{pexrav} fits.  Red triangles and small upper/lower limit arrows depict low absorption ($\rm log \thinspace N_{\rm H} \thinspace < 22$) objects, blue circles and large upper/lower limit arrows depict high absorption objects ($\rm log \thinspace N_{\rm H} \thinspace > 23$) and green squares and intermediate-sized upper/lower-limit arrows represent objects with intermediate absorptions between these two limits.  The dark red square shows the mean $R$ and $\Gamma$ for the whole sample.\label{Refl_vs_gamma}.  We plot any objects with $R<0.01$ as upper limits at $R=0.01$ (emphasized by the gray shaded area).}
\end{figure}

\begin{figure}
%\figurenum{1}
\centerline{
\includegraphics[width=8.0cm]{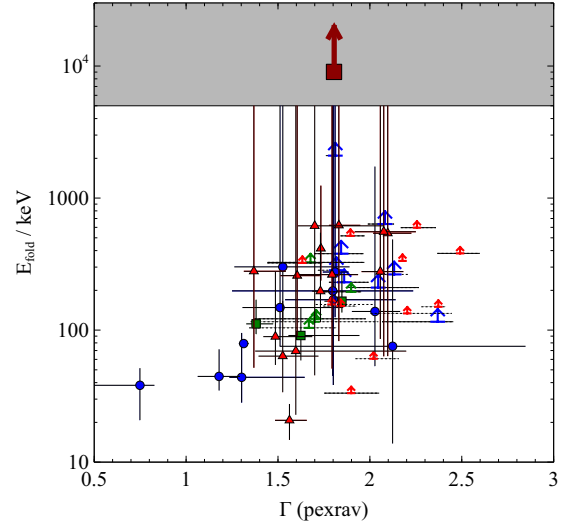}
}
\caption{
\small
Folding energy $E_{\rm fold}$ against photon index $\Gamma$ from our \textsc{pexrav} fits.  Red triangles and small upper/lower limit arrows depict low absorption ($\rm log \thinspace N_{\rm H} \thinspace < 22$) objects, blue circles and large upper/lower limit arrows depict high absorption objects ($\rm log \thinspace N_{\rm H} \thinspace > 23$) and green squares and intermediate-sized upper/lower-limit arrows represent objects with intermediate absorptions between these two limits.  The dark red square shows the mean $E_{\rm fold}$ and $\Gamma$ for the whole sample.  For objects for which the value of $E_{\rm fold}$ is above 5000~keV, we treat them as having an unobservable fold energy outside the BAT band and use the negative error bar to compute a lower limit on $E_{\rm fold}$.  The 5000~keV limit is indicated using the gray shaded area. \label{Efold_vs_gamma}}
\end{figure}

The hard X-ray BAT data provide an opportunity to constrain the reflection properties of the sample, since the Compton reflection hump peaks in the BAT band.  We present the reflection properties for the subset of objects with {\xmm} data.  We fit a \textsc{pexrav} model to the combined {\xmm} and BAT data, again renormalizing the BAT spectrum wherever possible (and linking BAT and PN normalizations).  In our \textsc{pexrav} fits, we allow the reflection fraction $R$, the photon index $\Gamma$, the folding energy $E_{\rm fold}$ and the normalization to vary, fix the redshift of the source and freeze all other parameters (abundance of elements heavier than Helium relative to solar abundances, iron abundance and cosine of the inclination angle) at their default values.  We ignore any data below 1.5~keV to avoid soft excesses or edges and ignore data between 5.5 and 7.5~keV to avoid the iron line and edge which are not modelled by \textsc{pexrav}.  We include either a simple absorption component or a partial covering absorber, depending on whether the basic fit from Table~\ref{table:fitresults} indicates that the 0.4--10~keV spectral shape is simple or complex, respectively.  For complex objects, we seed the absorbing column density with the value of $N_{\rm H}$ obtained from our 0.4--10~keV fits.   We do not impose any restriction on $\Gamma$ as done before in the power-law fits, since we wish to use $\Gamma$ here purely to constrain the spectral shape, and can then more easily probe any correlations between the different reflection parameters.  For objects where the error calculation on $R$, $E_{\rm fold}$ or $\Gamma$ fails, we perform a detailed contour-plot using the \textsc{steppar} command in \textsc{xspec} to better constrain the parameters.   We present the results in Table~\ref{table:reflection} and show plots of the three key variables in the reflection scenario, $R$, $\Gamma$ and $E_{\rm fold}$ in Figs.~\ref{Refl_vs_Efold}, \ref{Refl_vs_gamma} and \ref{Efold_vs_gamma}.

We find that reflection can be constrained in a large fraction of our sources, and that for the majority of the sample, some level of reflection is required ($R>0$).  The average value of the reflection parameter for the sample is $\langle R \rangle = 2.7 \pm 0.75$, indicating that reflection is important in this unbiased sample of AGN, in both absorbed and unabsorbed objects.  This large average value indicates that strong reflection involving light bending may be common ($R>1$) or that highly complex absorption could be important.  In our dataset, there are no `strong' reflection ($R>4$) sources with sufficient signal-to-noise ratio (when considering the combined {\xmm} and BAT data) to distinguish these two possibilities.  However, the soft excess has been linked to reflection (e.g., \citealt{2005MNRAS.358..211R}), and in a companion study we aim to explore how a comparison between the soft-excess strength and the reflection parameter may allow determination of which of the two scenarios, reflection or absorption, is favored in individual objects.  Inspection of Fig.~\ref{Refl_vs_Efold} reveals that absorbed objects (blue filled circles) exhibit similar ranges in reflection parameter as less absorbed sources.    The average fold energy is well above the maximum energy of BAT, indicating that we cannot reliably detect high-energy cut-offs using BAT data alone.  There is significant uncertainty in determining $E_{\rm fold}$, since when the fit yields a value outside the BAT band, we typically see very large error bars on the fit value. In Table~\ref{table:reflection}, we have shown objects with $E_{\rm fold}>5000$~keV (arbitrarily chosen) as lower limits, using their negative error bar to determine the lower limit in $E_{\rm fold}$.  For some objects with fold energies below 5000~keV, the positive error bar on $E_{\rm fold}$ is difficult to constrain due to the poor sampling in the BAT band.  For those objects, we do not plot a positive error bar but note that the fold energy is difficult to determine in those cases.  The average photon index from \textsc{pexrav} for the whole sample is 1.80 with a standard deviation of 0.32; the 1$\sigma$ range in photon indices falls within within the limits we previously imposed when performing our 0.4--10~keV fits.   There is no evidence of any correlations between any of the three parameters $R$, $E_{\rm fold}$ and $\Gamma$ (see Figs.~\ref{Refl_vs_gamma}, \ref{Efold_vs_gamma}).  Given the limited signal-to-noise ratio and bandpass of the BAT data, one finds that one cannot separately constrain the three components of the {\sc pexrav} model using BAT data alone, and that degeneracies exist between these three parameters.  However, inclusion of the 0.4--10~keV data constrains the fit better, and the BAT re-normalization removes one free parameter and provides better constraints on fit parameters in many cases.  Overall, our reflection fits show that reflection is present in some form in the majority of our sample.  The fold energy is generally outside the BAT range and we only observe a few definitive high-energy cut-offs in our BAT data.  Our determinations of $R$ and $E_{\rm fold}$ are given more credence by using the 0.4--10~keV to help lock down the overall broad-band X-ray spectral shape (via the photon index $\Gamma$).  

%Previous attempts to constrain these parameters using only BAT data have met with difficulty due to the degeneracy between parameters and displayed correlations between them (R. Mushotzky, private communication),

We also briefly contrast our results with those of \cite{2012ApJ...745..107W}, who fit a reflection model to a sample of Seyfert 1-1.5 AGN using a slightly different approach (fitting a model combination {\sc tbabs(ztbabs(cutoffpl + zbbody + zgauss + pexrav))} in {\sc xspec}) to simultaneously model the soft excess, iron line, power-law, absorption and reflection, with edges added where needed).  For the 8 objects that are common between our sample and theirs, they employ \emph{Suzaku} data alongside BAT data and fit them with the above model combination.  All of the common objects were therefore observed at different epochs in \cite{2012ApJ...745..107W} than presented here.  We note that some objects exhibit very similar reflection fractions in the two studies (e.g., NGC 4151, NGC 4593, NGC 5548, Mrk 841) but some objects display large discrepancies (e.g., NGC 4051, Mrk 766).  This could be in part be attributed to re-normalization employed here providing new constraints on the fit, but also due to intrinsic changes in the spectrum between observations.  NGC 4051 and Mrk 766 in particular are known for their strong variability in both spectral shape and intensity (see \S\ref{subsec:comptonthick} above), which is capable of producing such pronounced spectral changes.

Another interesting comparison can be made with the \cite{2011A&A...532A.102R} analysis of \emph{INTEGRAL} AGN detected in the 15--1000~keV band by the observatory's soft $\gamma$-ray imager instrument. Their work differs from ours in that they use only hard X-ray spectra ($>10$~keV) to calculate reflection parameters, whereas we restrict ourselves to the subset of objects with {\xmm} data below 10~keV to provide an additional constraint on reflection parameters.  Additionally,the BAT survey is flux-limited and taken across the whole sky, unlike the \emph{INTEGRAL} survey which consists of pointed observations of sources of interest. The 165 Seyfert galaxies in the \emph{INTEGRAL} sample show a very similar average photon index of $\Gamma \approx 1.8$ and an unobservable average hard X-ray cut-off ($E_{\rm C} \gtrsim 200$~keV).  They also show a range of reflection values, and in particular find that moderately-obscured Seyfert 2 nuclei (with $23 < \rm log \thinspace N_{\rm H} \thinspace < 24$) have a higher reflection component than other classes of AGN, with $\langle R \rangle = 2.2^{+4.5}_{-1.1}$ for this class of sources.  We do not notice any such trend in our sample and find a broad distribution of reflection amplitudes at every $\rm log \thinspace N_{\rm H}$ level, but our {\xmm}+BAT sample is significantly smaller than theirs (49 AGN vs. 165 AGN) and we therefore may not have sufficient statistics to make a detailed comparison. We defer more detailed discussion of this issue to a companion paper on the stacked spectrum of the Northern Galactic Cap BAT AGN (Vasudevan et al. 2013 in prep).

Both our results and those of \cite{2011A&A...532A.102R} may be subject to selection bias when using hard X-ray selection, since the effect of reflection is to boost the flux in the BAT band.  Using a simple {\sc pexrav} model in {\sc xspec}, we determine the BAT flux boost factor for different reflection strengths, assuming $\Gamma=1.8$ and $E_{\rm fold}=10^{6}$~keV.  If we divide the observed BAT fluxes of these sources by the boost factors appropriate for each source to predict what flux they would have without reflection, we find that 14$\%$ of the {\xmm}-BAT sample would drop below the flux limit of this survey (not including sources with upper-limiting values of $R$), and all of these sources have $R>1$.  This is an important consideration when attempting to determine average reflection properties using hard X-ray selected surveys.

%Average reflection: 2.73572357986 +/- 0.751597360011 stdev 5.3674787536
%Average reflection (log): 0.000383746056853 , [ 6.79902387075e-05 , 0.00216591438638 ]
%Average Efold: 385348.929277 +/- 68439.7762426 keV 488757.763702
%Average gamma: 1.79512144231 +/- 0.0451394940954 0.322360466383
%1-sigma Ranges:
% R from  -2.63175517373 to 8.10320233346
% R from log range 1 sigma: 2.63639052912e-10 to 558.570646206
% Efold from -103408.834425 to 874106.692979
% Gamma from 1.47276097592 to 2.11748190869

\section{Average Spectrum and log(N)-log(S)}
\label{sec:avgspecandlognlogs}

\subsection{Stacked Spectrum}
\label{subsec:averagespectrum}

\begin{figure*}
%\figurenum{1}
\centerline{
\includegraphics[width=8.0cm]{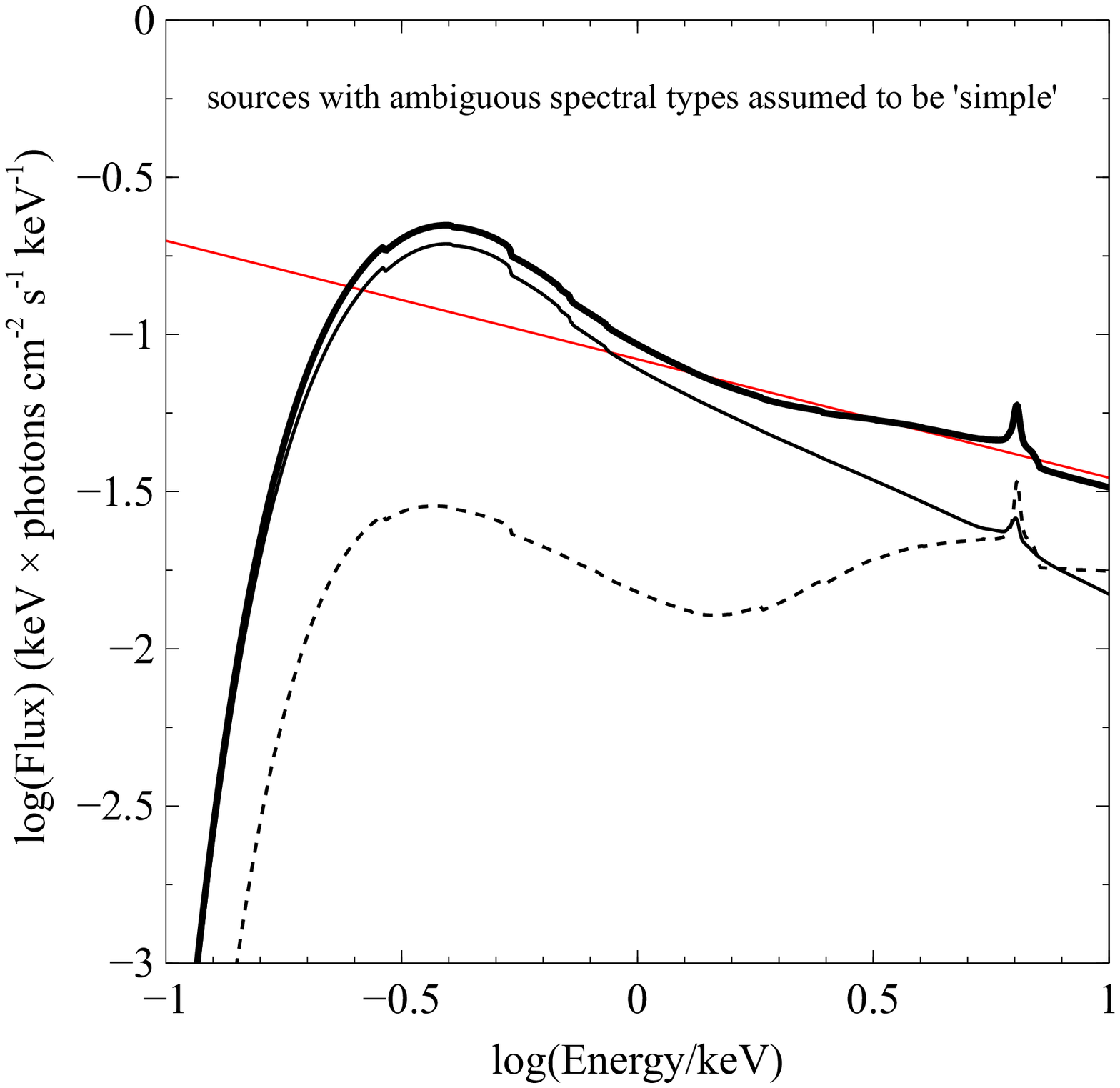}
\includegraphics[width=8.0cm]{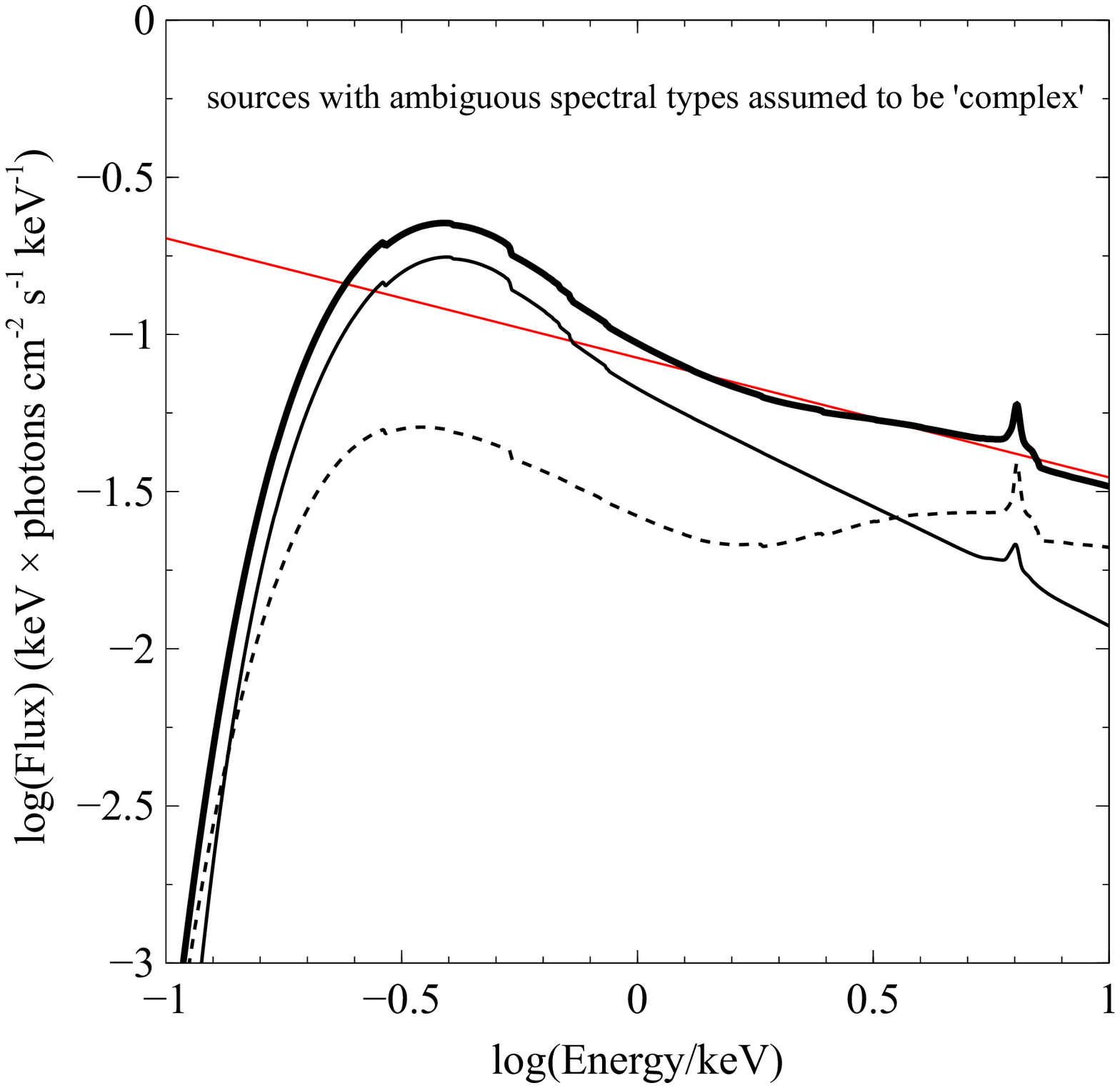}
}
\caption{
\small
Summed spectrum for the whole sample.  Left panel: the summed spectrum assuming a simple spectral type for any intermediate spectral-type objects.  Right panel: the summed spectrum assuming a complex spectral type for such intermediate spectral-type objects.  We stack the model spectra for all objects between 0.4--10~keV.  The thick black solid line shows the summed overall spectrum in both panels, and the thin solid black and thin dashed black lines show the contributions from simple and complex spectrum-type sources, respectively.  The red straight line in both panels shows a fit to the 1--10~keV data (yielding a best-fit photon index of $\Gamma \approx 1.37-1.38$, matching that from the X-ray background).  \label{fig:averagespectrum}}
\end{figure*}

We present the stacked/summed spectrum for all sources (excluding the five sources with insufficient counts to construct their spectra) in Fig.~\ref{fig:averagespectrum}, for comparison with the corresponding average spectrum generated from the 9-month catalog in W09.  While W09 compiled their fits from a variety of different sources in the literature in addition to their own fits, here we have performed a uniform analysis of all of the X-ray spectra for the objects in our sample.  We are therefore able to simply sum the model fits to all of the spectra to obtain the stacked spectrum, and avoid the need to employ the complex method outlined in W09.  We also distinguish, as done in W09, between the contributions due to simple power-law sources and complex (those with a partial-covering spectral shape) sources.  Due to the presence of 13 intermediate-spectral type objects, we produce two versions of the summed spectrum: one assuming all of the intermediate objects are best fit by the simple power-law, and one where all such objects are assumed to have a complex spectrum.   The relative importance of the two source types is clearly seen in Fig.~\ref{fig:averagespectrum}, but the two versions of the summed spectrum at energies above 0.2~keV show no appreciable differences.  Although AGN emission is thought to be responsible for the X-ray background (XRB), the XRB is only well-known for energies $E>0.6$~keV (\citealt{2002ApJ...576..188M}, \citealt{2003ApJ...583...70M}, \citealt{2007A&A...475..837S}, \citealt{2009ApJ...702..270G}).  If we fit a power-law to the 1--10~keV region, we find an average power-law slope of $\Gamma=1.37-1.38$, in line with the key finding from W09 that the slope of the XRB below 15~keV found from \emph{HEAO-1} ($\Gamma \approx 1.4$) can be reproduced by the summed emission from BAT AGN.  We confirm this measurement here with the deeper 58-month catalog, with a greater fraction of obscured AGN (compared to the 9-month catalog), but using a sample that covers 11\% of the sky instead of 74\% of the sky ( $|b|>15^{\circ}$) from W09.  In a companion paper, we extend this stacking analysis to include the BAT data above 10~keV and comment on the relevance of our results to X-ray background studies (Vasudevan et al. 2013 in prep).

Although we do not have data of sufficient quality to detect soft excesses and iron lines in our XRT or ASCA sources ($\sim 50$\% of the sample), we find that based solely stacking the objects with {\xmm} detections, we see both a clear soft excess and an iron line in the summed spectrum for the simple power-law sources (thin solid line in both panels of Fig.~\ref{fig:averagespectrum}).  The iron line is also notably prominent in the complex spectrum sources (thin dashed line).  It would be highly desirable to complete the {\xmm}-quality coverage for the entire sample to acquire better statistics on the prominence of this feature.  This discussion highlights the utility of completing the {\xmm} coverage of this region of the sky, or obtaining other high-quality data to understand the detailed features in AGN spectra in the Northern Galactic Cap.

%Average spectrum: detail what I did.  Soon.

%\section{Other properties}
\subsection{log(N)-log(S) diagram}
\label{subsec:logNlogS}

\begin{figure}
%\figurenum{1}
\centerline{
\includegraphics[width=8.0cm]{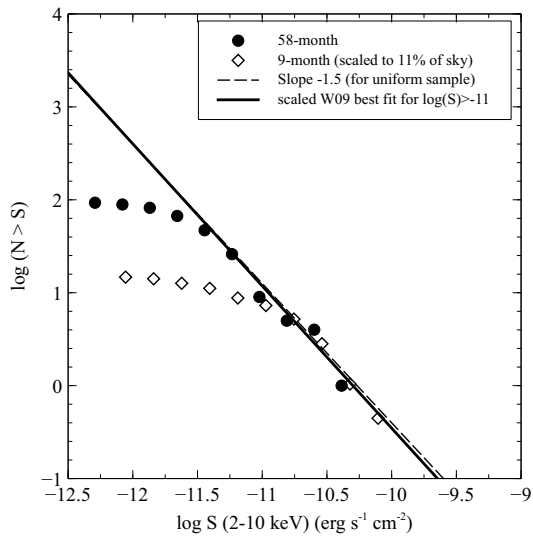}
}
\caption{
\small
Plot of log($N>S$) against log($S$), for 2--10~keV flux $S$ in $\rm erg \thinspace s^{-1} \thinspace cm^{-2}$, where log($N>S$) is the logarithm of the number of sources with flux greater than $S$.  The thin line shows a slope of $-1.5$ expected for a uniform distribution.  Our sample shows a slope consistent with this down to fluxes of log($S$)=$-11.6$. \label{fig:logNlogS}}
\end{figure}

In order to estimate the completeness of our sample in the 2--10~keV band, we present the 2--10~keV log($N$)-log($S$) plot for the sample in Fig.~\ref{fig:logNlogS}.  We adopt all of the same conventions as in W09; $N$ is the number of sources above a threshold 2--10 flux $S$.  The slope of the log($N$)-log($S$) relation is expected to be $-1.5$ for a uniform distribution of objects throughout a Euclidean volume (i.e., for low redshifts) with a broad luminosity distribution, and we find our results consistent with this down to a flux of log($S$)$\approx-11.6$ (by inspection; we do not attempt to fit these points due to the complications involved in fitting points with correlated errors).  This result contrasts with a completeness limit of log($S$)$\approx-11.0$ from the 9-month catalog, indicating that we have a complete census of objects up to fluxes $\sim 4$ times fainter than the 9-month catalog completeness limit.  This log($S$) completeness limit is consistent with the sensitivity expected from the all-sky survey.

Our Northern Galactic Cap sample covers $11$\% of the sky, in contrast to $74$\% coverage of the sky in the complete sample of W09.  In order to more easily compare the results from the 9-month and 58-month catalogs, we therefore scale the results obtained in W09 down to the expected numbers of objects for 11\% of the sky, and overplot the scaled 9-month catalog log($N$)-log($S$) relation using empty diamonds in Fig.~\ref{fig:logNlogS}, along with the best fit above log($S$)$\approx-11.0$ from W09.   We find that the objects above the W09 completeness limit constitute half of the sample in the 9-month catalog, whereas 65\% of our objects lie above our latest completeness limit of log($S$)$\approx-11.6$.

We can also contrast the properties of the source in the `complete' and `incomplete' fractions of the sample from the earlier and later catalogs, in their column density distribution and spectral shapes as done by W09.  For clarity, we state that `complete' refers to objects with fluxes log($S$)$>-11.0$ when discussing the 9-month catalog, but refers to objects with log($S$)$>-11.6$ when discussing the 58-month catalog.  In order to compare absorption distributions, we redefine any lower-limiting log($N_{\rm H}$)=20 values from W09 to log($N_{\rm H}$)=19 as done in W09.  In the 9-month catalog, the objects above completeness show both low absorption ($\langle \rm log \thinspace N_{\rm H} \thinspace \rangle = 20.9 \pm 0.2$) and a wide range of absorption ($ \rm 19.0 < log \thinspace N_{\rm H} \thinspace < 23.5$), whereas objects above completeness in our latest sample exhibit slightly higher absorption levels ($\langle \rm log \thinspace N_{\rm H} \thinspace \rangle = 21.4-21.6 \pm 0.2$) and a greater range ($ \rm 19.0 < log \thinspace N_{\rm H} \thinspace < 24.5$), including four Compton-thick candidates.   Below completeness, where we expect to be missing a substantial fraction of objects, we find that the incomplete 50\% of the 9-month sample has high absorption ($\langle \rm log \thinspace N_{\rm H} \thinspace \rangle = 22.5 \pm 0.2$) and a wide range ($ \rm 19.0 < log \thinspace N_{\rm H} \thinspace < 24.1$) including two almost-Compton-thick candidates, NGC 1365 and NGC 612, whereas our latest sample shows even higher absorptions below completeness ($\langle \rm log \thinspace N_{\rm H} \thinspace \rangle = 22.9-23.2 \pm 0.3$) and a wider range ($ \rm 19.0 < log \thinspace N_{\rm H} \thinspace < 25.3$) including the remaining four Compton-thick candidates in our sample.  There is larger uncertainty in the average $\rm log(N_{\rm H})$ for objects below completeness since there are more objects with ambiguous spectral types in this regime.  Since objects below completeness are by definition fainter, we expect them to have fewer counts in their observations, and therefore require longer exposures before the spectral shape can be properly constrained. 

In both the `complete' and `incomplete' regions of our sample, we have a wider range of absorption levels than seen in W09 and a marked increase in the number of Compton-thick candidates (subject to further investigation of their spectra with more physical models).  The broad-band spectral shape, as parameterized by the ratio $F_{\rm BAT}/F_{\rm 2-10 keV}$, also shows a marked variation.  In the 9-month catalog, W09 found indications that the missing sources in the sample were likely to be heavily absorbed since this flux ratio changed from an average $3.0 \pm 0.3$ above the completeness flux limit to $16.9 \pm 2.3$ in the incomplete part of the sample.  This effect is less pronounced in our latest sample: there is a change from $5.3 \pm 0.7$ to $12.4 \pm 1.9$ for the average value of this flux ratio, indicating that our sample is more homogenous than the previous 9-month catalog. 

Below completness, W09 estimate $\sim 3000$ missing objects at log($S$)$\approx-12$, which corresponds to $\sim 400$ sources if scaled to the 11\% of sky covered by our Northern Galactic Cap sample.  At the same flux level, we are missing $\sim 300$ sources.  Based upon the discussion of flux ratios above, it is likely that the missing sources are more heavily absorbed than those above the completeness limit, but since our sample is more homogenous than the 9-month catalog, it is difficult to predict the properties of the missing sources.

\section{Results from the Northern Galactic Cap sources in the 22-month BAT catalog}
\label{sec:22monthresultscompare}

Throughout this paper, all of our results pertain to the 58-month BAT catalog objects in the Northern Galactic Cap.  We choose this source list since the 58-month catalog is the deepest edition of the catalog for which the source list and BAT spectra are publicly available.  Using the deepest version of the catalog provides the largest possible sample, and therefore allows a more robust determination of the true column density distribution and luminosity distribution.  Ideally this requires 0.4--10~keV data of sufficient quality, but high-quality {\xmm} data is only available for $49\%$ of the objects in our 58-month catalog subsample.  However, the 13 new {\xmm} observations in this sky region from the proposal by PI Brandt were taken to complete {\xmm} coverage for the 22-month catalog \citep{2010ApJS..186..378T} in this sky region, and the observations that were successfully obtained provide {\xmm} coverage for $90\%$ of the $b>50^{\circ}$ 22-month BAT catalog sources.   We therefore present some key comparisons between results for the 9-month catalog (W09's uniform sample, 102 objects, 74\% of the sky), the 22-month catalog (39 objects, 11\% of the sky), and the 58-month catalog results (100 objects, 11\% of the sky) in Tables~\ref{table:cataloguecompare} and \ref{table:cataloguecompare_2} below, to illustrate the utility of the deepening sample, and to identify if any insight can be gained using a complete subsample with higher-quality data.  We contrast the flux limits, completeness limits, percentage of sources with ambiguous spectral types, and percentages of objects with different levels/types of absorption in Table~\ref{table:cataloguecompare}; we do the same for the luminosity distributions, frequency of spectral features, and percentage of `hidden/buried' sources in Table~\ref{table:cataloguecompare_2}.  A few sources in the 22-month catalog list do not occur in the 58-month catalog (due to intrinsic variability of those sources); for our 22-month catalog analysis we only use those sources that survive in the 58-month catalog.

The 22-month catalog flux limit is 1.8 times fainter than the 9-month catalog flux limit, and the 58-month catalog flux limit is 5 times fainter than that for the 9-month.  While the 58-month catalog completeness limit (from the $\rm \thinspace logN \thinspace - \thinspace log \thinspace S$ relation) is estimated to be 4 times fainter than the 9-month catalog, the 22-month catalog is only complete to fluxes 1.8 times fainter than the 9-month catalog.   Despite the better {\xmm} coverage for the 22-month catalog subsample, we find that there is still a large degree of uncertainty in the percentages of sources at different absorption levels and with different absorption types (simple/complex) due to four ambiguous spectral-type sources in the 22-month catalog list.  The fraction of complex absorption sources appears to change little between the catalogs, within the uncertainties.  However, there does seem to be a trend towards uncovering a higher fraction of absorbed sources, in particular Compton-thick objects, as the catalog gets deeper.

The luminosity distributions in all three catalogs show a trend for unabsorbed sources to have marginally brighter luminosities, and percentages of sources with iron K-$\alpha$ lines or soft excess appears relatively stable between the different catalogs.  We do not consider warm absorbers here, as this requires a more considered approach; \cite{2012ApJ...745..107W} outline such an approach using {\xmm} and \emph{Suzaku} data.  In summary, while the 22-month catalog objects in this sky region have better data quality on average, the smaller number of objects in that subset make the percentages presented above more prone to uncertainty.

\section{Discussion and Conclusions}
\label{sec:discussionsandconclusions}

We present a detailed X-ray spectral analysis of the non-beamed AGN in the Northern Galactic Cap of the 58-month BAT catalog, consisting of 100 AGN with $b>50^{\circ}$.  This field has excellent potential for further investigations due to a wide range of multi-wavelength data that is already available, and we propose the field as a low-redshift analog to the `deep field' observations at higher redshifts (e.g. CDFN/S, Lockman Hole). We present distributions of the redshift, luminosity, absorbing column density and other key quantities for the catalog.  We use a consistent approach to fit all data, using our semi-automated X-ray spectral fitting workflow, useful for fitting suites of models to large samples of AGN, producing consistent comparisons between models and determining the significance of various spectral components.  In summary, we find that:

\begin{itemize}
\item We probe to deeper redshifts with this representative subsample of 100 objects from the 58-month catalog ($\langle z \rangle = 0.043$ compared to $0.03$ from W09 and $0.03$ from \citealt{2011ApJ...728...58B})
\item The average X-ray luminosity found here ($\langle \rm log \thinspace L_{\rm 2-10 keV} \thinspace \rangle = 43.0$) is identical to that seen in W09 ($\langle \rm log \thinspace L_{\rm 2-10 keV} \thinspace \rangle = 43.0$), but with more pronounced tails in the distribution at low and high luminosities.  The average 14--195~keV BAT luminosity is $\langle \rm log \thinspace L_{\rm BAT} \thinspace \rangle= 43.5$, compared to $43.7$ from W09.
\item We uncover a broader absorbing column density distribution.  The obscured fraction ($\rm log \thinspace N_{\rm H} \thinspace \ge 22$) is $\sim 60$\%, an increase from the measurements for the 36-month catalog \citep{2011ApJ...728...58B} and the 9-month catalog (W09).   The obscured fraction broadly overlaps with complex spectrum (partial covering-type) sources, which constitute 43-56\% of the sample.
\item Thirteen objects have ambiguous spectral types in that a unique best-fit model could not be found, and both simple or complex absorption can describe the observed spectral shape.  These are typically objects where the existing (generally XRT) data does not have sufficient counts below 1~keV to distinguish between the two models.  For these objects, we have presented the results for both models in all subsequent analyses, and take into account the resulting uncertainty in model fit parameters.  The most significant uncertainty stemming from these ambiguous sources is the absorbing column density: two different model fits sometimes produce extremely different $\rm log(N_{\rm H})$ values.  This behaviour highlights the need to obtain better quality data for these sources.
\item We present the properties of iron lines, soft excesses and ionized absorbers detected in the 39 objects with $>4600$ counts in their spectra.  Iron lines are detected in 79\% of the {\xmm} objects, similar to 81\% found in W09.  Soft excesses are detected at a frequency of 31--33\% compared to W09's 41\%, and ionized absorption edges are detected in 18\% of our sample.  We present upper limiting iron line equivalent widths for sources where the feature was not detected as a significant addition to the spectrum, and confirm the X-ray Baldwin effect to the degree identified by e.g., \cite{2004MNRAS.347..316P} in our sample.
\item We introduce the concept of the soft-excess strength, defined as the luminosity in the soft excess (modelled as a black body) divided by the power-law luminosity between the `clean range' 1.5--6~keV.  In W09, a linear relation was found between the soft excess power and power-law luminosity.  We find a deviation from this description, suggesting that the soft excess is not simply powered by the power-law coronal emission, as the fraction of power seen in the soft excess drops slightly at higher power-law luminosities.  We also find a disjoint distribution of sources: one set in which a soft excess is well-detected and for which the soft excess fraction is $>0.1$, and another family of sources where the soft excess is not detected at all (indicated by stringent upper limits in the soft-excess strength).  This result suggests that the process responsible for producing the soft excess is not ubiquitous in AGN.
\item The fraction of unabsorbed ($\rm log \thinspace N_{\rm H} \thinspace < 22$) sources with ionized absorber edges is $\sim 32 \%$, lower than the $53 \%$ found for Seyfert 1--1.5 BAT AGN in \cite{2012ApJ...745..107W}.  However, given that our study and previous studies estimating the prevalence of ionized absorbers have small sample sizes, our result is broadly consistent with previous results.
\item The fraction of Compton-thick sources ($\rm log \thinspace N_{\rm H} \thinspace > 24.15$) in our sample is $\sim 9$\%, using a simple absorption model (\textsc{ztbabs}).  Other measures of Compton thickness, such as high Fe-K line equivalent width, high reflection $R$ or a flat photon index (pegging at $1.5$), were also investigated, which may add 2--6 Compton thick candidates, increasing the proportion of such sources to 11--15\%.  The true Compton-thick fraction involves estimating the number missed due to the attenuation of Compton-thick AGN spectra in the BAT band.  \cite{2011ApJ...728...58B} suggest that the true Compton-thick fraction could be 20\%.
%However, our preliminary attempts to include the effects of Compton scattering in our absorption determinations suggest that the true $N_{\rm H}$ values for these sources could be a factor $\sim 0.65$ lower than found using the simple model; this could push one-third to one-half of the Compton-thick sources out of this category, moving the proportion found closer to \cite{2011ApJ...728...58B}'s 4.6\%.
\item We identify seven new `hidden' sources, unidentified in W09, three by using newly obtained {\xmm} data.  The fraction of such sources in our sample is 13--14\%, lower than the proportion found in W09.
\item Reflection is found to be important in a large fraction of our sample, with the average value of the reflection fraction found to be $\langle R \rangle = 2.7 \pm 0.75$, suggestive of strong light bending or highly complex absorption.  The average fold energy of the sample is well outside the BAT band, but we do observe a well-defined high-energy cut-off in some sources.  The use of BAT data in conjunction with {\xmm} data allows reflection parameters to be better constrained, and our technique for re-normalizing the BAT data (using the BAT light curves) to the epoch during which the 0.4--10~keV data were taken removes a degree of freedom from the fit, further constraining the fit for some objects.
\item We present the summed spectrum for the sample.  The slope of the summed spectrum between 1--10~keV reproduces the X-ray background slope, as found in W09, but we find this with a deeper sample that exhibits a different absorption distribution.  Iron lines and soft excesses appear to be significant in the whole sample, but higher quality data are needed for about half the sample (that does not yet have {\xmm} coverage) to understand the frequency of these components properly.
\item The 2--10~keV log($N$)-log($S$) plot for the Northern Galactic Cap reveals completeness down to log($S$)$\approx-11.6$, a factor of $\sim 4$ fainter than the 9-month catalog.  A larger proportion of our sources lie above the completeness limit than in W09, and whilst the missing sources are expected to be more heavily absorbed and possibly Compton-thick due to their $F_{\rm BAT}/F_{\rm 2-10 keV}$ ratio, the absorption properties of our 58-month BAT catalog subsample seem to be more homogenous both above and below completeness than found in W09 for the 9-month catalog.
\item We consider the properties of the complete subsample of 39 objects drawn from the 22-month BAT catalog source list in this sky region, for which the {\xmm} coverage is 90\% (compared to 49\% for our full 58-month sample), to identify whether the improved data quality allows more robust determination of the subsample properties.  However, we do not find significantly more stringent constraints on fractions of complex spectra, luminosity distributions, or the proportions of sources with iron lines and soft excesses.  This is in part due to the smaller sample size of the 22-month subset and four sources in the 22-month catalog with ambiguous spectral types.  There is some evidence, however, for an increasing fraction absorbed (including Compton-thick sources) with survey depth.

\end{itemize}

In summary, we have analysed an unbiased sample that is representative of local AGN activity spanning a wide range of parameters (absorption, luminosity), with detailed information on the X-ray properties of the sample.   There is substantial scope to build a suite of multi-wavelength studies on this sample, to further our understanding of AGN accretion at $z < 0.2$.  We defer some of the following issues to later studies, for brevity.  The complete sample assembled here will provide a useful base for understanding long-term AGN variability, by analysing multi-epoch observations of these AGN; many such observations exist in the archives, and such a study would complement other such variability studies on AGN samples (e.g., \citealt{2004ApJ...611...93P}).  It would also be interesting to examine low-energy X-ray emission (below $\sim$1~keV) in buried AGNs for potential signatures of hot-gas emission due to nuclear starburst or broader galactic activity.  We again remind the reader of the existing multi-band coverage and therefore potential for generating broad-band SEDs and accretion rates, which we identify as a key priority for future work. We are currently working on companion studies related to understanding reflection in AGN; including a study of correlations between soft X-ray features (soft excesses, iron lines) and the Compton reflection hump.  Our previous studies using the BAT catalog (\citealt{2008arXiv0807.4695M}, W09, \citealt{2009MNRAS.399.1553V,2010MNRAS.402.1081V}) all reveal the complexities of determining accurate black-hole mass estimates for this sample; an object-specific approach is required to determine black-hole masses in a diverse AGN sample such as this, making this another priority for the future.

\section{Acknowledgements}
\label{sec:acknowledgements}

RVV would like to thank Jeremy Sanders for designing and providing extensive support for the \emph{Veusz} plotting package with which most of the plots in this paper were generated, and Mike Koss for his comments on the paper.   We thank Jack Tueller, Craig Markwardt and Neil Gehrels for their work on the proposal to extend the {\xmm} coverage of this region, and all of the BAT team for their work on the BAT catalog.  We also thank the anonymous referee for useful comments that improved the paper. This work makes use of data from {\xmm}, an ESA science mission with instruments and contributions directly funded by ESA Member States and NASA.  This research has made use of the NASA/IPAC Extragalactic Database (NED) which is operated by the Jet Propulsion Laboratory, California Institute of Technology, under contract with the National Aeronautics and Space Administration (NASA). 

\appendix
\section{APPENDIX: Moderately radio loud sources Mrk 463 and 3C 303.0}
\label{appendix:radioloud}

The objects Mrk 463 and 3C 303.0 were identified as having the highest radio loudness parameters in the sample ($RL \gtrsim -3$).  It is not straightforward to account for how this will affect the X-ray spectral fitting, since the radio emission could be due to a combination of star formation (e.g., \citealt{2002AJ....124..675C}) and jet emission.  In the case of Mrk 463, this is a known dual AGN system (\S\ref{radioloudness}) the radio images available on NED for this object \footnote{http://ned.ipac.caltech.edu/} reveal a non-standard morphology for the radio source which cannot be straightforwardly classified as one of the traditional Fanaroff-Riley classes (FR-I/FR-II).  There is little evidence for a significant component of X-ray jet emission in radio-loud non-blazar AGN \citep{1999ApJ...526...60S}, and indeed the X-ray analysis of Mrk 463 in \cite{2008MNRAS.386..105B} does not consider any X-ray component in the \emph{Chandra} or {\xmm} data.  For 3C 303.0, we inspect the archival \emph{Chandra} images and find that while the 2--10~keV emission has a negligible contribution from extended jet emission, the 0.5--2~keV emission may be contaminated by jet emission at a level of $17\%$ of the nuclear emission.  This is likely to influence our spectral fit for this object.  However, inspection of the spectrum reveals a smooth power law from 0.5--10~keV.  If the intrinsic nuclear spectrum is more heavily absorbed, the radio emission due to jets at soft energies may give the illusion of less absorption.  However, determining the precise degree of jet contamination is complex, and since these considerations only affect two of our objects, they will not influence the absorption distributions and other results significantly.

\section{APPENDIX: Sources with low counts}
\label{appendix:upperlimits}

The five sources in Table~\ref{table:upperlimits} had XRT data but lacked sufficient counts to construct a spectrum, and due to their poor data quality we have not included them when calculating sample-wide properties presented in this paper.  For completeness, we present a basic analysis of these datasets here.  We examined the counts in the 0.5--2~keV and 2--10~keV in the source and background regions for these objects.  When sources were detected at a $ \geq 2 \sigma$ level above background, we calculate basic fluxes and luminosities (in 0.5--2 and 2--10~keV bands) using \textsc{WebPIMMS} assuming a Galactic-absorbed powerlaw with a photon index of $1.9$; where the sources were not detected at greater than $2 \sigma$, we present 95\% confidence upper limits on fluxes and luminosities.  These upper limits were calculated using the \cite{1986ApJ...303..336G} prescription.  These sources would clearly benefit from longer, higher signal-to-noise ratio observations to identify whether there is any sustained X-ray emission.

We attempt to recover an estimate for the intrinsic column density in these sources, since the very faint X-ray fluxes may be indicative of heavily absorbed, potentially Compton-thick objects, when considered alongside the available BAT spectra.  However, we find that this approach produces uncertain results with large uncertainties on the inferred log($N_{\rm H}$), unless we are able to renormalize the BAT spectrum and thereby have a better estimate of the absolute ratio between the 2--10~keV and 14--195~keV fluxes (only possible for three of these objects).  These results are presented in the final column of Table~\ref{table:upperlimits} for completeness. We do not find any convincing hints of Compton-thick candidates amongst these five objects, but better data are needed.

%For NGC 4180, we find that a simple absorbed power-law model that reproduces the observed BAT and 0.5--2 keV luminosities requires an absorption log($N_{\rm H}$) of about $22.1$.   For NGC 4500, with BAT renorm yields 22.9 sys errors dominate - what is true gamma, etc.  MCG -01-33-063: 22.9 but poorly determined no renorm, CGCG 102-048 gives 23.2 with BAT renorm, 2MASX J13542913+1328068 23.4 but no renorm poorly determined.

%A number of these sources are included in an accepted {\xmm}  proposal to cover eight further objects in the Northern Galactic Cap.

\section{APPENDIX: Poor fits}
\label{appendix:poorfits}

%\subsection{Poor fits (with low null hypothesis probabilities)}

From the 95 with spectra with sufficient counts to attempt fitting, we identify 19 which have null-hypothesis probabilities of their best-fitting models less than $5 \times 10^{4}$, indicating that some key features of these spectra have not been modeled fully.  In this work, we do not attempt to account for all of the detail in each spectrum, but try to account for the key properties of the spectrum such as the photon index, intrinsic absorption, and the basic set of features discussed in \S\ref{subsec:detailedfeatures} and restrict ourselves to these features.  AGN spectra often exhibit much more complexity than the model combinations used here, and we examine if this complexity can account for the poor fits in the 19 sources with low null-hypothesis probabilities.  On inspection, these objects can be split into three categories: 1) those with few counts, 2) partial covering sources with un-modelled residuals at low energies, such as bumps or other types of structure below 1~keV, and 3) sources with broad iron lines with complex structure not modelled by the \textsc{zgauss} model, which were already discussed above. The soft features in the second category may be due to features in the host galaxies of these AGN (such as emission lines from photo-ionized/hot gas), but are not addressed further here.

\bibliographystyle{apj} %% 
\bibliography{northgalcap}

%\begin{itemize}{labelitemi}{$\bullet$}
%\item First item in the list
%\item Second item
%\item and so on
%\end{itemize}  

%\section{Summary}

%\section{Sample}

%\begin{figure}
%\figurenum{1}
%\centerline{
%\includegraphics[width=14.0cm]{./f1.eps}
%}
%\caption{
%\small
%Test caption}
%\end{figure}

%\begin{landscape}
%\rotate
%\begin{table*}
%{\tiny\begin{tabular}{l|l|l|l|l|l|l|l|l|l|l|l|l|l}
%{p{2cm}p{1.3cm}p{2.8cm}p{2.7cm}p{2.8cm}p{2.0cm}p{1.5cm}}
%{|l|l|l|l|l|l}
%\hline
%AGN & Redshift & RA & Dec & $l$ & $b$ & Instrument & Obs. ID & Obs. & Usable & Obs. & Usable & Optical\\
%    &          &    &     &     &     &            &  &  date       & counts &  time (ks)  & $\%$ of obs.  & Type  & BAT flux (SNR) \\
%\hline

%\hline
%\end{tabular}}

%Table \ref{table:observations}: Table of observations used for each object (continued).

%\caption{Table of observations used for each object.}
%\end{table*}
%\end{landscape}
%\clearpage

%Tables start here

\clearpage
%\LongTables
%\rotate
%\begin{landscape}
\begin{turnpage}
\begin{table}
%\hoffset=cm
%\newgeometry{top=7cm}
%\addtolength{\voffset}{2cm}
%\begin{deluxetable*}{l|l|l|l|l|l|l|l|l|l|l|l|l|l}
%\tablecolumns{13}
%\tabletypesize{\scriptsize}
%\tablewidth{0pt}

{\tiny\begin{tabular}{l|l|l|l|l|l|l|l|l|l|l|l|l|l}
%{p{2cm}p{1.3cm}p{2.8cm}p{2.7cm}p{2.8cm}p{2.0cm}p{1.5cm}}
%{|l|l|l|l|l|l}
%\startdata 
\hline
AGN & Redshift & RA & Dec & $l$ & $b$ & Instrument & Obs. ID & Obs. & Source & Obs. & Usable & Optical\\
    &          &    &     &     &     &            &                &  date       & Counts &  time (ks)  & $\%$ of obs.  & Type & BAT flux (SNR)\\
\hline
3C 234 & 0.1849 & 150.457 & 28.785 & 200.208 & 52.708 & XMM & 0405340101 & 2006-04-24 & 9214 & $39.9$ & $84$ & Sy1/Sy2 & 8.73 (5.10) \\
NGC 3227 & 0.0039 & 155.878 & 19.865 & 216.992 & 55.446 & XMM & 0101040301 & 2000-11-28 & 53612 & $40.1$ & $99$ & Sy1.5 & 112.78 (56.21) \\
%MCG +09-17-074 & 0.0238 & 158.563 & 52.871 & 158.130 & 53.838 & XRT & 00040951001 & 2010-02-28 & 3 & $5.6$ & -- & galaxy & 9.84 (6.52) \\
SDSS J104326.47+110524.2 & 0.0476 & 160.860 & 11.089 & 234.761 & 55.932 & XRT & 00040954001 & 2010-10-29 & 2085 & $9.8$ & -- & Sy1 & 14.85 (4.84) \\
MCG +06-24-008 & 0.0259 & 161.203 & 38.181 & 182.222 & 61.326 & XRT & 00040955004 & 2010-10-31 & 133 & $4.4$ & -- & galaxy & 13.69 (5.04) \\
UGC 05881 & 0.0206 & 161.679 & 25.932 & 208.222 & 62.148 & XRT & 00037314002 & 2008-07-03 & 217 & $8.8$ & -- & Sy2 & 20.94 (10.42) \\
Mrk 417 & 0.0328 & 162.379 & 22.964 & 214.722 & 62.143 & XMM & 0312191501 & 2006-06-15 & 1788 & $14.3$ & $75$ & Sy2 & 33.63 (15.22) \\
2MASX J10523297+1036205 & 0.0878 & 163.137 & 10.606 & 237.757 & 57.525 & XRT & 00037131004 & 2008-07-13 & 190 & $16.2$ & -- & Sy1 & 17.52 (6.63) \\
Mrk 728 & 0.0356 & 165.258 & 11.047 & 239.366 & 59.481 & XMM & 0103861801 & 2002-05-23 & 21479 & $9.7$ & $100$ & Sy1.9 & 11.90 (5.17) \\
FBQS J110340.2+372925 & 0.0739 & 165.918 & 37.490 & 181.567 & 65.101 & XRT & 00039831001 & 2009-07-12 & 1049 & $8.2$ & -- & Sy1 & 8.86 (5.38) \\
2MASX J11053754+5851206 & 0.1930 & 166.407 & 58.856 & 145.639 & 53.357 & XRT & 00040957001 & 2010-08-30 & 910 & $11.2$ & -- & QSO/BLAGN & 5.52 (4.97) \\
CGCG 291-028 & 0.0477 & 166.496 & 58.946 & 145.481 & 53.320 & XRT & 00040957001 & 2010-08-30 & 46 & $11.2$ & -- & Sy2 & 4.37 (4.97) \\
{\bf IC 2637} & 0.0292 & 168.457 & 9.586 & 245.603 & 61.057 & XMM & 0601780201 & 2009-12-20 & 31401 & $15.1$ & $100$ & Sy1.5 & 15.16 (6.40) \\
MCG +09-19-015 & 0.0703 & 168.833 & 54.389 & 149.169 & 57.569 & XRT & 00031290001 & 2008-11-07 & 94 & $4.6$ & -- & Sy2 & 9.08 (5.58) \\
PG 1114+445 & 0.1438 & 169.277 & 44.226 & 164.675 & 64.494 & XMM & 0109080801 & 2002-05-14 & 42913 & $43.5$ & $97$ & Sy1 & 8.93 (4.93) \\
ARP 151 & 0.0211 & 171.401 & 54.383 & 147.029 & 58.546 & XRT & 00037369002 & 2009-02-15 & 3076 & $8.1$ & -- & Sy1 & 17.81 (11.05) \\
{\bf 1RXS J1127+1909} & 0.1055 & 171.818 & 19.156 & 230.898 & 69.111 & XMM & 0601780301 & 2009-05-29 & 25011 & $11.9$ & $100$ & Sy1.8 & 16.84 (8.59) \\
UGC 06527 & 0.0274 & 173.170 & 52.949 & 147.136 & 60.320 & XMM & 0200430501 & 2004-05-02 & 976 & $12.7$ & $95$ & Sy2 & 8.09 (7.32) \\
IC 2921 & 0.0437 & 173.205 & 10.296 & 251.453 & 64.978 & XRT & 00038055002 & 2010-07-23 & 781 & $8.6$ & -- & Sy1 & 17.51 (7.12) \\
{\bf NGC 3758} & 0.0299 & 174.123 & 21.596 & 226.837 & 72.080 & XMM & 0601780401 & 2009-06-14 & 30020 & $11.9$* & $100$ & Sy1 & 11.46 (8.16) \\
SBS 1136+594 & 0.0601 & 174.786 & 59.198 & 139.255 & 55.580 & XRT & 00035265001 & 2005-12-14 & 3521 & $9.2$ & -- & Sy1.5 & 20.85 (12.12) \\
Mrk 744 & 0.0089 & 174.928 & 31.909 & 191.584 & 73.704 & XMM & 0204650301 & 2004-05-24 & 10239 & $29.5$ & $42$ & Sy1.8 & 20.49 (8.70) \\
PG 1138+222 & 0.0632 & 175.317 & 21.939 & 227.137 & 73.238 & XRT & 00037371001 & 2008-10-26 & 2693 & $9.1$ & -- & Sy1 & 17.19 (8.13) \\
2E 1139.7+1040 & 0.1505 & 175.570 & 10.394 & 255.411 & 66.650 & XRT & 00036986001 & 2008-07-23 & 25 & $1.6$ & -- & Sy1 & 15.75 (7.21) \\
{\bf KUG 1141+371} & 0.0381 & 176.125 & 36.886 & 174.169 & 72.821 & XMM & 0601780501 & 2009-05-23 & 17251 & $19.8$ & $66$ & Sy1 & 15.47 (5.93) \\
MCG+10-17-061 & 0.0099 & 176.387 & 58.977 & 138.147 & 56.162 & XRT & 00038346001 & 2009-02-17 & 435 & $8.9$ & -- & galaxy & 15.99 (9.36) \\
2MASX J11475508+0902284 & 0.0688 & 176.980 & 9.041 & 260.274 & 66.498 & XRT & 00040962001 & 2010-07-02 & 1357 & $8.9$ & -- & Sy1.5 & 11.06 (6.38) \\
MCG +05-28-032 & 0.0230 & 177.190 & 29.642 & 198.919 & 75.999 & XRT & 00090176001 & 2009-10-20 & 139 & $2.7$ & -- & LINER & 23.79 (10.20) \\
2MASX J11491868-0416512 & 0.0845 & 177.327 & -4.280 & 275.039 & 55.192 & XRT & 00038057001 & 2009-11-09 & 634 & $5.0$ & -- & Sy1 & 11.83 (5.99) \\
MCG -01-30-041 & 0.0188 & 178.160 & -5.208 & 277.014 & 54.678 & XRT & 00037373002 & 2009-08-04 & 44 & $6.0$ & -- & Sy1.8 & 13.69 (5.64) \\
NGC 3998 & 0.0035 & 179.484 & 55.453 & 138.172 & 60.064 & XMM & 0090020101 & 2001-05-09 & 37289 & $13.2$* & $94$ & Sy1/LINER & 16.89 (9.83) \\
CGCG 041-020 & 0.0360 & 180.242 & 6.806 & 270.116 & 66.407 & XMM & 0312191701 & 2006-06-26 & 7098 & $12.9$ & $100$ & Sy2 & 21.18 (9.61) \\
MRK 1310 & 0.0194 & 180.310 & -3.678 & 279.540 & 56.896 & XRT & 00035361001 & 2005-12-25 & 1789 & $6.3$ & -- & Sy1 & 13.23 (6.11) \\
NGC 4051 & 0.0023 & 180.791 & 44.531 & 148.882 & 70.085 & XMM & 0157560101 & 2002-11-22 & 147191 & $51.9$* & $86$ & Sy1.5 & 37.62 (24.17) \\
Ark 347 & 0.0224 & 181.124 & 20.316 & 242.830 & 77.289 & XRT & 00035599002 & 2006-07-04 & 141 & $10.4$ & -- & Sy2 & 29.15 (11.73) \\
PG 1202+281 & 0.1653 & 181.176 & 27.902 & 205.968 & 79.613 & XMM & 0109080101 & 2002-05-30 & 57522 & $17.9$ & $100$ & Sy1.2 & 9.99 (5.79) \\
{\bf UGC 7064} & 0.0250 & 181.180 & 31.177 & 188.499 & 79.034 & XMM & 0601780601 & 2009-12-26 & 15499 & $39.5$ & $88$ & Sy1.9 & 13.64 (5.28) \\
2MASX J12055599+4959561 & 0.0631 & 181.483 & 49.999 & 140.753 & 65.525 & XRT & 00040963001 & 2010-04-26 & 204 & $7.8$ & -- & BLAGN & 12.37 (5.30) \\
{\bf NGC 4102} & 0.0028 & 181.596 & 52.711 & 138.079 & 63.072 & XMM & 0601780701 & 2009-10-30 & 2085 & $31.8$ & $41$ & LINER & 28.55 (14.07) \\
B2 1204+34 & 0.0791 & 181.887 & 33.879 & 174.707 & 78.396 & XRT & 00037315003 & 2008-10-20 & 110 & $8.6$ & -- & Sy2 & 16.91 (5.04) \\
{\bf Mrk 198} & 0.0242 & 182.309 & 47.059 & 142.750 & 68.413 & XMM & 0601780801 & 2009-11-07 & 15537 & $26.9$ & $92$ & Sy2 & 21.94 (12.32) \\
NGC 4138 & 0.0030 & 182.374 & 43.685 & 147.305 & 71.404 & XMM & 0112551201 & 2001-11-26 & 7674 & $15.0$ & $100$ & Sy1.9 & 30.67 (14.92) \\
NGC 4151 & 0.0033 & 182.637 & 39.406 & 155.074 & 75.064 & XMM & 0112310101 & 2000-12-21 & 247586 & $33.0$ & $100$ & Sy1.5 & 533.09 (275.00) \\
{\bf KUG 1208+386} & 0.0228 & 182.686 & 38.336 & 157.646 & 75.920 & XMM & 0601780901 & 2009-06-14 & 3691 & $15.8$ & $71$ & Sy1 & 21.63 (12.63) \\
NGC 4180 & 0.0070 & 183.262 & 7.038 & 276.792 & 67.940 & XRT & 00036654002 & 2007-11-30 & 7 & $3.0$ & -- & AGN & 14.44 (6.04) \\
2MASX J12135456-0530193 & 0.0660 & 183.477 & -5.506 & 286.006 & 56.128 & XRT & 00040964003 & 2010-11-21 & 534 & $4.0$ & -- & Sy1 & 11.94 (5.18) \\
Was 49b & 0.0640 & 183.574 & 29.529 & 194.394 & 81.484 & ASCA & 73036000 & 1995-05-22 & 1693 & $39.6$ & -- & Sy2/binary AGN & 15.40 (7.96) \\
NGC 4235 & 0.0080 & 184.292 & 7.191 & 279.183 & 68.468 & XMM & 0204650201 & 2004-06-09 & 13865 & $13.1$ & $100$ & Sy1 & 31.39 (14.07) \\
Mrk 202 & 0.0210 & 184.479 & 58.660 & 131.140 & 57.931 & ASCA & 77076000 & 09-11-1999 & 4519 & $20.0$ & -- & Sy1 & 8.07 (4.82) \\
Mrk 766 & 0.0129 & 184.611 & 29.813 & 190.681 & 82.271 & XMM & 0304030101 & 2005-05-23 & 454079 & $95.5$ & $91$ & Sy1.5 & 21.42 (14.64) \\
NGC 4258 & 0.0015 & 184.740 & 47.304 & 138.319 & 68.842 & XMM & 0059140101 & 2001-05-06 & 10415 & $12.7$ & $100$ & Sy1.9/LINER & 23.92 (11.71) \\

\hline
%\enddata
\end{tabular}}
\caption{Observations \label{table:observations}}
\tablecomments{Table of observations used for each object.  New observations are indicated using bold type. The positions quoted are for the identified soft X-ray (0.4--10~keV) counterpart to the BAT source. The total number of counts in the source region at energies 0.4--10~keV (in all detectors used for fitting, i.e. {\sc pn} + {\sc mos1} + {\sc mos2} for {\xmm}) are presented. Asterisks indicate that pile-up was corrected for these {\xmm} sources. BAT fluxes are provided in units of $10^{-12} \rm erg \thinspace s^{-1} \thinspace cm^{-2}$, along with the signal-to-noise ratio of the BAT detection (in brackets).  For {\xmm} observations, corrections for flaring required excising part of the observation; this was not done for \emph{ASCA} or \emph{XRT} observations where the whole observations were used, and the `Usable $\%$ of obs.' column therefore only applies to {\xmm} datasets. }
%\end{deluxetable*}
\end{table}
%\end{landscape}
\end{turnpage}

\begin{turnpage}
%\begin{landscape}
%\rotate
\begin{table*}
{\tiny\begin{tabular}{l|l|l|l|l|l|l|l|l|l|l|l|l|l}
%{p{2cm}p{1.3cm}p{2.8cm}p{2.7cm}p{2.8cm}p{2.0cm}p{1.5cm}}
%{|l|l|l|l|l|l}
\hline
AGN & Redshift & RA & Dec & $l$ & $b$ & Instrument & Obs. ID & Obs. & Source & Obs. & Usable & Optical\\
    &          &    &     &     &     &            &                &  date       & Counts &  time (ks)  & $\%$ of obs.    & Type  & BAT flux (SNR) \\
\hline
{\bf Mrk 50} & 0.0234 & 185.851 & 2.679 & 286.395 & 64.647 & XMM & 0601781001 & 2009-07-09 & 64116 & $11.9$* & $100$ & Sy1 & 23.95 (10.18) \\
NGC 4388 & 0.0084 & 186.445 & 12.662 & 279.123 & 74.335 & XMM & 0110930301 & 2002-07-07 & 3629 & $18.8$ & $78$ & Sy2 & 275.78 (110.73) \\
NGC 4395 & 0.0011 & 186.455 & 33.546 & 162.095 & 81.534 & XMM & 0112521901 & 2002-05-31 & 11194 & $15.9$ & $97$ & Sy1.9 & 26.08 (14.28) \\
NGC 4500 & 0.0104 & 187.842 & 57.965 & 128.094 & 58.962 & XRT & 00040965001 & 2010-04-25 & 24 & $4.5$ & -- & starburst galaxy? & 8.69 (5.78) \\
Ark 374 & 0.0630 & 188.016 & 20.158 & 269.447 & 81.739 & XMM & 0301450201 & 2005-07-09 & 95204 & $25.5$ & $100$ & Sy2 & 14.84 (6.17) \\
NGC 4579 & 0.0051 & 189.432 & 11.818 & 290.398 & 74.355 & XMM & 0112840101 & 2003-06-12 & 79300 & $23.7$ & $100$ & LINER & 10.41 (4.93) \\
NGC 4593 & 0.0090 & 189.914 & -5.344 & 297.483 & 57.403 & XMM & 0109970101 & 2000-07-02 & 278937 & $28.1$ & $80$ & Sy1 & 88.68 (33.45) \\
NGC 4619 & 0.0231 & 190.436 & 35.063 & 136.975 & 81.799 & ASCA & 75081000 & 1997-06-03 & 3179 & $33.4$ & -- & Sy1 & 6.64 (5.59) \\
NGC 4686 & 0.0167 & 191.669 & 54.534 & 124.432 & 62.581 & XMM & 0554500101 & 2008-06-28 & 1350 & $30.3$ & $62$ & XBONG & 27.88 (13.12) \\
2MASX J13000533+1632151 & 0.0800 & 195.024 & 16.537 & 314.092 & 79.220 & XMM & 0149170701 & 2003-07-14 & 1136 & $6.4$ & $69$ & Sy1? & 14.83 (5.31) \\
MCG -01-33-063 & 0.0263 & 195.080 & -8.086 & 306.739 & 54.720 & XRT & 00041773002 & 2010-12-08 & 10 & $2.6$ & -- & galaxy & 10.32 (4.92) \\
MRK 0783 & 0.0672 & 195.746 & 16.407 & 317.528 & 78.950 & XRT & 00037318001 & 2008-05-09 & 679 & $5.8$ & -- & Sy1.5 & 17.93 (9.00) \\
SWIFT J1303.9+5345 & 0.0299 & 196.000 & 53.791 & 118.811 & 63.237 & XMM & 0312192001 & 2006-06-23 & 64141 & $11.9$* & $100$ & Sy1 & 34.56 (18.53) \\
NGC 4941 & 0.0037 & 196.055 & -5.552 & 308.806 & 57.174 & ASCA & 74040000 & 1996-07-19 & 567 & $17.1$ & -- & Sy2 & 19.41 (7.41) \\
NGC 4939 & 0.0104 & 196.059 & -10.338 & 308.096 & 52.405 & XRT & 00031153005 & 2008-03-07 & 65 & $6.6$ & -- & Sy2 & 25.43 (8.38) \\
SWIFT J1309.2+1139 & 0.0251 & 197.274 & 11.633 & 318.766 & 73.960 & XMM & 0312192101 & 2006-06-27 & 2365 & $16.4$ & $98$ & XBONG & 55.64 (21.63) \\
2MASX J13105723+0837387 & 0.0527 & 197.738 & 8.627 & 317.846 & 70.932 & XRT & 00041174001 & 2010-08-20 & 50 & $8.8$ & -- & Sy2 & 10.94 (5.19) \\
II SZ 010 & 0.0343 & 198.274 & -11.128 & 311.463 & 51.384 & XRT & 00037378001 & 2009-02-25 & 1855 & $4.8$ & -- & Sy1 & 14.55 (5.71) \\
NGC 5033 & 0.0029 & 198.365 & 36.594 & 98.060 & 79.448 & XMM & 0094360501 & 2002-12-18 & 34064 & $11.9$ & $100$ & Sy1.9 & 6.95 (5.27) \\
UGC 08327 NED02 & 0.0366 & 198.823 & 44.407 & 108.983 & 72.069 & XRT & 00037093003 & 2007-09-20 & 152 & $2.4$ & -- & Sy2 & 16.79 (10.92) \\
NGC 5106 & 0.0319 & 200.246 & 8.980 & 325.355 & 70.554 & XRT & 00038063001 & 2009-11-23 & 73 & $6.9$ & -- & AGN & 13.97 (5.51) \\
{\bf NGC 5231} & 0.0218 & 203.951 & 2.999 & 328.569 & 63.640 & XMM & 0601781201 & 2010-01-30 & 13934 & $17.9$ & $100$ & Sy2 & 17.24 (7.46) \\
NGC 5252 & 0.0230 & 204.567 & 4.543 & 331.299 & 64.803 & XMM & 0152940101 & 2003-07-18 & 84375 & $67.3$ & $78$ & Sy1.9 & 111.13 (42.40) \\
Mrk 268 & 0.0399 & 205.297 & 30.378 & 52.466 & 78.630 & XMM & 0554500701 & 2008-07-20 & 6135 & $28.5$ & $76$ & Sy2/gal. pair & 18.97 (10.14) \\
NGC 5273 & 0.0035 & 205.535 & 35.654 & 74.348 & 76.246 & XMM & 0112551701 & 2002-06-14 & 25885 & $17.1$ & $100$ & Sy1.9 & 14.15 (5.41) \\
CGCG 102-048 & 0.0269 & 206.065 & 19.567 & 3.747 & 75.726 & XRT & 00037319001 & 2008-06-25 & 3 & $0.3$ & -- & Sy1.9 & 20.01 (7.62) \\
NGC 5290 & 0.0086 & 206.330 & 41.713 & 89.276 & 71.714 & XRT & 00038067001 & 2008-09-05 & 589 & $5.6$ & -- & Sy2 & 19.14 (8.05) \\
2MASX J13462846+1922432 & 0.0840 & 206.618 & 19.379 & 4.296 & 75.190 & XRT & 00090327001 & 2010-04-04 & 629 & $5.3$ & -- & galaxy & 10.71 (6.54) \\
{\bf UM 614} & 0.0327 & 207.470 & 2.079 & 334.600 & 61.305 & XMM & 0601781301 & 2010-01-31 & 3132 & $19.8$ & $34$ & Sy1 & 16.18 (6.77) \\
2MASX J13542913+1328068 & 0.0635 & 208.621 & 13.467 & 353.206 & 69.912 & XRT & 00040970001 & 2010-06-01 & 39 & $3.4$ & -- & galaxy & 9.70 (5.00) \\
2MASX J13553383+3520573 & 0.1016 & 208.892 & 35.350 & 67.887 & 74.047 & XRT & 00040971001 & 2010-08-26 & 135 & $7.2$ & -- & galaxy & 7.31 (4.91) \\
Mrk 464 & 0.0501 & 208.973 & 38.575 & 77.329 & 72.322 & XMM & 0072340701 & 2002-12-10 & 11416 & $8.1$ & $100$ & Sy1.5 & 20.26 (10.35) \\
Mrk 463 & 0.0504 & 209.011 & 18.371 & 5.821 & 72.747 & XMM & 0094401201 & 2001-12-22 & 4022 & $26.8$ & $90$ & Sy1/dual AGN & 11.61 (5.22) \\
NGC 5506 & 0.0062 & 213.313 & -3.208 & 339.150 & 53.809 & XMM & 0201830201 & 2004-07-11 & 122599 & $21.6$ & $100$ & Sy1.9 & 242.76 (95.00) \\
NGC 5548 & 0.0172 & 214.499 & 25.137 & 31.960 & 70.495 & XMM & 0089960301 & 2001-07-09 & 1875711 & $95.8$ & $97$ & Sy1.5 & 73.57 (32.55) \\
H 1419+480 & 0.0723 & 215.376 & 47.790 & 88.564 & 62.886 & XMM & 0094740201 & 2002-05-27 & 59341 & $21.5$ & $55$ & Sy1.5 & 19.42 (10.55) \\
NGC 5610 & 0.0169 & 216.095 & 24.615 & 31.299 & 68.973 & XRT & 00090180002 & 2009-11-18 & 232 & $5.8$ & -- & Sy2 & 19.30 (9.15) \\
Mrk 813 & 0.1105 & 216.854 & 19.831 & 19.700 & 66.861 & XRT & 00035307003 & 2007-01-10 & 1873 & $6.2$ & -- & Sy1 & 12.29 (6.23) \\
Mrk 1383 & 0.0865 & 217.278 & 1.285 & 349.218 & 55.125 & XMM & 0102040501 & 2000-07-28 & 22461 & $17.6$ & $100$ & Sy1 & 17.66 (7.16) \\
NGC 5674 & 0.0249 & 218.469 & 5.459 & 355.897 & 57.383 & XRT & 00040977001 & 2010-09-10 & 320 & $6.0$ & -- & Sy1.9 & 13.61 (5.71) \\
NGC 5683 & 0.0362 & 218.719 & 48.662 & 86.979 & 60.615 & XRT & 00038070002 & 2010-11-12 & 175 & $10.4$ & -- & Sy1 & 13.39 (7.99) \\
{\bf Mrk 817} & 0.0315 & 219.093 & 58.794 & 100.299 & 53.478 & XMM & 0601781401 & 2009-12-13 & 61638 & $14.2$* & $75$ & Sy1.5 & 27.39 (15.65) \\
2MASX J14391186+1415215 & 0.0714 & 219.799 & 14.256 & 11.223 & 61.795 & XRT & 00037321001 & 2008-04-27 & 163 & $4.7$ & -- & XBONG & 15.79 (6.53) \\
Mrk 477 & 0.0377 & 220.159 & 53.504 & 93.037 & 56.819 & ASCA & 73028000 & 1995-12-04 & 871 & $21.4$ & -- & Sy1 & 13.44 (7.71) \\
3C 303.0 & 0.1412 & 220.761 & 52.027 & 90.528 & 57.501 & ASCA & 73008000 & 1995-05-22 & 3131 & $18.2$ & -- & Sy1.5/FR-II & 8.12 (4.90) \\
2MASX J14530794+2554327 & 0.0490 & 223.283 & 25.910 & 37.352 & 62.817 & XRT & 00037322001 & 2008-02-01 & 1613 & $4.2$ & -- & Sy1 & 23.43 (10.56) \\
Mrk 841 & 0.0364 & 226.006 & 10.437 & 11.209 & 54.631 & XMM & 0112910201 & 2001-01-13 & 128688 & $10.1$ & $100$ & Sy1 & 36.08 (14.55) \\
MRK 1392 & 0.0361 & 226.485 & 3.708 & 2.754 & 50.264 & XRT & 00037323001 & 2008-12-23 & 764 & $5.8$ & -- & Sy1 & 18.81 (7.70) \\
2MASX J15064412+0351444 & 0.0377 & 226.682 & 3.863 & 3.137 & 50.211 & XRT & 00036622001 & 2007-12-19 & 472 & $9.4$ & -- & Sy2 & 16.77 (7.70) \\
{\bf NGC 5899} & 0.0085 & 228.763 & 42.050 & 69.398 & 57.216 & XMM & 0651850501 & 2011-02-13 & 14459 & $22.9$ & $100$ & Sy2 & 20.95 (9.67) \\

\hline
\end{tabular}}

Table \ref{table:observations}: Table of observations used for each object (continued).

%\caption{Table of observations used for each object.}
\end{table*}
\end{turnpage}
%\end{landscape}
\clearpage

\clearpage
\begin{turnpage}
%\newgeometry{top=7cm}
\begin{table*}
{\footnotesize\begin{tabular}{l|l|l|l|l|l|l|l|l|l|l|l|l|l}
%{p{2cm}p{1.3cm}p{2.8cm}p{2.7cm}p{2.8cm}p{2.0cm}p{1.5cm}}
%{|l|l|l|l|l|l}
\hline
Model identifier & \textsc{xspec} model string & Description \\
\hline
Simple power-law models\\
\hline
S1 & \textsc{tbabs(powerlaw)} & Power-law with Galactic absorption only \\
S2 & \textsc{tbabs(ztbabs(powerlaw))} & Absorbed power-law with Galactic and intrinsic (neutral) absorption\\
S3 & \textsc{tbabs(ztbabs(powerlaw+zgauss))} & As for S2, with a Fe K-$\alpha$ line at (default) 6.4~keV\\
S4 & \textsc{tbabs(ztbabs(powerlaw+zbbody))} & As for S2, with a soft excess modelled as a black body\\
S5 & \textsc{tbabs(ztbabs(zedge(powerlaw)))} & As for S2, with an edge at 0.73~keV (default) to model a warm absorber\\
S6 & \textsc{tbabs(ztbabs((powerlaw+zbbody+zgauss)))} & As for S2, with both a soft excess and Fe K-$\alpha$ line  \\
S7 & \textsc{tbabs(ztbabs(zedge(powerlaw+zgauss)))} & Absorbed power-law with warm absorber edge and Fe K-$\alpha$ line \\
S8 & \textsc{tbabs(ztbabs(zedge(powerlaw+zgauss+zbbody)))} & Absorbed power-law with warm absorber edge, Fe K-$\alpha$ line and soft excess\\
S9 & \textsc{tbabs(ztbabs(zedge(zedge(powerlaw))))} & Absorbed power-law with two warm absorber edges at 0.73 and 0.87~keV (default energies)\\
S10 & \textsc{tbabs(ztbabs(zedge(zedge(powerlaw+zgauss))))} & Absorbed power-law with two warm absorber edges and a Fe K-$\alpha$ line\\
S11 & \textsc{tbabs(ztbabs(zedge(zedge(powerlaw+zgauss+zbbody))))} & Absorbed power-law with two warm absorber edges, Fe K-$\alpha$ line and soft excess\\
\hline
Complex models (partial covering)\\
\hline
C1 & \textsc{tbabs(zpcfabs(powerlaw))} & Partially-covered absorbed power-law with Galactic absorption\\
C2 & \textsc{tbabs(zpcfabs(powerlaw+zgauss))} & As for C1, including a Fe K-$\alpha$ line at (default) 6.4~keV\\
\hline

\hline
\end{tabular}}
\caption{\label{table:modelcombinations} Model combinations used}
%\tablecomments{}
\end{table*}
\end{turnpage}
\clearpage

%\begin{landscape}
%\rotate
%\newgeometry{top=12cm}
\clearpage
\begin{turnpage}
\begin{table*}
{\tiny
\begin{tabular}{l|l|l|l|l|l|l|l|l|l|l|l|l|l|l|l|l}
%{p{2cm}p{1.3cm}p{2.8cm}p{2.7cm}p{2.8cm}p{2.0cm}p{1.5cm}}
%{|l|l|l|l|l|l}
\hline
AGN & Model ($\chi^{2}/\rm d.o.f$, $P(null\thinspace hyp.)$) & $N_{\rm H}^{\rm Gal}$ & $N_{\rm H}$ (covering fraction) & $\Gamma$ & $F_{\rm 0.5-2keV}$ & $F_{\rm 2-10keV}$ & $L^{\rm int}_{\rm 0.5-2keV}$ & $L^{\rm int}_{\rm 2-10keV}$ & $L_{\rm 14-195 keV}$ & $RL = L_{\rm 5 GHz}/L_{\rm 2-10 keV}^{\rm int}$ \\
    &                    (1)                                      &  (2)    &         (3)            &      (4)        &        (5)          &               (6)                    & (7) &  (8) & (9)  &             (10)          \\

\hline
3C 234 & C2 (658.26/361, 0.000) & $1.76$ & $23.81^{+0.02}_{-0.02}\thinspace(0.98)$ & $2.20_{-0.01}$ & 0.94 & 12.54 & 44.70 & $44.64^{+0.02}_{-0.00}$ & 44.9 (44.6$\dagger$) & $<-5.331$ \\
NGC 3227 & C2 (1988.13/1775, 0.000) & $1.99$ & $23.02^{+0.01}_{-0.01}\thinspace(0.92)$ & $1.50^{+0.01}$ & 3.83 & 84.25 & 41.18 & $41.58^{+0.01}_{-0.02}$ & 42.6 (42.7) & $-3.917$ \\
SDSS J104326.47+110524.2 & S2+BAT (80.53/89, 0.728) & $2.47$ & $20.82^{+0.19}_{-0.32}$ & $1.83^{+0.10}_{-0.09}$ & 29.18 & 51.99 & 43.26 & $43.43^{+0.10}_{-0.10}$ & 43.9 (43.8) & $<-5.403$ \\
MCG +06-24-008 & C1+BAT (4.72/7, 0.694) & $1.26$ & $23.81^{+0.55}_{-1.26}\thinspace(0.79^{+0.21}_{-0.48})$ & $1.50^{+0.46}$ & 9.04 & 46.06 & 42.81 & $43.20^{+0.90}_{-0.81}$ & 43.3 (43.3) & $-4.895$ \\
MCG +06-24-008 & S2+BAT (6.04/8, 0.643) & $1.26$ & $22.96^{+0.31}_{-1.05}$ & $1.50^{+0.38}$ & 0.18 & 34.11 & 42.46 & $42.85^{+0.33}_{-0.47}$ & 43.3 (43.3) & $-4.671$ \\
UGC 05881 & C1+BAT (7.81/13, 0.856) & $2.51$ & $24.39^{+0.15}_{-0.14}\thinspace(0.92^{+0.03}_{-0.09})$ & $2.05^{+0.15}_{-0.31}$ & 22.63 & 33.48 & 43.44 & $43.47^{+0.48}_{-43.47}$ & 43.3 (43.6$\dagger$) & $-5.245$ \\
Mrk 417 & C1+BAT (129.03/81, 0.001) & $1.88$ & $24.08^{+0.02}_{-0.02}\thinspace(0.99)$ & $1.87^{+0.04}_{-0.04}$ & 0.51 & 16.35 & 43.36 & $43.51^{+0.07}_{-0.07}$ & 43.9 (43.9$\dagger$) & $<-5.798$ \\
2MASX J10523297+1036205 & C1+BAT (10.65/12, 0.559) & $2.29$ & $24.41^{+0.23}_{-0.18}\thinspace(0.89^{+0.07}_{-0.09})$ & $1.63^{+0.40}_{-1.63}$ & 6.89 & 19.25 & 44.08 & $44.38^{+0.62}_{-0.56}$ & 44.5 (45.0$\dagger$) & $-5.277$ \\
Mrk 728 & S3 (616.39/633, 0.674) & $2.02$ & -- & $1.77^{+0.03}_{-0.02}$ & 21.34 & 37.28 & 42.80 & $43.03^{+0.02}_{-0.03}$ & 43.5 (43.5) & $-4.872$ \\
FBQS J110340.2+372925 & S2+BAT (52.36/49, 0.345) & $1.64$ & $<20.95$ & $1.69^{+0.15}_{-0.12}$ & 16.84 & 33.94 & 43.37 & $43.64^{+0.14}_{-0.15}$ & 44.1 (44.2$\dagger$) & $-4.442$ \\
2MASX J11053754+5851206 & C1+BAT (17.93/39, 0.998) & $0.61$ & $22.26^{+0.19}_{-0.29}\thinspace(0.61^{+0.07}_{-0.15})$ & $2.19^{+0.01}_{-0.31}$ & 10.35 & 16.65 & 44.33 & $44.27^{+0.22}_{-0.18}$ & 44.8 (44.5) & $<-4.961$ \\
2MASX J11053754+5851206 & S2+BAT (25.06/40, 0.969) & $0.61$ & $<20.86$ & $1.71^{+0.15}_{-0.08}$ & 10.22 & 18.76 & 44.01 & $44.27^{+0.11}_{-0.12}$ & 44.8 (44.5) & $<-4.930$ \\
CGCG 291-028 & S2+BAT (8.81/7, 0.266) & $0.61$ & $23.87^{+0.17}_{-0.30}$ & $2.20*$ & 0.00 & 6.63 & 43.34 & $43.28^{+0.47}_{-0.68}$ & 43.4 (43.5) & $<-4.342$ \\
IC 2637 & S6 (804.05/786, 0.320) & $2.23$ & -- & $1.69^{+0.04}_{-0.04}$ & 16.97 & 29.92 & 42.53 & $42.75^{+0.04}_{-0.05}$ & 43.5 (43.2$\dagger$) & $-3.998$ \\
MCG +09-19-015 & C1+BAT (3.28/6, 0.773) & $0.78$ & $<25.23\thinspace(0.64*)$ & $1.95^{+0.25}_{-0.43}$ & 9.90 & 19.37 & 43.51 & $43.61^{+0.74}_{-43.61}$ & 44.0 (43.9$\dagger$) & $<-5.191$ \\
MCG +09-19-015 & S2+BAT (3.63/7, 0.821) & $0.78$ & $<23.39$ & $1.80^{+0.36}_{-0.29}$ & 0.25 & 17.69 & 43.26 & $43.46^{+0.70}_{-43.46}$ & 44.0 (43.9$\dagger$) & $<-5.091$ \\
PG 1114+445 & S10 (1254.56/1133, 0.007) & $1.77$ & -- & $1.50^{+0.01}$ & 5.52 & 22.33 & 43.66 & $44.06^{+0.01}_{-0.01}$ & 44.7 (44.7) & $<-5.010$ \\
ARP 151 & C1+BAT (107.26/117, 0.729) & $1.08$ & $21.73^{+0.59}_{-1.99}\thinspace(0.16^{+0.84}_{-0.15})$ & $1.78^{+0.07}_{-0.06}$ & 51.10 & 92.65 & 42.75 & $42.96^{+0.09}_{-0.09}$ & 43.3 (43.4$\dagger$) & $-5.569$ \\
ARP 151 & S2+BAT (108.53/118, 0.722) & $1.08$ & $<20.57$ & $1.75^{+0.05}_{-0.04}$ & 51.39 & 91.73 & 42.72 & $42.95^{+0.05}_{-0.06}$ & 43.3 (43.4$\dagger$) & $-5.567$ \\
1RXS J1127+1909 & S7 (781.23/796, 0.639) & $1.40$ & $<19.90$ & $1.50*$ & 15.67 & 46.75 & 43.70 & $44.09^{+0.01}_{-0.01}$ & 44.7 (44.6$\dagger$) & $-4.600$ \\
UGC 06527 & C1 (84.84/35, 0.000) & $1.09$ & $24.05^{+0.03}_{-0.03}\thinspace(0.99)$ & $2.20_{-0.03}$ & 0.39 & 5.45 & 42.96 & $42.90^{+0.06}_{-0.04}$ & 43.2 (43.2) & $-3.923$ \\
IC 2921 & C1+BAT (52.41/38, 0.060) & $3.21$ & $21.85^{+0.15}_{-0.24}\thinspace(0.89^{+0.10}_{-0.08})$ & $2.09^{+0.11}_{-0.21}$ & 11.38 & 26.66 & 43.08 & $43.08^{+0.21}_{-0.11}$ & 43.9 (43.8) & $<-5.121$ \\
NGC 3758 & S4 (767.93/756, 0.374) & $2.00$ & $21.05^{+0.14}_{-0.18}$ & $1.96^{+0.04}_{-0.04}$ & 33.91 & 51.62 & 42.96 & $43.01^{+0.05}_{-0.05}$ & 43.4 (43.4$\dagger$) & $-5.089$ \\
SBS 1136+594 & C1+BAT (134.38/124, 0.247) & $0.94$ & $22.39^{+0.22}_{-0.22}\thinspace(0.36^{+0.08}_{-0.12})$ & $2.17^{+0.03}_{-0.06}$ & 49.69 & 63.58 & 43.80 & $43.75^{+0.09}_{-0.06}$ & 44.3 (43.8$\dagger$) & $<-5.554$ \\
SBS 1136+594 & S2+BAT (151.40/125, 0.054) & $0.94$ & $<20.46$ & $2.05^{+0.04}_{-0.03}$ & 50.85 & 57.38 & 43.65 & $43.69^{+0.04}_{-0.05}$ & 44.3 (43.8$\dagger$) & $<-5.530$ \\
Mrk 744 & C2 (517.06/538, 0.734) & $2.03$ & $22.59^{+0.02}_{-0.02}\thinspace(0.97)$ & $1.50^{+0.02}$ & 3.04 & 66.74 & 41.73 & $42.13^{+0.02}_{-0.03}$ & 42.6 (42.6) & $-4.590$ \\
PG 1138+222 & C1+BAT (116.81/99, 0.107) & $2.02$ & $24.52^{+0.43}_{-0.21}\thinspace(0.79^{+0.05}_{-0.07})$ & $2.02^{+0.06}_{-0.05}$ & 39.89 & 49.93 & 44.28 & $44.33^{+0.15}_{-0.16}$ & 44.2 (44.5$\dagger$) & $-5.714$ \\
2E 1139.7+1040 & C1+BAT (10.23/7, 0.176) & $3.72$ & $23.87^{+0.77}_{-0.58}\thinspace(0.91^{+0.07}_{-0.35})$ & $1.95*$ & 2.95 & 14.47 & 44.35 & $44.44^{+0.42}_{-44.44}$ & 45.0 (44.7$\dagger$) & $<-5.392$ \\
KUG 1141+371 & C2 (736.77/689, 0.101) & $1.65$ & $23.16^{+0.37}_{-0.33}\thinspace(0.24^{+0.10}_{-0.10})$ & $1.80^{+0.04}_{-0.04}$ & 9.92 & 19.42 & 42.64 & $42.85^{+0.09}_{-0.08}$ & 43.7 (43.2$\dagger$) & $<-5.030$ \\
KUG 1141+371 & S3 (748.57/690, 0.060) & $1.65$ & -- & $1.75^{+0.03}_{-0.03}$ & 10.04 & 18.44 & 42.53 & $42.78^{+0.03}_{-0.03}$ & 43.7 (43.2$\dagger$) & $<-4.982$ \\
MCG+10-17-061 & C1+BAT (30.58/23, 0.133) & $1.31$ & $22.91^{+0.18}_{-0.23}\thinspace(0.85^{+0.07}_{-0.07})$ & $2.13^{+0.07}_{-0.12}$ & 8.06 & 35.19 & 42.05 & $42.03^{+0.20}_{-0.15}$ & 42.5 (42.1$\dagger$) & $-4.810$ \\
2MASX J11475508+0902284 & S5 (80.10/52, 0.007) & $1.97$ & -- & $1.78^{+0.08}_{-0.08}$ & 18.60 & 40.09 & 43.45 & $43.65^{+0.08}_{-0.08}$ & 44.1 (44.1) & $-5.132$ \\
MCG +05-28-032 & C1+BAT (14.55/15, 0.485) & $1.56$ & $22.71^{+0.20}_{-0.28}\thinspace(1.00^{+0.00}_{-1.00})$ & $1.51^{+0.12}_{-1.51}$ & 1.17 & 51.53 & 42.48 & $42.87^{+0.17}_{-0.16}$ & 43.5 (43.6$\dagger$) & $-4.550$ \\
MCG +05-28-032 & S2+BAT (14.54/16, 0.558) & $1.56$ & $22.70^{+0.25}_{-0.28}$ & $1.51^{+0.12}_{-1.51}$ & 1.16 & 51.54 & 42.48 & $42.87^{+0.17}_{-0.16}$ & 43.5 (43.6$\dagger$) & $-4.550$ \\
2MASX J11491868-0416512 & C1+BAT (34.15/30, 0.275) & $2.18$ & $24.23^{+0.52}_{-0.20}\thinspace(0.90^{+0.04}_{-0.07})$ & $2.01^{+0.13}_{-0.12}$ & 18.03 & 35.92 & 44.50 & $44.56^{+0.30}_{-0.31}$ & 44.3 (44.7$\dagger$) & $<-5.851$ \\
MCG -01-30-041 & C1+BAT (5.28/6, 0.509) & $2.16$ & $24.16^{+0.18}_{-0.16}\thinspace(0.99^{+0.01}_{-0.02})$ & $1.96^{+0.20}_{-0.26}$ & 1.37 & 14.61 & 42.99 & $43.08^{+0.42}_{-0.54}$ & 43.0 (43.3$\dagger$) & $-5.101$ \\
NGC 3998 & S2 (640.90/675, 0.823) & $1.01$ & $20.37^{+0.12}_{-0.17}$ & $1.84^{+0.03}_{-0.02}$ & 75.99 & 121.10 & 41.32 & $41.49^{+0.03}_{-0.03}$ & 41.7 (41.7) & $-3.799$ \\
CGCG 041-020 & C1 (334.61/315, 0.214) & $1.18$ & $23.10^{+0.02}_{-0.02}\thinspace(0.99)$ & $1.50^{+0.04}$ & 0.33 & 51.59 & 42.96 & $43.36^{+0.04}_{-0.04}$ & 43.8 (44.0$\dagger$) & $-5.011$ \\
MRK 1310 & C1+BAT (71.21/74, 0.570) & $2.50$ & $24.23^{+0.24}_{-0.18}\thinspace(0.79^{+0.06}_{-0.10})$ & $1.96^{+0.07}_{-0.07}$ & 38.85 & 61.82 & 43.20 & $43.29^{+0.20}_{-0.21}$ & 43.1 (43.5$\dagger$) & $-5.619$ \\
NGC 4051 & S8 (2277.83/1202, 0.000) & $1.15$ & -- & $1.53*$ & 47.69 & 61.74 & 40.77 & $40.86^{+0.01}_{-0.01}$ & 41.7 (41.7) & $-4.394$ \\
Ark 347 & C1+BAT (16.63/16, 0.410) & $2.30$ & $23.66^{+0.16}_{-0.15}\thinspace(0.96^{+0.03}_{-0.03})$ & $1.58^{+0.15}_{-1.58}$ & 0.85 & 20.04 & 42.44 & $42.78^{+0.23}_{-0.27}$ & 43.5 (43.5$\dagger$) & $-5.022$ \\
PG 1202+281 & S6 (1044.22/903, 0.001) & $1.77$ & -- & $1.88^{+0.02}_{-0.02}$ & 24.14 & 33.52 & 44.28 & $44.39^{+0.02}_{-0.02}$ & 44.9 (44.8) & $<-5.215$ \\

\hline

\end{tabular}}
\caption{\label{table:fitresults} Basic fit results}
\tablecomments{\footnotesize  (1) Best-fitting model (chi-squared/number of degrees of freedom and null hypothesis probability), (2) Galactic neutral hydrogen column density in units of $10^{20} \rm cm^{-2}$ (3) Intrinsic column density $N_{\rm H}$: `--' indicates an insignificant column density below $10^{19} \thinspace \rm cm^{-2}$ (if partial covering model used, the covering fraction is provided in brackets: errors below $5 \times 10^{-3}$ are not shown, and * denotes that the covering fraction was poorly constrained), (4) Photon index $\Gamma$ (* denotes that $\Gamma$ was frozen in the fit; limits of $1.5<\Gamma<2.2$ were imposed on the fit on physical grounds), (5,6) Fluxes in 0.5--2~keV and 2--10~keV bands in units of $10^{-13} \rm erg s^{-1} cm^{-2}$, (7,8) Absorption-corrected 0.5--2~keV and 2--10~keV luminosities, quoted as $log(L_{\rm 2-10keV}/\rm erg s^{-1})$, (9) BAT luminosity quoted as $log(L_{\rm 2-10keV}/\rm erg s^{-1})$, (10) Logarithm of radio loudness parameter derived using 5 GHz fluxes or upper limits from the FIRST survey and 2--10~keV intrinsic luminosities, with $RL$ as defined in \cite{2003ApJ...583..145T}.}

\end{table*}
%\end{landscape}

%\begin{landscape}
%\rotate
\begin{table*}
{\tiny
\begin{tabular}{l|l|l|l|l|l|l|l|l|l|l|l|l|l|l|l|l}
%{p{2cm}p{1.3cm}p{2.8cm}p{2.7cm}p{2.8cm}p{2.0cm}p{1.5cm}}
%{|l|l|l|l|l|l}
\hline
AGN & Model ($\chi^{2}/\rm d.o.f$, $P(null\thinspace hyp.)$) & $N_{\rm H}^{\rm Gal}$ & $N_{\rm H}$ (covering fraction) & $\Gamma$ & $F_{\rm 0.5-2keV}$ & $F_{\rm 2-10keV}$ & $L^{\rm int}_{\rm 0.5-2keV}$ & $L^{\rm int}_{\rm 2-10keV}$ & $L_{\rm 14-195 keV}$ & $RL = L_{\rm 5 GHz}/L_{\rm 2-10 keV}^{\rm int}$ \\
    &                    (1)                                      &  (2)    &         (3)            &      (4)        &        (5)          &               (6)                    & (7) &  (8) & (9)  &             (10)          \\

\hline
UGC 7064 & C2 (876.84/648, 0.000) & $1.35$ & $23.00^{+0.02}_{-0.02}\thinspace(0.95^{+0.01}_{-0.01})$ & $1.88^{+0.02}_{-0.07}$ & 1.24 & 22.97 & 42.51 & $42.67^{+0.06}_{-0.06}$ & 43.3 (43.1$\dagger$) & $-4.540$ \\
2MASX J12055599+4959561 & S5 (3.35/5, 0.646) & $1.98$ & $<21.26$ & $1.64^{+0.53}_{-1.64}$ & 3.29 & 8.96 & 42.62 & $42.92^{+0.24}_{-0.45}$ & 44.1 (41.6$\dagger$) & $-4.597$ \\
NGC 4102 & C1+BAT (143.35/80, 0.000) & $1.68$ & $24.51^{+0.04}_{-0.03}\thinspace(0.99)$ & $2.20_{-0.02}$ & 2.76 & 5.41 & 41.75 & $41.69^{+0.03}_{-0.02}$ & 41.7 (41.6$\dagger$) & $-3.789$ \\
B2 1204+34 & C1+BAT (8.50/13, 0.810) & $1.36$ & $25.35^{+0.56}_{-1.08}\thinspace(0.70^{+0.21}_{-0.15})$ & $1.50^{+0.37}$ & 4.59 & 11.66 & 43.35 & $43.74^{+0.68}_{-0.40}$ & 44.4 (44.4$\dagger$) & $-4.362$ \\
Mrk 198 & C2 (649.78/654, 0.539) & $1.66$ & $23.00^{+0.01}_{-0.01}\thinspace(0.99)$ & $1.61^{+0.04}_{-0.04}$ & 0.65 & 51.25 & 42.66 & $42.98^{+0.06}_{-0.05}$ & 43.5 (43.6$\dagger$) & $-4.735$ \\
NGC 4138 & C2 (408.63/332, 0.003) & $1.25$ & $23.12^{+0.02}_{-0.02}\thinspace(0.99)$ & $1.61^{+0.05}_{-0.06}$ & 0.51 & 53.98 & 40.89 & $41.21^{+0.09}_{-0.09}$ & 41.8 (41.8) & $<-5.648$ \\
NGC 4151 & C2 (11118.00/2780, 0.000) & $2.30$ & $23.23^{+0.00}_{-0.00}\thinspace(0.97)$ & $2.20_{-0.00}$ & 29.49 & 409.58 & 42.33 & $42.28^{+0.00}_{-0.00}$ & 43.1 (43.1) & $-4.189$ \\
KUG 1208+386 & C1+BAT (287.72/230, 0.006) & $1.84$ & $23.16^{+0.04}_{-0.04}\thinspace(0.97^{+0.01}_{-0.00})$ & $1.72^{+0.04}_{-0.04}$ & 0.71 & 29.53 & 42.50 & $42.75^{+0.07}_{-0.07}$ & 43.4 (43.3$\dagger$) & $-4.668$ \\
2MASX J12135456-0530193 & S2+BAT (32.90/27, 0.200) & $3.54$ & $<21.01$ & $1.96^{+0.24}_{-0.18}$ & 18.64 & 26.25 & 43.34 & $43.43^{+0.20}_{-0.06}$ & 44.1 (44.1) & $-4.911$ \\
Was 49b & C1 (120.87/139, 0.864) & $1.77$ & $23.59^{+0.18}_{-0.20}\thinspace(0.87^{+0.04}_{-0.05})$ & $1.50^{+0.12}$ & 1.50 & 14.58 & 43.05 & $43.45^{+0.15}_{-0.22}$ & 44.2 (44.3) & $<-5.226$ \\
NGC 4235 & C2 (502.43/514, 0.634) & $1.45$ & $21.47^{+0.10}_{-0.12}\thinspace(0.82^{+0.08}_{-0.05})$ & $1.62^{+0.05}_{-0.05}$ & 9.19 & 28.52 & 41.28 & $41.60^{+0.06}_{-0.06}$ & 42.7 (42.7) & $-4.489$ \\
NGC 4235 & S3 (511.43/515, 0.536) & $1.45$ & $21.22^{+0.04}_{-0.05}$ & $1.57^{+0.04}_{-0.04}$ & 9.19 & 28.93 & 41.25 & $41.61^{+0.04}_{-0.04}$ & 42.7 (42.7) & $-4.487$ \\
Mrk 202 & C1 (207.46/164, 0.012) & $1.44$ & $23.80^{+0.74}_{-0.30}\thinspace(0.33^{+0.62}_{-0.33})$ & $1.87^{+0.10}_{-0.08}$ & 14.03 & 23.00 & 42.32 & $42.48^{+0.82}_{-0.22}$ & 42.9 (42.9) & $<-5.162$ \\
Mrk 766 & S6 (3841.54/2358, 0.000) & $1.78$ & $20.45^{+0.08}_{-0.06}$ & $1.50^{+0.00}$ & 34.29 & 66.57 & 42.16 & $42.38^{+0.00}_{-0.00}$ & 42.9 (42.6$\dagger$) & $-3.960$ \\
NGC 4258 & C1 (803.72/450, 0.000) & $1.60$ & $23.26^{+0.01}_{-0.01}\thinspace(0.98)$ & $2.14^{+0.02}_{-0.02}$ & 2.43 & 65.31 & 40.84 & $40.82^{+0.03}_{-0.03}$ & 41.1 (41.0) & $-5.622$ \\
Mrk 50 & S4 (902.32/879, 0.285) & $1.58$ & -- & $1.95^{+0.02}_{-0.02}$ & 109.79 & 126.46 & 43.14 & $43.19^{+0.02}_{-0.02}$ & 43.5 (43.4) & $<-5.778$ \\
NGC 4388 & C1+BAT (346.01/217, 0.000) & $2.58$ & $23.81^{+0.02}_{-0.02}\thinspace(0.99)$ & $1.82^{+0.02}_{-0.02}$ & 2.50 & 76.57 & 42.50 & $42.68^{+0.04}_{-0.04}$ & 43.6 (43.6) & $-4.829$ \\
NGC 4395 & C2 (503.48/476, 0.185) & $1.85$ & $23.01^{+0.02}_{-0.02}\thinspace(0.98)$ & $1.50^{+0.02}$ & 0.77 & 59.40 & 39.90 & $40.32^{+0.02}_{-0.03}$ & 40.8 (40.9) & $-5.573$ \\
Ark 374 & S8 (1124.99/1015, 0.009) & $2.75$ & -- & $2.14^{+0.03}_{-0.02}$ & 31.55 & 30.06 & 43.51 & $43.45^{+0.02}_{-0.02}$ & 44.2 (43.7$\dagger$) & $-5.705$ \\
NGC 4579 & S3 (1568.92/1153, 0.000) & $2.97$ & $20.44^{+0.08}_{-0.10}$ & $2.01^{+0.02}_{-0.02}$ & 25.25 & 36.41 & 41.20 & $41.30^{+0.02}_{-0.02}$ & 41.8 (41.8) & $-3.773$ \\
NGC 4593 & S6 (1803.06/1562, 0.000) & $1.89$ & $20.57^{+0.08}_{-0.09}$ & $1.83^{+0.01}_{-0.01}$ & 256.45 & 377.74 & 42.71 & $42.82^{+0.01}_{-0.01}$ & 43.2 (43.1) & $-5.809$ \\
NGC 4619 & C1 (158.50/162, 0.563) & $1.39$ & $21.08^{+0.95}_{-0.65}\thinspace(0.93^{+0.07}_{-0.79})$ & $1.61^{+0.20}_{-1.61}$ & 4.60 & 11.85 & 41.83 & $42.14^{+0.08}_{-0.21}$ & 42.9 (43.0) & $-4.572$ \\
NGC 4619 & S2 (158.56/163, 0.584) & $1.39$ & $21.01^{+0.32}_{-0.72}$ & $1.61^{+0.15}_{-1.61}$ & 4.62 & 11.93 & 41.83 & $42.15^{+0.07}_{-0.13}$ & 42.9 (43.0) & $-4.572$ \\
NGC 4686 & C1+BAT (119.36/61, 0.000) & $1.35$ & $23.89^{+0.03}_{-0.03}\thinspace(0.99)$ & $2.00^{+0.05}_{-0.05}$ & 0.23 & 8.47 & 42.38 & $42.44^{+0.10}_{-0.10}$ & 43.2 (42.7$\dagger$) & $-4.972$ \\
2MASX J13000533+1632151 & C1+BAT (67.97/84, 0.898) & $1.98$ & $22.51^{+0.10}_{-0.07}\thinspace(0.95^{+0.03}_{-0.02})$ & $1.50^{+0.17}$ & 1.49 & 20.43 & 43.14 & $43.53^{+0.12}_{-0.16}$ & 44.4 (44.3) & $-4.023$ \\
2MASX J13000533+1632151 & S2+BAT (76.32/85, 0.738) & $1.98$ & $22.41^{+0.07}_{-0.06}$ & $1.50^{+0.11}$ & 1.38 & 19.78 & 43.12 & $43.51^{+0.08}_{-0.13}$ & 44.4 (44.3) & $-4.015$ \\
MRK 0783 & S2+BAT (42.52/34, 0.150) & $1.89$ & $21.16^{+0.19}_{-0.24}$ & $1.50^{+0.07}$ & 14.32 & 45.81 & 43.28 & $43.68^{+0.05}_{-0.08}$ & 44.3 (44.4$\dagger$) & $-3.996$ \\
SWIFT J1303.9+5345 & C1 (1069.08/1048, 0.318) & $1.69$ & $23.66^{+0.17}_{-0.17}\thinspace(0.33^{+0.07}_{-0.06})$ & $1.85^{+0.02}_{-0.02}$ & 92.89 & 162.47 & 43.46 & $43.62^{+0.06}_{-0.05}$ & 43.9 (44.0$\dagger$) & $-5.948$ \\
SWIFT J1303.9+5345 & S4 (1090.34/1045, 0.161) & $1.69$ & -- & $1.71^{+0.02}_{-0.02}$ & 92.85 & 155.66 & 43.28 & $43.49^{+0.03}_{-0.03}$ & 43.9 (44.0$\dagger$) & $-5.833$ \\
NGC 4941 & C1+BAT (57.40/64, 0.707) & $2.17$ & $24.12^{+0.12}_{-0.17}\thinspace(0.97^{+0.03}_{-0.07})$ & $1.84^{+0.23}_{-0.23}$ & 2.13 & 12.75 & 41.30 & $41.47^{+0.23}_{-0.24}$ & 41.8 (41.8) & $-4.572$ \\
NGC 4939 & C1+BAT (16.78/9, 0.052) & $3.30$ & $23.81^{+0.16}_{-0.13}\thinspace(0.98^{+0.01}_{-0.01})$ & $1.50^{+0.13}$ & 0.36 & 15.44 & 41.72 & $42.12^{+0.20}_{-0.17}$ & 42.8 (42.9$\dagger$) & $<-5.576$ \\
SWIFT J1309.2+1139 & C1+BAT (227.36/106, 0.000) & $1.93$ & $24.00^{+0.02}_{-0.02}\thinspace(1.00)$ & $1.50^{+0.03}$ & 0.10 & 22.19 & 42.85 & $43.24^{+0.05}_{-0.05}$ & 43.9 (44.0$\dagger$) & $-5.394$ \\
2MASX J13105723+0837387 & S2+BAT (14.22/7, 0.047) & $2.18$ & $23.63^{+0.26}_{-0.19}$ & $1.99*$ & 0.00 & 9.12 & 43.19 & $43.26^{+0.39}_{-0.60}$ & 43.9 (43.8) & $<-4.342$ \\
II SZ 010 & S2+BAT (316.86/76, 0.000) & $2.65$ & $<20.52$ & $2.20_{-0.00}$ & 50.40 & 47.83 & 43.17 & $43.11^{+0.03}_{-0.03}$ & 43.6 (43.1$\dagger$) & $<-5.510$ \\
NGC 5033 & S3 (837.13/834, 0.463) & $1.06$ & $<20.08$ & $1.70^{+0.03}_{-0.02}$ & 22.92 & 44.60 & 40.63 & $40.91^{+0.03}_{-0.03}$ & 41.1 (41.2) & $-4.401$ \\
UGC 08327 NED02 & S2+BAT (22.85/17, 0.154) & $1.48$ & $23.16^{+0.15}_{-0.17}$ & $2.02^{+0.14}_{-0.13}$ & 0.13 & 62.03 & 43.48 & $43.53^{+0.27}_{-0.27}$ & 43.7 (43.7$\dagger$) & $-4.732$ \\
NGC 5106 & C1+BAT (10.83/10, 0.371) & $1.71$ & $24.13^{+0.17}_{-0.39}\thinspace(0.93^{+0.07}_{-0.13})$ & $1.78^{+0.42}_{-1.78}$ & 6.21 & 25.21 & 43.33 & $43.54^{+0.58}_{-43.54}$ & 43.5 (44.0$\dagger$) & $-3.910$ \\
NGC 5231 & C2 (607.80/582, 0.222) & $1.88$ & $22.57^{+0.02}_{-0.02}\thinspace(0.99)$ & $1.68^{+0.05}_{-0.05}$ & 3.30 & 61.69 & 42.60 & $42.88^{+0.07}_{-0.07}$ & 43.3 (43.5$\dagger$) & $-4.860$ \\
NGC 5252 & C2 (2827.17/2270, 0.000) & $2.14$ & $22.72^{+0.01}_{-0.01}\thinspace(0.96)$ & $1.50^{+0.00}$ & 3.50 & 94.79 & 42.72 & $43.13^{+0.01}_{-0.01}$ & 44.1 (44.2) & $-4.704$ \\
Mrk 268 & C2 (505.01/339, 0.000) & $1.37$ & $23.72^{+0.02}_{-0.02}\thinspace(0.99)$ & $1.89^{+0.04}_{-0.04}$ & 0.56 & 24.65 & 43.33 & $43.48^{+0.05}_{-0.05}$ & 43.8 (44.0$\dagger$) & $-3.991$ \\
NGC 5273 & S10 (1054.39/841, 0.000) & $0.92$ & -- & $1.50^{+0.00}$ & 13.14 & 59.75 & 40.79 & $41.21^{+0.01}_{-0.01}$ & 41.6 (41.5) & $-5.089$ \\
NGC 5290 & S2+BAT (30.61/31, 0.486) & $0.94$ & $22.05^{+0.08}_{-0.09}$ & $1.50^{+0.03}$ & 8.37 & 67.64 & 41.66 & $42.05^{+0.04}_{-0.06}$ & 42.5 (42.8$\dagger$) & $-4.807$ \\
2MASX J13462846+1922432 & S2+BAT (35.21/35, 0.458) & $1.84$ & $<21.07$ & $1.84^{+0.09}_{-0.08}$ & 17.04 & 28.48 & 43.52 & $43.68^{+0.11}_{-0.12}$ & 44.3 (44.1$\dagger$) & $<-5.851$ \\

\hline
\end{tabular}}
Table \ref{table:fitresults}: Basic fit results (continued)
%\caption{\tiny Table \ref{fitresults} Basic fit results (continued).}
\end{table*}
%\end{landscape}

%\begin{landscape}
%\rotate
\begin{table*}
{\tiny
\begin{tabular}{l|l|l|l|l|l|l|l|l|l|l|l|l|l|l|l|l}
%{p{2cm}p{1.3cm}p{2.8cm}p{2.7cm}p{2.8cm}p{2.0cm}p{1.5cm}}
%{|l|l|l|l|l|l}
\hline
AGN & Model ($\chi^{2}/\rm d.o.f$, $P(null\thinspace hyp.)$) & $N_{\rm H}^{\rm Gal}$ & $N_{\rm H}$ (covering fraction) & $\Gamma$ & $F_{\rm 0.5-2keV}$ & $F_{\rm 2-10keV}$ & $L^{\rm int}_{\rm 0.5-2keV}$ & $L^{\rm int}_{\rm 2-10keV}$ & $L_{\rm 14-195 keV}$ & $RL = L_{\rm 5 GHz}/L_{\rm 2-10 keV}^{\rm int}$ \\
    &                    (1)                                      &  (2)    &         (3)            &      (4)        &        (5)          &               (6)                    & (7) &  (8) & (9)  &             (10)          \\

\hline
UM 614 & S3 (135.25/157, 0.895) & $1.79$ & $21.11^{+0.11}_{-0.11}$ & $1.50^{+0.08}$ & 4.11 & 14.18 & 42.10 & $42.53^{+0.03}_{-0.07}$ & 43.6 (43.3$\dagger$) & $-4.761$ \\
2MASX J13553383+3520573 & S2+BAT (11.33/8, 0.183) & $1.23$ & $<23.38$ & $2.20_{-0.41}$ & 0.46 & 20.89 & 43.97 & $43.91^{+0.46}_{-0.35}$ & nan (44.3) & $-4.911$ \\
Mrk 464 & C1 (397.20/390, 0.390) & $1.42$ & $23.98^{+0.34}_{-0.43}\thinspace(0.49^{+0.28}_{-0.22})$ & $1.72^{+0.03}_{-0.03}$ & 11.98 & 25.80 & 43.14 & $43.39^{+0.37}_{-0.18}$ & 44.1 (44.1) & $-4.158$ \\
Mrk 463 & C2 (419.98/162, 0.000) & $2.03$ & $23.82^{+0.03}_{-0.03}\thinspace(0.96^{+0.01}_{-0.01})$ & $2.20_{-0.02}$ & 0.97 & 5.99 & 43.20 & $43.15^{+0.03}_{-0.05}$ & 43.8 (43.8) & $-2.554$ \\
NGC 5506 & C2 (2029.06/1850, 0.002) & $4.08$ & $22.63^{+0.01}_{-0.01}\thinspace(0.99)$ & $1.69^{+0.01}_{-0.01}$ & 28.21 & 668.44 & 42.55 & $42.83^{+0.02}_{-0.02}$ & 43.3 (43.4) & $-4.118$ \\
NGC 5548 & S11 (3263.75/2420, 0.000) & $1.55$ & -- & $1.71^{+0.00}_{-0.00}$ & 207.31 & 391.76 & 43.17 & $43.40^{+0.00}_{-0.01}$ & 43.7 (43.6) & $-5.242$ \\
H 1419+480 & S7 (1143.32/1051, 0.024) & $1.64$ & $<20.01$ & $1.83^{+0.03}_{-0.02}$ & 40.14 & 71.63 & 43.75 & $43.95^{+0.02}_{-0.03}$ & 44.4 (44.4) & $-5.316$ \\
NGC 5610 & S2+BAT (21.63/14, 0.087) & $1.90$ & $22.89^{+0.13}_{-0.16}$ & $1.63^{+0.12}_{-0.11}$ & 0.37 & 40.74 & 42.24 & $42.55^{+0.22}_{-0.22}$ & 43.1 (43.2$\dagger$) & $-4.153$ \\
Mrk 813 & S2+BAT (75.65/76, 0.490) & $2.59$ & -- & $1.99^{+0.06}_{-0.05}$ & 39.27 & 49.43 & 44.11 & $44.18^{+0.05}_{-0.05}$ & 44.6 (44.2$\dagger$) & $-5.186$ \\
Mrk 1383 & S4 (427.24/371, 0.023) & $2.60$ & -- & $2.06^{+0.05}_{-0.05}$ & 83.27 & 80.88 & 44.22 & $44.17^{+0.05}_{-0.05}$ & 44.5 (44.5) & $-5.481$ \\
NGC 5674 & C1+BAT (7.92/17, 0.968) & $2.48$ & $23.05^{+0.19}_{-0.12}\thinspace(0.95^{+0.05}_{-0.07})$ & $2.14^{+0.06}_{-0.25}$ & 4.15 & 48.52 & 43.07 & $43.04^{+0.26}_{-0.19}$ & 43.3 (43.3) & $-4.975$ \\
NGC 5674 & S2+BAT (8.11/18, 0.977) & $2.48$ & $22.94^{+0.10}_{-0.14}$ & $2.09^{+0.11}_{-0.38}$ & 0.59 & 48.37 & 43.01 & $43.01^{+0.37}_{-0.29}$ & 43.3 (43.3) & $-4.948$ \\
NGC 5683 & C1+BAT (6.97/10, 0.729) & $2.87$ & $22.56^{+0.42}_{-0.48}\thinspace(0.68^{+0.12}_{-0.28})$ & $2.15^{+0.05}_{-0.31}$ & 2.19 & 5.10 & 42.28 & $42.24^{+0.28}_{-0.25}$ & 43.6 (43.6) & $<-4.471$ \\
Mrk 817 & S6 (587.32/562, 0.222) & $1.15$ & $<20.12$ & $2.09^{+0.02}_{-0.02}$ & 157.81 & 143.66 & 43.56 & $43.50^{+0.02}_{-0.03}$ & 43.8 (43.5$\dagger$) & $-4.939$ \\
2MASX J14391186+1415215 & S2+BAT (20.57/15, 0.151) & $1.42$ & $22.50^{+0.21}_{-0.21}$ & $1.57^{+0.15}_{-1.57}$ & 1.68 & 29.06 & 43.25 & $43.59^{+0.17}_{-0.24}$ & 44.3 (44.3$\dagger$) & $<-5.212$ \\
Mrk 477 & C1 (82.02/87, 0.631) & $1.05$ & $23.80^{+0.17}_{-0.17}\thinspace(0.95^{+0.03}_{-0.05})$ & $1.74^{+0.24}_{-1.74}$ & 1.53 & 16.26 & 43.01 & $43.25^{+0.09}_{-0.51}$ & 43.6 (43.7) & $-3.768$ \\
3C 303.0 & S2 (156.02/148, 0.310) & $1.71$ & $21.23^{+0.22}_{-0.38}$ & $1.73^{+0.13}_{-0.10}$ & 9.15 & 21.20 & 43.79 & $44.03^{+0.12}_{-0.14}$ & 44.6 (44.6) & $-2.969$ \\
2MASX J14530794+2554327 & C1+BAT (79.55/72, 0.254) & $3.26$ & $22.02^{+0.17}_{-0.22}\thinspace(0.59^{+0.05}_{-0.07})$ & $2.20_{-0.05}$ & 51.05 & 79.43 & 43.73 & $43.66^{+0.06}_{-0.06}$ & 44.1 (43.7$\dagger$) & $<-5.682$ \\
Mrk 841 & S6 (1053.49/941, 0.006) & $2.22$ & -- & $2.13^{+0.02}_{-0.02}$ & 139.24 & 127.78 & 43.65 & $43.58^{+0.02}_{-0.02}$ & 44.0 (44.0) & $<-5.811$ \\
MRK 1392 & C1+BAT (31.35/35, 0.645) & $3.80$ & $23.56^{+0.36}_{-0.35}\thinspace(0.57^{+0.16}_{-0.25})$ & $2.02^{+0.13}_{-0.13}$ & 17.94 & 32.62 & 43.13 & $43.19^{+0.29}_{-0.29}$ & 43.8 (43.4$\dagger$) & $-4.560$ \\
2MASX J15064412+0351444 & C1 (14.31/17, 0.645) & $3.73$ & $22.33^{+0.54}_{-0.19}\thinspace(0.89^{+0.11}_{-0.31})$ & $1.50^{+0.67}$ & 4.06 & 36.52 & 42.71 & $43.10^{+0.48}_{-0.49}$ & 43.7 (--) & $<-5.301$ \\
2MASX J15064412+0351444 & S2 (14.38/18, 0.704) & $3.73$ & $22.23^{+0.20}_{-0.14}$ & $1.50^{+0.37}$ & 3.37 & 36.37 & 42.70 & $43.09^{+0.24}_{-0.30}$ & 43.7 (--) & $<-5.298$ \\
NGC 5899 & C2 (645.03/605, 0.126) & $1.80$ & $23.24^{+0.01}_{-0.01}\thinspace(0.99)$ & $1.84^{+0.04}_{-0.04}$ & 0.37 & 52.86 & 42.02 & $42.20^{+0.06}_{-0.06}$ & 42.5 (42.5) & $-4.700$ \\

\hline
\end{tabular}}
\\
Table \ref{table:fitresults}: Basic fit results (continued)
%\caption{\tiny Fit results (XRT objects) (continued).}
%\label{table:fitresults}
\end{table*}
%\end{landscape}
%\restoregeometry
\end{turnpage}
\clearpage

\clearpage
\begin{turnpage}
%\begin{landscape}
%\rotate
%\newgeometry{top=7cm}
\begin{table*}
{\tiny
\begin{tabular}{l|l|l|l|l|l|l|l|l|l|l|l|l|l|l|l}
%{p{2cm}p{1.3cm}p{2.8cm}p{2.7cm}p{2.8cm}p{2.0cm}p{1.5cm}}
%{|l|l|l|l|l|l}
\hline
AGN & Model & $E_{\rm FeK}$ (1) & $EQW_{\rm FeK}$ (2) & $E_{\rm softex}$ (3) & $S_{\rm softex}$ (4) & $L_{\rm BB}$ (5) & $\tau_{\rm [O VII]}$ (6) & $E_{\rm [O VII]}$ (7) & $\tau_{\rm [O VIII]}$ (8) & $E_{\rm [O VIII]}$ (9) \\
\hline
3C 234 & C2 & $6.40*$ & $0.117^{+0.052}_{-0.060}$ &  -- & -- & -- &  -- & -- & -- & --  \\
NGC 3227 & C2 & $6.40^{+0.01}_{-0.01}$ & $0.230^{+0.028}_{-0.010}$ &  -- & -- & -- &  -- & -- & -- & --  \\
Mrk 728 & S3 & $6.36^{+0.06}_{-6.36}$ & $0.201^{+0.070}_{-0.055}$ &  -- & $<0.011$ & -- &  $<0.017$ & -- & -- & --  \\
IC 2637 & S6 & $6.40*$ & $0.256^{+0.146}_{-0.158}$ &  $0.203^{+0.020}_{-0.026}$ & $0.110^{+0.041}_{-0.039}$ & $0.005^{+0.002}_{-0.002}$ &  $<0.043$ & -- & -- & --  \\
PG 1114+445 & S10 & $6.40^{+0.06}_{-0.05}$ & $0.141^{+0.036}_{-0.020}$ &  -- & $<0.007$ & -- &  $2.131^{+0.139}_{-0.127}$ & $0.73$ & $0.534^{+0.100}_{-0.105}$ & $0.87$  \\
1RXS J1127+1909 & S7 & $6.40*$ & $0.076^{+0.044}_{-0.035}$ &  -- & $<0.012$ & -- &  $0.794^{+0.071}_{-0.069}$ & $0.73$ & -- & --  \\
NGC 3758 & S4 & -- & $<0.142$ &  $0.082^{+0.011}_{-0.016}$ & $0.167^{+0.137}_{-0.140}$ & $0.014^{+0.014}_{-0.009}$ &  $<0.032$ & -- & -- & --  \\
Mrk 744 & C2 & $6.48^{+0.08}_{-0.08}$ & $0.135^{+0.139}_{-0.089}$ &  -- & -- & -- &  -- & -- & -- & --  \\
KUG 1141+371 & C2 & $6.40*$ & $0.148^{+0.082}_{-0.111}$ &  -- & -- & -- &  -- & -- & -- & --  \\
KUG 1141+371 & S3 & $6.40*$ & $0.394^{+0.200}_{-0.196}$ &  -- & $<0.021$ & -- &  $<0.100$ & -- & -- & --  \\
NGC 3998 & S2 & -- & $<0.058$ &  -- & $<0.023$ & -- &  $<0.000$ & -- & -- & --  \\
CGCG 041-020 & C1 & -- & $<0.117$ &  -- & -- & -- &  -- & -- & -- & --  \\
NGC 4051 & S8 & $6.30^{+0.04}_{0.00}$ & $0.621^{+0.132}_{-0.103}$ &  $0.122*$ & $0.913^{+0.027}_{-0.026}$ & $0.000^{+0.000}_{-0.000}$ &  $0.103^{+0.022}_{-0.020}$ & $0.73$ & -- & --  \\
PG 1202+281 & S6 & $6.90_{-0.26}$ & $0.324^{+0.186}_{-0.230}$ &  $0.100*$ & $0.184^{+0.023}_{-0.022}$ & $0.356^{+0.046}_{-0.036}$ &  $<0.019$ & -- & -- & --  \\
UGC 7064 & C2 & $6.40*$ & $0.432^{+0.037}_{-0.045}$ &  -- & -- & -- &  -- & -- & -- & --  \\
Mrk 198 & C2 & $6.40*$ & $0.138^{+0.036}_{-0.034}$ &  -- & -- & -- &  -- & -- & -- & --  \\
NGC 4138 & C2 & $6.41*$ & $0.075^{+0.029}_{-0.033}$ &  -- & -- & -- &  -- & -- & -- & --  \\
NGC 4151 & C2 & $6.41^{+0.00}_{-0.00}$ & $0.345^{+0.008}_{-0.014}$ &  -- & -- & -- &  -- & -- & -- & --  \\
NGC 4235 & C2 & $6.42^{+0.07}_{-0.05}$ & $0.303^{+0.067}_{-0.071}$ &  -- & -- & -- &  -- & -- & -- & --  \\
NGC 4235 & S3 & $6.42^{+0.06}_{-0.05}$ & $0.275^{+0.074}_{-0.056}$ &  -- & $<0.011$ & -- &  $<0.080$ & -- & -- & --  \\
Mrk 766 & S6 & $6.30^{+0.01}_{0.00}$ & $0.721^{+0.084}_{-0.066}$ &  $0.088*$ & $0.640^{+0.025}_{-0.026}$ & $0.010^{+0.000}_{-0.000}$ &  $<0.048$ & -- & -- & --  \\
NGC 4258 & C1 & -- & $<0.061$ &  -- & -- & -- &  -- & -- & -- & --  \\
Mrk 50 & S4 & -- & $<0.091$ &  $0.108*$ & $0.212^{+0.029}_{-0.027}$ & $0.027^{+0.003}_{-0.003}$ &  $<0.032$ & -- & -- & --  \\
NGC 4395 & C2 & $6.40*$ & $0.443^{+0.209}_{-0.084}$ &  -- & -- & -- &  -- & -- & -- & --  \\
Ark 374 & S8 & $6.43^{+0.21}_{-6.43}$ & $0.537^{+0.104}_{-0.106}$ &  $0.112*$ & $0.251^{+0.027}_{-0.025}$ & $0.061^{+0.007}_{-0.004}$ &  $0.017^{+0.029}_{-0.017}$ & $0.73$ & -- & --  \\
NGC 4579 & S3 & $6.66^{+0.08}_{-0.08}$ & $1.093^{+0.079}_{-0.109}$ &  -- & $<0.007$ & -- &  $<0.000$ & -- & -- & --  \\
NGC 4593 & S6 & $6.40^{+0.03}_{-0.03}$ & $0.114^{+0.033}_{-0.041}$ &  $0.088*$ & $0.289^{+0.031}_{-0.030}$ & $0.015^{+0.002}_{-0.001}$ &  $<0.045$ & -- & -- & --  \\
SWIFT J1303.9+5345 & C1 & -- & $<0.037$ &  -- & -- & -- &  -- & -- & -- & --  \\
SWIFT J1303.9+5345 & S4 & -- & $<0.068$ &  $0.155^{+0.015}_{-0.017}$ & $0.111^{+0.025}_{-0.023}$ & $0.026^{+0.005}_{-0.005}$ &  $<0.048$ & -- & -- & --  \\
NGC 5033 & S3 & $6.43^{+0.03}_{-0.03}$ & $0.317^{+0.044}_{-0.052}$ &  -- & $<0.006$ & -- &  $<0.000$ & -- & -- & --  \\
NGC 5231 & C2 & $6.40*$ & $0.120^{+0.048}_{-0.043}$ &  -- & -- & -- &  -- & -- & -- & --  \\
NGC 5252 & C2 & $6.40*$ & $0.235^{+0.046}_{-0.033}$ &  -- & -- & -- &  -- & -- & -- & --  \\
Mrk 268 & C2 & $6.38^{+0.04}_{-0.04}$ & $0.176^{+0.052}_{-0.040}$ &  -- & -- & -- &  -- & -- & -- & --  \\
NGC 5273 & S10 & $6.42^{+0.11}_{-0.11}$ & $0.686^{+0.164}_{-0.076}$ &  -- & $<0.007$ & -- &  $1.439^{+0.118}_{-0.109}$ & $0.73$ & $0.769^{+0.108}_{-0.103}$ & $0.87$  \\
Mrk 464 & C1 & -- & $<0.122$ &  -- & -- & -- &  -- & -- & -- & --  \\
NGC 5506 & C2 & $6.42^{+0.05}_{-0.03}$ & $0.140^{+0.012}_{-0.014}$ &  -- & -- & -- &  -- & -- & -- & --  \\
NGC 5548 & S11 & $6.41^{+0.01}_{-0.01}$ & $0.101^{+0.007}_{-0.006}$ &  $0.100*$ & $0.084^{+0.002}_{-0.002}$ & $0.016^{+0.000}_{-0.000}$ &  $0.221^{+0.009}_{-0.009}$ & $0.73$ & $0.103^{+0.009}_{-0.006}$ & $0.87$  \\
H 1419+480 & S7 & $6.40*$ & $0.320^{+0.079}_{-0.069}$ &  -- & $<0.003$ & -- &  $0.480^{+0.052}_{-0.047}$ & $0.73$ & -- & --  \\
Mrk 1383 & S4 & -- & $<0.196$ &  $0.116^{+0.009}_{-0.010}$ & $0.304^{+0.062}_{-0.053}$ & $0.399^{+0.058}_{-0.061}$ &  $<0.059$ & -- & -- & --  \\
Mrk 817 & S6 & $6.40*$ & $0.190^{+0.160}_{-0.184}$ &  $0.100*$ & $0.338^{+0.061}_{-0.058}$ & $0.095^{+0.020}_{-0.009}$ &  $<0.027$ & -- & -- & --  \\
Mrk 841 & S6 & $6.30^{+0.09}_{0.00}$ & $0.562^{+0.209}_{-0.212}$ &  $0.100*$ & $0.357^{+0.028}_{-0.027}$ & $0.118^{+0.009}_{-0.006}$ &  $<0.008$ & -- & -- & --  \\
NGC 5899 & C2 & $6.35^{+0.05}_{-0.06}$ & $0.156^{+0.035}_{-0.034}$ &  -- & -- & -- &  -- & -- & -- & --  \\

\hline
\end{tabular}}
\caption{\label{table:fitresults_features} Fit results - detailed features (iron K-$\alpha$ lines, soft excesses and warm absorber signatures) for objects with $>4600$ counts in the fit spectra}
\tablecomments{\footnotesize All upper limits correspond to no significant detections of the corresponding component, and therefore, those components are not included in the best-fit model. (1) Energy of \textsc{zgauss} component used to model iron K-$\alpha$ line, with * denoting a frozen value, (2) Equivalent width of putative iron K-$\alpha$ line, (3) Energy of \textsc{zbbody} component used to model a soft excess, (4) Strength of soft excess, defined as the luminosity in the \textsc{zbbody} component ($L_{\rm BB}$) divided by the 1.5--6~keV luminosity in the underlying power-law continuum ($L_{\rm 1.5-6 keV}$), i.e., ($L_{\rm BB}/L_{\rm 1.5-6 keV}$), (5) The \textsc{zbbody} component luminosity in units of $10^{44} \rm erg \thinspace s^{-1}$, (6) Optical depth of a putative $\rm[O VII]$ edge (warm absorber signature) near 0.73~keV, (7) Energy of $\rm[O VII]$ edge, frozen unless errors quoted, (8) Optical depth of a putative $\rm[O VIII]$ edge (warm absorber signature) near 0.87~keV, (9) Energy of $\rm[O VIII]$ edge, frozen unless errors quoted.}

\end{table*}
%\end{landscape}
\end{turnpage}
\clearpage

\clearpage
%\begin{turnpage}
%\begin{landscape}
%\rotate
\begin{table*}
{\tiny
\begin{tabular}{l|l|l|l|l|l|l|l|l|l|l|l|l|l|l|l}
%{p{2cm}p{1.3cm}p{2.8cm}p{2.7cm}p{2.8cm}p{2.0cm}p{1.5cm}}
%{|l|l|l|l|l|l}
\hline
AGN & Partial covering? (1) & BAT renormed? (2) & $R$ (3) & $E_{\rm fold}$ (4) & $\Gamma_{\rm \textsc{pexrav}}$ (5) & $\Delta{\Gamma}$ (6) \\
\hline
3C 234 & Y & Y & $<0.58$ & $138^{+1594}_{-85}$ & $2.03^{+0.14}_{-0.13}$ & $-0.17$ \\
NGC 3227 & Y & -- & $12.86^{+3.10}_{-3.14}$ & $>636$ & $2.08^{+0.05}_{-0.09}$ & $0.58$ \\
Mrk 417 & Y & Y & $<0.45$ & $38^{+13}_{-17}$ & $0.75^{+0.08}_{-0.31}$ & $-1.31$ \\
Mrk 728 & -- & -- & $0.07^{+3.10}_{-0.07}$ & $616_{-570}$ & $1.70^{+0.38}_{-0.09}$ & $-0.07$ \\
IC 2637 & -- & Y & $1.09^{+2.38}_{-0.91}$ & $>156$ & $1.79^{+0.24}_{-0.09}$ & $0.10$ \\
PG 1114+445 & -- & -- & $0.95^{+5.81}_{-0.95}$ & $69_{-47}$ & $1.60^{+0.60}_{-0.22}$ & $0.10$ \\
1RXS J1127+1909 & -- & Y & $1.07^{+1.71}_{-0.99}$ & $259_{-168}$ & $1.60^{+0.12}_{-0.09}$ & $0.10$ \\
UGC 06527 & Y & -- & $<16.61$ & $75^{+411}_{-62}$ & $2.12^{+0.72}_{-0.61}$ & $-0.08$ \\
NGC 3758 & -- & Y & $2.87^{+2.34}_{-1.55}$ & $277_{-192}$ & $2.06^{+0.13}_{-0.10}$ & $0.10$ \\
Mrk 744 & Y & -- & $1.62^{+3.33}_{-1.62}$ & $>122$ & $1.71^{+0.27}_{-0.34}$ & $0.21$ \\
KUG 1141+371 & -- & Y & $<0.30$ & $263_{-212}$ & $1.79^{+0.14}_{-0.18}$ & $0.05$ \\
NGC 3998 & -- & -- & $<0.98$ & $622_{-539}$ & $1.83^{+0.12}_{-0.05}$ & $-0.01$ \\
CGCG 041-020 & Y & Y & $15.43^{+4.77}_{-7.73}$ & $>116$ & $2.37^{+0.08}_{-0.30}$ & $0.87$ \\
NGC 4051 & -- & -- & $32.67^{+8.39}_{-6.84}$ & $>381$ & $2.49^{+0.11}_{-0.12}$ & $0.96$ \\
PG 1202+281 & -- & -- & $4.61^{+3.57}_{-2.26}$ & $556_{-492}$ & $2.08^{+0.17}_{-0.16}$ & $0.19$ \\
UGC 7064 & Y & Y & $2.03^{+3.56}_{-1.71}$ & $>326$ & $1.68^{+0.29}_{-0.23}$ & $-0.21$ \\
NGC 4102 & Y & Y & $<0.37$ & $>264$ & $2.13^{+0.09}_{-0.07}$ & $-0.07$ \\
Mrk 198 & Y & Y & $0.51^{+0.65}_{-0.37}$ & $91^{+57}_{-32}$ & $1.62^{+0.32}_{-0.07}$ & $0.01$ \\
NGC 4138 & Y & -- & $0.05^{+2.85}_{-0.05}$ & $148_{-73}$ & $1.51^{+0.34}_{-0.20}$ & $-0.10$ \\
NGC 4151 & Y & -- & $<0.01$ & $79^{+4}_{-4}$ & $1.31^{+0.03}_{-0.03}$ & $-0.89$ \\
KUG 1208+386 & Y & Y & $<0.83$ & $44^{+51}_{-16}$ & $1.30^{+0.34}_{-0.07}$ & $-0.42$ \\
NGC 4235 & -- & -- & $0.06^{+2.96}_{-0.06}$ & $89^{+188}_{-35}$ & $1.49^{+0.20}_{-0.07}$ & $-0.08$ \\
NGC 4235 & Y & -- & $0.95^{+3.22}_{-0.95}$ & $64^{+102}_{-30}$ & $1.53^{+0.19}_{-0.13}$ & $-0.10$ \\
Mrk 766 & -- & Y & $8.12^{+3.57}_{-2.38}$ & $21^{+7}_{-6}$ & $1.56^{+0.09}_{-0.08}$ & $0.06$ \\
NGC 4258 & Y & -- & $0.43^{+2.33}_{-0.43}$ & $>284$ & $1.82^{+0.06}_{-0.10}$ & $-0.32$ \\
Mrk 50 & -- & -- & $4.79^{+0.96}_{-0.91}$ & $>334$ & $2.18^{+0.02}_{-0.02}$ & $0.23$ \\
NGC 4388 & Y & -- & $0.22^{+0.21}_{-0.22}$ & $>2096$ & $1.81^{+0.03}_{-0.05}$ & $-0.25$ \\
NGC 4395 & Y & -- & $<0.71$ & $45^{+27}_{-10}$ & $1.18^{+0.11}_{-0.12}$ & $-0.32$ \\
Ark 374 & -- & Y & $1.12^{+2.62}_{-0.85}$ & $542_{-478}$ & $2.10^{+0.13}_{-0.08}$ & $-0.04$ \\
NGC 4579 & -- & -- & $3.70^{+4.28}_{-2.44}$ & $>61$ & $2.02^{+0.14}_{-0.10}$ & $0.01$ \\
NGC 4593 & -- & -- & $0.90^{+0.82}_{-0.65}$ & $>517$ & $1.89^{+0.08}_{-0.05}$ & $0.06$ \\
NGC 4686 & Y & Y & $<0.37$ & $>230$ & $1.86^{+0.13}_{-0.08}$ & $0.10$ \\
2MASX J13000533+1632151 & -- & -- & $<416.96$ & $>104$ & $1.67^{+0.15}_{-0.33}$ & $0.17$ \\
2MASX J13000533+1632151 & Y & -- & $<0.00$ & $>116$ & $1.70^{+0.35}_{-0.35}$ & $0.20$ \\
SWIFT J1303.9+5345 & -- & Y & $0.25^{+0.40}_{-0.25}$ & $197^{+350}_{-82}$ & $1.73^{+0.05}_{-0.04}$ & $0.02$ \\
SWIFT J1303.9+5345 & Y & Y & $<0.00$ & $275_{-150}$ & $1.81^{+0.07}_{-0.07}$ & $-0.04$ \\
SWIFT J1309.2+1139 & Y & Y & $3.81^{+6.92}_{-1.85}$ & $301_{-185}$ & $1.53^{+0.36}_{-0.26}$ & $-0.38$ \\
NGC 5033 & -- & -- & $3.85^{+3.70}_{-2.49}$ & $>33$ & $1.90^{+0.15}_{-0.15}$ & $0.20$ \\
NGC 5231 & Y & Y & $0.45^{+0.48}_{-0.45}$ & $>195$ & $1.90^{+0.05}_{-0.15}$ & $0.22$ \\
NGC 5252 & Y & -- & $<0.49$ & $111^{+58}_{-18}$ & $1.38^{+0.09}_{-0.05}$ & $-0.12$ \\
Mrk 268 & Y & Y & $7.42^{+3.54}_{-4.02}$ & $>379$ & $1.84^{+0.12}_{-0.15}$ & $-0.04$ \\
NGC 5273 & -- & -- & $<2.62$ & $279_{-227}$ & $1.37^{+0.17}_{-0.05}$ & $-0.13$ \\
UM 614 & -- & Y & $7.66^{+3.85}_{-2.18}$ & $>323$ & $1.63^{+0.11}_{-0.19}$ & $0.13$ \\
Mrk 464 & Y & -- & $3.52^{+8.16}_{-3.52}$ & $170_{-131}$ & $1.80^{+0.34}_{-0.26}$ & $0.08$ \\
Mrk 463 & Y & -- & $7.52^{+31.68}_{-7.29}$ & $198_{-153}$ & $1.80^{+0.44}_{-0.55}$ & $-0.40$ \\
NGC 5506 & Y & -- & $1.24^{+0.70}_{-0.24}$ & $166^{+107}_{-30}$ & $1.85^{+0.02}_{-0.10}$ & $0.16$ \\
NGC 5548 & -- & -- & $0.99^{+0.24}_{-0.23}$ & $415^{+827}_{-178}$ & $1.73^{+0.02}_{-0.02}$ & $0.03$ \\
H 1419+480 & -- & -- & $0.87^{+1.38}_{-0.87}$ & $>152$ & $1.84^{+0.08}_{-0.10}$ & $0.01$ \\
Mrk 1383 & -- & -- & $2.23^{+4.56}_{-2.23}$ & $>134$ & $2.20^{+0.24}_{-0.20}$ & $0.14$ \\
Mrk 817 & -- & Y & $<0.10$ & $>150$ & $2.37^{+0.04}_{-0.09}$ & $0.28$ \\
Mrk 841 & -- & -- & $4.12^{+2.21}_{-1.63}$ & $>597$ & $2.26^{+0.10}_{-0.09}$ & $0.13$ \\
NGC 5899 & Y & -- & $2.22^{+3.21}_{-2.00}$ & $>210$ & $2.05^{+0.22}_{-0.18}$ & $0.20$ \\

\hline
\end{tabular}}
\caption{\label{table:reflection} Reflection fits}
\tablecomments{\small Reflection fit results for objects with {\xmm} data, fit in conjunction with BAT data.  We fit an absorbed \textsc{pexrav} model to all the objects. We inspect the best-fit model from basic fits to 0.4--10~keV data (Table \ref{table:fitresults}) to select whether partial covering absorption or standard absorption should be used, indicated in column (1).  (2) - if the {\xmm} spectrum was taken during the time of assembly of the BAT catalog, the BAT spectrum was renormalized as detailed in \S\ref{subsec:batrenormsection}; if this was done, this is indicated by a 'Y' in this column. (3) - the \textsc{pexrav} reflection fraction $R$.  If below 0.01, we only present the upper limit and assume that the reflection fraction is zero. (4) - The \textsc{pexrav} fold energy in~keV. All values above 5000~keV are presented as 5000~keV (well outside the BAT band) along with the lower limit determined from the negative error bar. (5) - the photon index $\Gamma_{\textsc{pexrav}}$ from \textsc{pexrav}, with no limits imposed. (6) - the difference between the \textsc{pexrav} $\Gamma$ and that from the 0.4--10~keV basic fits presented in Table \ref{table:fitresults}.}

\end{table*}
%\end{turnpage}
\clearpage

\clearpage
\begin{turnpage}
%\newgeometry{top=9cm}
\begin{table*}
{\footnotesize
\begin{tabular}{l|l|l|l|l|l|l|l|l|l|l|l|l|l|l|l|l}
%{p{2cm}p{1.3cm}p{2.8cm}p{2.7cm}p{2.8cm}p{2.0cm}p{1.5cm}}
%{|l|l|l|l|l|l}
\hline
Catalog & Flux limit & Completeness limit & Ambiguous sources & $\rm log(N_{\rm H }>22)$ &  $\rm log(N_{\rm H }>23)$  &  $\rm log(N_{\rm H }>24.15)$ & Simple & Complex \\
 (1) & (2) & (3) & (4)  & (5) & (6) & (C-thick) (7) & abs. ($\langle \rm log \thinspace N_{\rm H } \rangle$,$\sigma$) (8) & abs. ($\langle \rm log \thinspace N_{\rm H }\rangle$,$\sigma$) (9) \\
\hline
9-month	& -10.70 & -11.0 & 0\% & 55\% & 33\% & 0\% ($<6\%$)	& 45\% (20.58,0.74) & 55\% (23.03,0.71) \\
22-month & -10.96 & -11.25	& 10\%	& 59-64	\% & 49--54\% & 5\% ($<18\%$) & 36--46\% (20.47--20.56, 0.86--0.90) & 54--64\% (23.28--23.4, 0.57--0.68) \\
58-month & -11.40 & -11.6 & 13\% $\dagger$ & 57--61\%& 41--45\% & 9\% ($<15\%$) & 38--50\% (20.67--20.80, 1.12--1.18) & 43--56\% (23.27--23.55,0.71--0.95) \\
\hline
\end{tabular}}
\caption{\label{table:cataloguecompare} Comparing BAT catalogs: flux limits, completeness and absorption properties}
\tablecomments{(1) - Catalog, (2) - Logarithm of BAT flux limit (14--195~keV) in $\rm erg \thinspace cm^{-2} \thinspace s^{-1}$, (3) - Completeness limit, given as log(S) for 2--10~keV flux S in units of $\rm erg \thinspace cm^{-2} \thinspace s^{-1}$, (4) - percentage of sources with ambiguous spectral types, (5) - percentage of sources with $\rm log(N_{\rm H})>22$, (6) - percentage of sources with $\rm log(N_{\rm H})>23$, (7) - percentage of Compton-thick sources, with $\rm log(N_{\rm H})>24.15$ (upper limits are based on consideration of the other Compton-thickness metrics discussed in \S\ref{subsec:comptonthick}), (8) - percentage of simple absorption sources, with average column density and standard deviation (9) - percentage of complex absorption sources, with average column density and standard deviation.  Ranges in these values are due to sources with ambiguous spectral types. $\dagger$ - An additional 5\% of our 58-month sources do not have enough counts to construct a spectrum, so these are not classified into any of the categories shown here.}
\end{table*}
%\end{landscape}
%\restoregeometry

%\newgeometry{top=7cm}
\begin{table*}
{\small
\begin{tabular}{l|l|l|l|l|l|l|l|l|l|l|l|l|l|l|l|l}
%{p{2cm}p{1.3cm}p{2.8cm}p{2.7cm}p{2.8cm}p{2.0cm}p{1.5cm}}
%{|l|l|l|l|l|l}
\hline
Catalog & $\langle \rm log \thinspace L_{\rm 2-10 keV} \thinspace \rangle$,$\sigma$ & $\langle \rm log \thinspace L_{\rm 2-10 keV} \thinspace \rangle$,$\sigma$ & $\langle \rm log \thinspace L_{\rm 2-10 keV}\thinspace \rangle$,$\sigma$ & Fe K-$\alpha$ & Soft excess & Hidden/buried\\
(1) & (all) (2) & ($\rm log \thinspace N_{\rm H} \thinspace <22$) (3) & ($\rm log \thinspace N_{\rm H} \thinspace >22$) (4) & \% (5) & \% (6) & \% (7) \\
\hline
9-month	& 43.01 (0.87)	&	43.42 (0.79)	&		42.67 (0.78) & 81\% & 41\% & 24\% \\
22-month &	42.70 (0.93)	&	42.80--42.84 (0.90-0.95)   &	42.60--42.65 (0.93--0.95) & 75\% & 32--36\% & 28\% \\
58-month &	43.00 (0.91-0.92)  &	43.02--43.07 (0.96-0.98)   &	42.91--42.97 (0.89) & 79\% & 31--33\% & 13--14\% \\
\hline
\end{tabular}}
\caption{\label{table:cataloguecompare_2} Comparing BAT catalogs: luminosity and prevalence of X-ray features}
\tablecomments{(1) - Catalog, (2) - logarithm of average intrinsic 2--10~keV luminosity in units of $\rm erg \thinspace s^{-1}$, (3) - logarithm of intrinsic 2--10~keV luminosity for sources with $\rm log \thinspace N_{\rm H}<22$ in units of $\rm erg \thinspace s^{-1}$, (4) - logarithm of intrinsic 2--10~keV luminosity for sources with $\rm log \thinspace N_{\rm H}>22$ in units of $\rm erg \thinspace s^{-1}$, (5) - percentage of sources with significant detections of iron lines (for 22-month and 58-month percentages, we only use sources with $>4600$ counts in their spectra), (6) - percentage of sources with soft excesses (for 22-month and 58-month percentages, we only use sources with $>4600$ counts in their spectra), (7) - percentage of hidden/buried sources.  Standard deviations are quoted for all log(L) values in this table, and ranges in the averages or standard deviations represent uncertainties due to ambiguous spectral types.}

\end{table*}
%\end{landscape}
%\restoregeometry
\end{turnpage}
\clearpage

%\begin{landscape}
%\rotate
%\newgeometry{top=7cm}
\clearpage
\begin{turnpage}
\begin{table*}
{\footnotesize
\begin{tabular}{l|l|l|l|l|l|l|l|l|l|l|l|l|l|l|l|l}
%{p{2cm}p{1.3cm}p{2.8cm}p{2.7cm}p{2.8cm}p{2.0cm}p{1.5cm}}
%{|l|l|l|l|l|l}
\hline
AGN & $N_{\rm H}^{\rm Gal}$ & $\rm F_{0.5-2keV}$ & $\rm F_{0.5-2keV}^{\rm int}*$ & $\rm F_{2-10keV}$ & $\rm F_{2-10keV}^{\rm int}*$ & $\rm F_{0.5-2keV}^{extrap.}\dagger$ & log($L_{\rm 0.5-2keV}^{\rm obs}$) & log($L_{\rm 2-10keV}^{\rm obs}$) & $L_{\rm 14-195 keV}$ & $\rm log (N^{\rm (est)}_{\rm H})$\\
 & (1) & (2) & (3) & (4)  & (5) & (6) & (7) & (8) & (9) & (10) \\
\hline
%MCG +09-17-074 & 1.310 & $<0.176$ & $<0.183$ & $<0.345$ & $<0.345$ & $<0.246$ & $<40.36$ & $<40.63$ & -- (350'' away) & --\\ 
%SDSS-C4-DR3 3043 & 0.787 & $<0.784$ & $<0.803$ & $<1.89$ & $<1.89$ & $<1.37$ & $<41.96$ & $<42.33$ & 44.0 (207'' away) & \\ 
%NGC 3718 & 0.999 & $<0.656$ & $<0.677$ & $<2.1$ & $<2.11$ & $<1.51$ & $<39.20$ & $<39.70$ & 41.3 (1541'' away) & \\ 
NGC 4180 & 1.390 & $0.256$ & $0.267$ & $<0.842$ & $<0.843$ & $<0.599$ & $39.45$ & $<39.95$ & 42.2 & $\sim 22.1$\\ 
NGC 4500 & 0.830 & $0.285$ & $0.292$ & $2.02$ & $2.03$ & $1.46$ & $39.83$ & $40.68$ & 42.3 & $\sim 22.9$ \\ 
MCG -01-33-063 & 2.800 & $<0.168$ & $<0.183$ & $3.01$ & $3.01$ & $2.05$ & $<40.45$ & $41.66$ & 43.2 & $\sim 22.9 \dagger \dagger$\\ 
%attempted fit 2E 1139.7+1040 & 3.720 & $<2.92$ & $<3.27$ & $<4.24$ & $<4.25$ & $<2.82$ & $<43.29$ & $<43.40$ &\\ 
%attempted fit 2MASX J10081910+3729039 & 1.310 & $<0.79$ & $<0.822$ & $<14.7$ & $<14.7$ & $<10.5$ & $<41.71$ & $<42.97$ & \\ 
%no bat data 2MASX J10085494+3739297 & 1.310 & $<7.08$ & $<7.37$ & $<12.3$ & $<12.3$ & $<8.73$ & $<42.69$ & $<42.91$ &  \\ 
%attempted fit B2 1210+33 & 1.080 & $<1.63$ & $<1.69$ & $<2.79$ & $<2.8$ & $<2$ & $<45.94$ & $<46.16$ & \\ 
CGCG 102-048 & 1.680 & $<0.261$ & $<0.275$ & $3.18$ & $3.19$ & $2.24$ & $<40.65$ & $41.71$ & 43.5 & $\sim 23.2$\\ 
%No bat data IC 2515 & 1.440 & $<0.142$ & $<0.148$ & $<6.32$ & $<6.33$ & $<4.49$ & $<40.09$ & $<41.72$ &\\ 
%Attempted fit MCG -01-30-041 & 2.160 & $<0.762$ & $<0.814$ & $<3.98$ & $<3.99$ & $<2.77$ & $<40.80$ & $<41.49$& \\ 
%Mrk 477 & 1.060 & $<0.206$ & $<0.213$ & $<0.743$ & $<0.744$ & $<0.534$ & $<40.84$ & $<41.38$ \\ 
2MASX J13542913+1328068 & 1.730 & $<0.555$ & $<0.585$ & $7.05$ & $7.06$ & $4.96$ & $<41.74$ & $42.82$ & 44.0 & $\sim 23.4 \dagger \dagger$ \\ 

\hline
\end{tabular}}
\caption{\label{table:upperlimits} XRT observations with few counts}
\tablecomments{\footnotesize Fluxes and luminosities for XRT observations with too few counts to construct a spectrum.  (1) Galactic absorption in units of $\rm 10^{20} cm^{-2}$. (2),(3),(4),(5) Fluxes in 0.5--2~keV and 2--10~keV bands. Full fluxes are provided if there is a $2 \sigma$ detection of the source in the source region; otherwise we present 95 per cent upper limits (see e.g., \citealt{1986ApJ...303..336G}). Fluxes determined from count-rates using \textsc{WebPIMMS} assuming Galactic absorption and a power-law with index $\Gamma=1.9$. All fluxes are in units of $10^{-13} \rm erg \thinspace s^{-1} \thinspace cm^{-2}$. * These fluxes have been corrected for Galactic absorption. $\dagger$ This is the expected observed 0.5--2~keV flux extrapolated from the observed 2--10~keV count rate using \textsc{WebPIMMS} assuming a power-law index of $\Gamma=1.9$.  (7), (8) Logarithm of luminosities in the 0.5--2~keV 2--10~keV, corrected for absorption, in units of $\rm erg \thinspace s^{-1}$, with upper limits provided for $< 2 \sigma$ detections. (9) log of BAT luminosity in $\rm erg \thinspace s^{-1}$. (10) Estimated column density; we omit errors since systematic errors dominate for these objects with few counts $\dagger \dagger$ BAT renormalisation was not possible for these sources, giving greater uncertainty on the column density estimates.}
 %(9) This is the logarithm of the estimated intrinsic absorption (in $\rm cm^{-2}$) required on such a power-law to reproduce the \emph{observed} 0.5--2keV flux, given the observed 2--10 keV count-rate. ** For NGC 4500, inclusion of the BAT data with a very poor-counts XRT spectrum suggests $N_{\rm H}\sim10^{24} \rm cm^{-2}$.\label{table:upperlimitresults}}
\end{table*}
%\end{landscape}
%\restoregeometry
\end{turnpage}
\clearpage

\end{document}